\documentclass[fleqn,usenatbib]{mnras}

\usepackage{newtxtext,newtxmath}
\usepackage{bm} % For bold italic mathematical expressions
\usepackage[T1]{fontenc}

\DeclareRobustCommand{\VAN}[3]{#2}
\let\VANthebibliography\thebibliography
\def\thebibliography{\DeclareRobustCommand{\VAN}[3]{##3}\VANthebibliography}

\usepackage{graphicx}	% Including figure files
\usepackage{amsmath}	% Advanced maths commands
\usepackage{xcolor} % Coloured text for marking revisions
\usepackage{soul} % Strikethrough for marking revisiions
\title[Centaurus A satellite classifications]{Classifying the satellite plane membership of Centaurus A's dwarf galaxies using orbital alignment constraints}

\author[K. J. Kanehisa et al.]{
Kosuke Jamie Kanehisa$^{1,2,3}$\thanks{E-mail: kkanehisa@aip.de},
Marcel S. Pawlowski$^{1}$,
Oliver M\"{u}ller$^{4,5}$,
Sangmo Tony Sohn$^{6}$
\\
$^{1}$Leibniz-Institut für Astrophysik Potsdam (AIP), An der Sternwarte 16, 14482 Potsdam, Germany\\
$^{2}$Institut für Physik und Astronomie, Universität Potsdam, Karl-Liebknecht-Straße 24/25, D-14476 Potsdam, Germany\\
$^{3}$Department of Physics, University of Surrey, Guildford GU2 7XH, UK\\
$^{4}$Google Switzerland, 8002 Zurich, Switzerland\\
$^{5}$Institute of Physics, Laboratory of Astrophysics, Ecole Polytechnique Fédérale de Lausanne (EPFL), 1290 Sauverny, Switzerland\\
$^{6}$Space Telescope Science Institute, 3700 San Martin Drive, Baltimore, MD 21218, USA
}

\date{Accepted XXX. Received YYY; in original form ZZZ}

\pubyear{2022}

\begin{document}
\label{firstpage}
\pagerange{\pageref{firstpage}--\pageref{lastpage}}
\maketitle

\begin{abstract}
The flattened, possibly co-rotating plane of satellite galaxies around Centaurus A, if more than a fortuitous alignment, adds to the pre-existing tension between the well-studied Milky Way and M31 planes and the $\Lambda$CDM model of structure formation.
It was recently reported that the Centaurus A satellite plane (CASP) may be rotationally supported, but a further understanding of the system's kinematics is elusive in the absence of full three-dimensional velocities. 
We constrain the transverse velocities of 27 satellites that would rotationally stabilise the Centaurus A plane, and classify the satellites by whether their possible orbits are consistent with the CASP.
Five satellites are identified to be unlikely to participate in the plane, two of which are clearly non-members.
Despite their previously reported line-of-sight velocity trend suggestive of a common co-rotating motion, 17 out of 22 potential CASP members are consistent with either orbital direction within both the full range of possible kinematics as well as when limiting orbits to those within the plane.
On the other hand, disregarding the 5 off-plane satellites found to be inconsistent with CASP membership enhances the significance of the CASP's line-of-sight velocity trend fivefold.
Our results are robust with different mass estimates of the Centaurus A halo, and the adoption of either spherical or triaxial NFW potentials.
\end{abstract}

\begin{keywords}
galaxies: dwarf -- galaxies: kinematics and dynamics -- galaxies: groups: individual: NGC5128 -- proper motions
\end{keywords}

%%%%%%%%%%%%%%%%%%%%%%%%%%%%%%%%%%%%%%%%%%%%%%%%%%

%%%%%%%%%%%%%%%%% BODY OF PAPER %%%%%%%%%%%%%%%%%%

\section{Introduction}
\label{sec:s1}

In the mid-1970s, \citet{Lynden-Bell1976mw} and \citet{Kunkel1976mw} made a curious discovery -- the 6 then-known dwarf satellite galaxies of the Milky Way, combined with several globular clusters, formed a thin, planar distribution intersecting with the Galactic centre. Since then, many new confirmed satellites were also found to align with this Vast Polar Structure (VPOS) \citep[e.g.][]{Kroupa2005mw, Metz2009mw, Pawlowski2015predict, Pawlowski2016mw} as well as a notable fraction of globular clusters and streams (\citealt{Pawlowski2012vpos, Pawlowski2014vpos}, though see \citealt{Riley2020vpos}). Proper motions of the most luminous satellites further revealed that the majority of the VPOS members were co-orbiting around the Milky Way along its defined plane \citep{Pawlowski2017gaia, Fritz2018gaia, Pawlowski2020gaia, Li2021gaia}.

Such highly correlated satellite distributions are not unique to the Milky Way. More recently, a similar structure was discovered around M31 \citep{McConnachie2006m31, Koch2006m31, Metz2007m31}. 15 out of 27 satellites form a thin disc a mere 13 kpc in height, now referred to as the Great Plane of Andromeda \citep[GPoA;][]{Pawlowski2018review}. While proper motion measurements are lacking for most of its constituents, the GPoA's fortuitously edge-on view from the Sun permits an alternative approach -- the line-of-sight velocities of its member satellites revealed a velocity trend indicative of a common co-rotational motion \citep{Ibata2013m31}. Among the three M31 satellites with proper motions, two demonstrate orbital poles consistent with the GPoA \citep{Pawlowski2021m31, Brunthaler2007m31}.

The presence of satellite planes around both major host galaxies in the Local Group is peculiar at best in $\Lambda$CDM cosmology, wherein the chaotic hierarchical clustering of dark haloes is thought to produce near-isotropic satellite galaxy systems \citep{Kroupa2005mw}. When straightforwardly comparing the observed VPOS and GPoA with CDM simulations, analogs which simultaneously match their high degree of flattening and kinematic correlation are rare \citep{Pawlowski2014conflict, Ibata2014shadows, Shao2019eagle, Pawlowski2020gaia, Samuel2021fire}.

Despite their apparent dearth of simulated counterparts, many argue that the two Local Group planes are simply a result of a shared evolutionary history. An ancient high-mass ratio merger of M31 has been raised as a possible origin of both the VPOS and GPoA \citep{Fouquet2012tidal, Hammer2013tidal}, while the evacuation of the Local Void may contribute to the formation of such flattened structures throughout the Local Volume \citep{Libeskind2019void}.

Moreover, a notable degree of anisotropy is inherent in satellite distributions within a CDM universe -- preferential directions of satellite infall are defined by the velocity shear tensor and hence the orientation of local filaments \citep{Libeskind2014accretion}, as well as the triaxial mass distributions of the host galaxies' dark haloes \citep{Zentner2005anisotropic}. In addition, luminous satellites do not trace the full distribution of the underlying substructure, but instead descendants of the most massive subhaloes upon infall \citep{Libeskind2005subhalo}.

And yet, these anisotropic accretion processes are self-consistently included in modern CDM simulations, and it is unclear whether the numerous proposed solutions to the so-called planes-of-satellites problem (see \citealt{Pawlowski2018review} for a recent review) can fully account for the VPOS and GPoA's high degree of phase-space correlation.

Further insight into this debate may be gained by studying satellite systems beyond the Local Group, whose individual evolutionary histories are not so closely interlinked with the LG planes. In the Local Volume ($D < 10$ Mpc), there are indications that M81 \citep{Chiboucas2013m81}, M83 \citep{Muller2018m83} and M101 \citep{Muller2017m101} may also host flattened dwarf distributions. In addition, \citet{Ibata2014sdss, Ibata2015sdss} reported an excess of satellite pairs with anti-correlated velocities indicative of co-rotation in the local Universe (though whether this detection reflects an underlying ubiquity of stable satellite planes is a topic of contention, see \citealt{Phillips2015sdss, Cautun2015sdss}).

Our work focuses on the Centaurus group, one of the best-studied congregations of galaxies beyond the Local Group. It is hosted by Centaurus A (NGC5128), a massive elliptical galaxy at a distance of 3.68 Mpc (\citealt{Tully2009distance}, though see \citealt{Harris2010distance} for an alternative estimate of 3.8 Mpc). The system is thought to lie within a dark halo nearly an order of magnitude more massive than that of the Milky Way or M31, with a virial mass of $M_{\mathrm{vir}} = 8 \times 10^{12} M_{\odot}$ and a corresponding radius of $R_{\mathrm{vir}} = 409$ kpc \citep{Tully2015groups}.

\citet{Tully2015planes} initially reported that of the 29 then-known satellites of Centaurus A, all but 2 dwarfs appeared to be distributed along two parallel planes. Follow-up work by \citet{Muller2016testing} verified their conclusions, but found that with the addition of 9 newly confirmed satellites, a single, thicker plane reminiscent of the VPOS and GPoA became statistically favourable. This single-plane model, now generally accepted in lieu of the previous interpretation, is referred to as the Centaurus A satellite plane \citep[CASP;][]{Pawlowski2018review}.

Much like the GPoA, the CASP is oriented nearly edge-on to the heliocentric line-of-sight \citep{Muller2019galaxy}. Satellites for which accurate radial velocities are available demonstrate a projected velocity trend indicative of co-rotation -- satellites to the south of the CASP's on-sky minor axis approach, while those to the north recede \citep{Muller2018whirling, Muller2021coherent}. Furthermore, recent estimates of the satellite distribution's circular velocity about the CASP's minor axis indicate that the plane may be rotationally supported \citep{Muller2021mass} -- and thus, a long-lived structure instead of an ephemeral alignment.

In CDM simulations, satellite systems that match the CASP's degree of flattening and co-rotation are rare, only hosted by $0.2$ per cent of Centaurus A analogs \citep{Muller2021coherent}. A significant fraction of this tension arises from the CASP's apparent kinematic coherence. However, previous works \citep{Muller2018whirling, Muller2021coherent} only consider the projected phase-space distribution of Centaurus A's satellites due to the lack of proper motions beyond the Local Group.

In the absence of direct measurements, it is possible to predict the proper motions of individual satellites under the assumption that the planar distribution they form is a long-lived feature. This approach was first taken by \citet{Pawlowski2013predict}, who predicted proper motions of the less-luminous satellite galaxies of the Milky Way for which proper motions were not already available. These predictions were updated in \citet{Pawlowski2014predict} and \citet{Pawlowski2015predict} with the discovery and confirmation of additional satellites around the Milky Way, and compared to Gaia DR2 \citep{GaiaDR2} proper motions in \citet{Fritz2018gaia}.

Recent work by \citet{Hodkinson2019m31} extended this analysis to the GPoA, wherein proper motions were only available for two objects -- the central M31 and the satellite galaxy IC10 (a sample now including NGC147 and NGC185 with \citealt{Sohn2020m31}). They identified the range of proper motions (with the assumption of a long-lived plane) for 19 plausibly on-plane satellites as reported in \citet{Ibata2013m31}, integrating satellite orbits forward over their next 10 periapses.

To be recognized as orbiting within the GPoA, each satellite was required to remain within a maximum distance from M31 (as to avoid capture by the Milky Way's potential), a maximum orthogonal separation from the satellite plane, and maintain a strong alignment between its angular momentum and the GPoA's minor axis. Upon comparing their predictions with the measured proper motion of IC10, \citet{Hodkinson2019m31} ruled out the satellite as a GPoA member, reporting that no proper motion within the observational uncertainties could permit an orbit aligned with the GPoA over an extended duration.

As a result of their positions beyond the Local Group, Centaurus A and its surrounding satellite galaxies are unlikely to obtain reliable proper motion measurements in the near future. None the less, transverse velocity (TV) constraints may still offer insights into the system's dynamical properties in the absence of observations. For instance, the CASP's on-sky velocity trend suggestive of co-rotation is the dominant contributor to the rarity of simulated CDM analogs (21 out of 28 satellites appear to co-orbit in the same sense, see \citealt{Muller2021coherent}), but it is unclear whether this necessarily corresponds to a excess of prograde orbits in phase-space. In addition, the presence (or lack thereof) of a highly correlated subsample of satellites around Centaurus A -- much like M31's clear distinction between on-plane and off-plane satellites -- has not yet been demonstrated.

In this work, we integrate the possible orbits of constituents of the Centaurus A system, consisting here of 27 satellite galaxies with available TRGB distances and radial velocities published in \citet{Muller2021coherent}. Under the assumption that the satellites follow bound and long-lived orbits, we constrain their range of transverse velocities (in proper motion units) that produce a rotationally supported plane-of-satellites. The fraction of TVs that results in CASP-aligned orbits serves as a rough tracer of the satellites' likelihood of orbiting within the plane -- in this manner, we identify a sample of satellites that are inconsistent with participating in the CASP. In addition, we determine the likely three-dimensional orbital sense of each satellite to constrain the number of co-rotating CASP members.

\section{Method}
\label{sec:s2}

\subsection{Observational data}
\label{sec:s2_data}

\begin{table}
	\centering
	\caption{Data from \citet{Muller2021coherent} for all Centaurus A satellites with both TRGB distances and velocity measurements. Centaurus A is marked in bold on the first row. On-sky positions are given in equatorial coordinates, heliocentric distances are shown with corresponding errors, and heliocentric recession velocities are given along the line-of-sight.}
	\begin{tabular}{lllll}
	    \hline
		Name & $\alpha$ ($^\circ$) & $\delta$ ($^\circ$) & $D$ (Mpc) & $v_{\mathrm{rad}}$ (km s$^{-1}$) \\
		\hline
        \textbf{Cen A} & 201.3667 & -43.0167 & $3.68 \pm 0.05$ & $556 \pm 10$\\
        ESO269-037 & 195.8875 & -46.5842 & $3.15 \pm 0.09$ & $744 \pm 2$\\
        NGC4945 & 196.3583 & -49.4711 & $3.72 \pm 0.03$ & $563 \pm 3$\\
        ESO269-058 & 197.6333 & -46.9908 & $3.75 \pm 0.02$ & $400 \pm 18$\\
        KK189 & 198.1883 & -41.8320 & $4.21 \pm 0.17$ & $752.0 \pm 6.8$\\
        ESO269-066 & 198.2875 & -44.8900 & $3.75 \pm 0.03$ & $784 \pm 31$\\
        NGC5011C & 198.2958 & -43.2656 & $3.73 \pm 0.03$ & $647 \pm 96$\\
        KKs54 & 200.3829 & -31.8864 & $3.75 \pm 0.10$ & $622.1 \pm 11.4$\\
        KK196 & 200.4458 & -45.0633 & $3.96 \pm 0.11$ & $741 \pm 15$\\
        NGC5102 & 200.4875 & -36.6297 & $3.74 \pm 0.39$ & $464 \pm 18$\\
        KK197 & 200.5086 & -42.5359 & $3.84 \pm 0.04$ & $638.8 \pm 4.0$\\
        KKs55 & 200.5500 & -42.7311 & $3.85 \pm 0.07$ & $550.2 \pm 7.0$\\
        dw1322-39 & 200.6336 & -39.9060 & $2.95 \pm 0.05$ & $656.3 \pm 9.7$\\
        dw1323-40b & 200.9809 & -40.8361 & $3.91 \pm 0.61$ & $497.0 \pm 12.4$\\
        dw1323-40a & 201.2233 & -40.7612 & $3.73 \pm 0.15$ & $450.0 \pm 14.2$\\
        KK203 & 201.8681 & -45.3524 & $3.78 \pm 0.25$ & $409.0 \pm 18.2$\\
        ESO324-024 & 201.9042 & -41.4806 & $3.78 \pm 0.09$ & $514 \pm 18$\\
        NGC5206 & 203.4292 & -48.1511 & $3.21 \pm 0.01$ & $583 \pm 6$\\
        NGC5237 & 204.4083 & -42.8475 & $3.33 \pm 0.02$ & $361 \pm 4$\\
        NGC5253 & 204.9792 & -31.6400 & $3.55 \pm 0.03$ & $407 \pm 3$\\
        dw1341-43 & 205.4032 & -43.8553 & $3.53 \pm 0.04$ & $634.1 \pm 12.3$\\
        KKs57 & 205.4079 & -42.5797 & $3.84 \pm 0.47$ & $511.3 \pm 16.8$\\
        KK211 & 205.5208 & -45.2050 & $3.68 \pm 0.14$ & $600 \pm 31$\\
        dw1342-43 & 205.6837 & -43.2548 & $2.90 \pm 0.13$ & $505.1 \pm 7.5$\\
        ESO325-011 & 206.2500 & -41.8589 & $3.40 \pm 0.05$ & $544 \pm 1$\\
        KKs58 & 206.5031 & -36.3289 & $3.36 \pm 0.1$ & $476.5 \pm 5.2$\\
        KK221 & 207.1917 & -46.9974 & $3.82 \pm 0.07$ & $507 \pm 13$\\
        ESO383-087 & 207.3250 & -36.0614 & $3.19 \pm 0.03$ & $326 \pm 2$\\
		\hline
	\end{tabular}
	\label{tab:s2_data}
\end{table}

At time of writing, Centaurus A has 42 confirmed satellites and an additional 30 awaiting confirmation \citep{Muller2019galaxy}. However, radial velocities were only available for 16 satellites \citep{Muller2018whirling} until recently. In this work, we use new velocity measurements for 12 additional dwarfs published in \citet{Muller2021properties} for a total of 28 confirmed satellites with radial velocities. Among these, KKs59 lacks a corresponding TGRB distance, and is thus disregarded -- producing a working sample of 27 satellites with accurate distance and radial velocity measurements presented in Table~\ref{tab:s2_data}.

Four phase-space coordinates are known for each satellite -- namely, its projected position in equatorial coordinates $\alpha$ and $\delta$, distance $D$, and radial velocity $v_{\mathrm{los}}$. We adopt a Cartesian coordinate system aligned with standard Galactic coordinates centred upon Centaurus A as the preferred coordinate system in this paper.

\subsection{Defining the Centaurus A plane}
\label{sec:s2_casp}

Previous definitions of the unimodally distributed Centaurus A plane (CASP) in the literature are either limited to an on-sky projection \citep{Muller2018whirling, Muller2021coherent} or consider a separate sample of satellites that does not require known velocities \citep{Muller2019galaxy}, motivating a separate three-dimensional definition of the CASP for this work. Since -- unlike the M31 system -- no subsample of highly correlated satellites has been reported around Centaurus A, we assume this plane definition is equally weighted by all 27 satellites in our sample.

In principle, a three-dimensional satellite plane can be defined by three parameters -- its normal vector $\hat{\bm{n}}$, its root-mean-square (rms) plane height $\Delta$, and its offset $s$ from the host galaxy along $\hat{\bm{n}}$. The former can be obtained through an unweighted tensor-of-inertia fit as demonstrated in \citet{Metz2007m31} and \citet{Pawlowski2013predict}. Denoting the position vectors of individual satellites as $\bm{r}_i$, the moment of inertia tensor $\mathbfss{T}_0$ is defined for $N$ satellites as
\begin{equation}
    \mathbfss{T}_0 = \sum_{i=1}^N \left[ (\bm{r}_i - \bm{r}_0)^2 \cdot \mathbfss{I} - (\bm{r}_i - \bm{r}_0) \cdot (\bm{r}_i - \bm{r}_0)^T \right],
    \label{eq:s2_toi}
\end{equation}
where $\mathbfss{I}$ is the identity matrix. We initially assume a plane passing through the coordinate origin, i.e. through Centaurus A, such that $\bm{r}_0 = 0$. $\mathbfss{T}_0$ has three eigenvalues, $\lambda_1 \leq \lambda_2 \leq \lambda_3$. The eigenvectors corresponding to $\lambda_1$, $\lambda_2$, and $\lambda_3$ describe the major, intermediate, and minor axes of the satellite distribution (with unit vectors $\hat{a}$, $\hat{b}$, $\hat{n}$) respectively.

Next, we uniformly generate 50,000 offset realisations within the range $[-50, 50]$ kpc. Each offset realisation $s_i$ results in a rms spread $\Delta_i$ of the satellite distribution along $\hat{n}$, given by
\begin{equation}
    \Delta_i^2 = \frac{1}{N} \sum_{j=1}^{N} \left|\, (\bm{r}_j \cdot \hat{n} - \bm{r}_0) + s_i \,\right| ^2.
    \label{eq:s2_rms}
\end{equation}
The offset $s_i$ for which $\Delta_i$ is minimized is taken as the CASP's offset from Centaurus A, and the plane's rms height is given by the corresponding $\Delta_i$. The magnitudes of the major and intermediate axes are found by applying equation~(\ref{eq:s2_rms}) along $\hat{\bm{a}}$ and $\hat{\bm{b}}$ respectively.

The spatial distribution of our sample of 27 satellites has major-to-minor, major-to-intermediate axis ratios of $c/a=0.39$ and $b/a=0.79$ respectively, and is defined by the following unit vectors:
\begin{equation}
    \begin{split}
        &\hat{\bm{n}} = [-0.876, -0.430,  0.218] \\
        &\hat{\bm{a}} = [-0.456,  0.886, -0.087] \\ 
        &\hat{\bm{b}} = [0.156, 0.175, 0.972].
    \end{split}
    \label{eq:s2_axes}
\end{equation}
The CASP itself has a rms plane height of $\Delta = 134.1$ kpc and a semi-major axis of 306.3 kpc, as well as an offset along $\hat{n}$ of 6.3 kpc. The plane is oriented nearly edge-on to the heliocentric line-of-sight at $10.7^{\circ}$. These results are comparable to \citet{Muller2019galaxy}'s estimates of $c/a=0.41$ and a plane height of 133 kpc for their sample of 45 satellites with TRGB distances. It is worth noting that the current census of Centaurus A's satellites \citep[e.g.][]{Muller2015survey, Muller2017survey} is incomplete to the south of the CASP due to the Galactic disk, potentially cutting off a section of the plane -- the CASP has the potential to be more spatially extended than estimated here.

\subsection{Modelling the Centaurus A potential}
\label{sec:s2_model}

A satellite galaxy orbiting within the Centaurus A system experiences the gravitational potential of three notable sources: the dark halo in which the system resides, the central baryonic component corresponding to the host galaxy, and close encounters with other satellites. We disregard the last in this work, as modelling individual satellite-satellite interactions -- each with a range of initial TVs -- is computationally unfeasible. Furthermore, interactions between individual satellites are expected to be relatively rare, owing to the long orbital periods of Centaurus A's satellites on a scale of $1 - 10$ Gyr.

Between the remaining two sources, the halo mass dominates over that of the central galaxy by nearly two orders of magnitude. As a result, the gravitational potential due to Centaurus A's halo will exert a dominant impact on the orbits of the system's satellite galaxies.

Using a mean line-of-sight velocity dispersion of $121 \pm 30\,\mathrm{km}\,\mathrm{s^{-1}}$ for 16 of Centaurus A's satellites, \citet{Tully2015groups} estimates the virial mass of its halo to lie around $M_{\mathrm{vir}} = 8 \times 10^{12} M_{\odot}$. A majority of recent works that require a mass estimate adopt this value \citep[e.g.][]{Tully2015planes, Muller2018whirling, Muller2019galaxy, Muller2021coherent}, as do we in this paper. Adopting a Hubble constant of $H_0 = 71\,\mathrm{km}\,\mathrm{s^{-1}}\,\mathrm{Mpc^{-1}}$ \citep{Komatsu2011} and a characteristic overdensity of 200 times the critical density $\rho_{\mathrm{crit}}$, we obtain a virial radius of $R_{\mathrm{vir}} = 409$ kpc.

To model Centaurus A's dark matter halo, we adopt a spherical Navarro-Frenk-White profile \citep[NFW;][]{Navarro1996nfw}. To fully characterise such a potential, however, an additional concentration parameter $c$ is required. As this is difficult to constrain from available observations alone, we assume Centaurus A's halo concentration lies on the median mass-concentration relation from \citet{Diemer2019conc}, as re-calibrated by \citet{Ishiyama2021uchuu}. At $z = 0$, our adopted halo mass corresponds to a median concentration of $c = 6.5$.

We additionally test a triaxial implementation -- argued to better represent haloes in a CDM universe \citep[e.g.][]{Despali2014triaxial} -- to evaluate the robustness of our results with halo morphology. In cosmological simulations, the shape of cold haloes generally varies by radius, transitioning from sphericality into a triaxial regime beyond a scale radius $r_a$ \citep{Hayashi2007triaxial}. Following the approach outlined in \citet{Vogelsberger2008triaxial}, the radius $r$ for a spherical NFW halo is replaced with a general radius $\tilde{r}$ such that
\begin{equation}
    \tilde{r} = \frac{(r_a + r) r_E}{r_a + r_E}.
    \label{eq:s2_genradius}
\end{equation}
The ellipsoidal radius $r_E$ is calculated as
\begin{equation}
    r_E^2 = \left(\frac{r_x}{a}\right)^2 + \left(\frac{r_y}{b}\right)^2 + \left(\frac{r_z}{c}\right)^2,
    \label{eq:s2_ellipradius}
\end{equation}
where $r_x$, $r_y$, and $r_z$ are Cartesian components of the satellite's position along the major, intermediate, and minor axes of the triaxial halo. The axes' magnitudes $a$, $b$, and $c$ are subject to the normalisation condition $a^2 + b^2 + c^2 = 3$, ensuring that $\tilde{r} \rightarrow r_E$ for $r \gg r_a$ and $\tilde{r} \rightarrow r$ for $r \ll r_a$. Here, we set $r_a = 2 R_s = 2 R_{\mathrm{vir}} / c$ as suggested by \citet{Vogelsberger2008triaxial} for CDM haloes of galaxy mass.

For our triaxial model, we assume the visible distribution of satellites traces the underlying halo morphology perfectly such that $(c/a)_{\mathrm{halo}} = 0.39$ and $(b/a)_{\mathrm{halo}} = 0.79$. In cosmological simulations, however, the spatial distribution of luminous satellites is generally more triaxial than their host's dark subhalo distribution \citep{Libeskind2005subhalo}, which in turn traces the underlying halo morphology well. As a result, our spherical and triaxial models of the Centaurus A halo should be taken as extreme bounds for its morphology. 

While the system's overall potential is dominated by that of its dark halo, the central baryonic component still contributes to the potential experienced by orbits passing near the host galaxy. We use \citet{Muller2021mass}'s estimate for Centaurus A's baryonic mass of $M_{\mathrm{bar}} = 2 \times 10^{11} M_{\odot}$. The galaxy potential itself is modelled by a simple Plummer sphere \citep{Plummer1911model} with a core size of 15 kpc.

\subsection{Initial conditions for integration}
\label{sec:s2_iniconds}

For each satellite in our sample, we generate a range of initial transverse velocity components to integrate from. Specifically, we specify a transverse velocity range in proper motion units of $[-40, 40]\,\mu\mathrm{as}\,\mathrm{yr}^{-1}$ for equatorial axes $\mu_{\alpha*}$ (corrected for declination such that $\mu_{\alpha*} = \mu_{\alpha} \cos\delta$) and $\mu_{\delta}$. This limit corresponds to an absolute transverse velocity of around $700\,\mathrm{km}\,\mathrm{s^{-1}}$ at 3.7 Mpc, which far exceeds our halo's characteristic velocity of $290\,\mathrm{km}\,\mathrm{s^{-1}}$. As Centaurus A lacks a TV measurement, we express satellite transverse velocities with respect to that of their host galaxy. For the remainder of this paper, we express transverse velocities in angular velocity space owing to their independence of satellite distance uncertainties.

For each realisation, we apply a corresponding transverse velocity component to the observed radial velocity and integrate the satellite forward over a period of 5 Gyr. Our approach does not take the effects of dynamical friction into account, leading to inaccurate results over a extended duration. Furthermore, close approaches or major mergers with a sufficiently massive galaxy -- an event which may have occurred as recently as 2 Gyr ago \citep{Wang2020merger} -- may cause the system's state to deviate significantly from our simulation. On the other hand, our chosen integration period is much shorter than that of \citet{Hodkinson2019m31}, who integrate the M31 satellites over 10 periapses.

However, we point out that our aim is not to evaluate the stability of the CASP over time -- a question which would also need to examine the plane's current age -- but instead to classify satellites by whether they are dynamically consistent with the plane \emph{at present time}. Using a short integration time ($\sim 2\,\mathrm{Gyr})$, we risk misidentifying satellites which are coincidentally lie in the CASP but are not aligned with the plane over their entire orbit. Conversely, while \citet{Hodkinson2019m31} only assumes a spherical halo potential, we also explore triaxial morphologies wherein precession over multiple orbital periods can misalign even current CASP members. As such, our TV constraints do not constitute a prediction that the given satellite will follow the modelled orbit unperturbed -- only that satellites within the constraints are consistent with CASP membership at present time.

\begin{figure}
    \centering
	\includegraphics[width=0.42\textwidth]{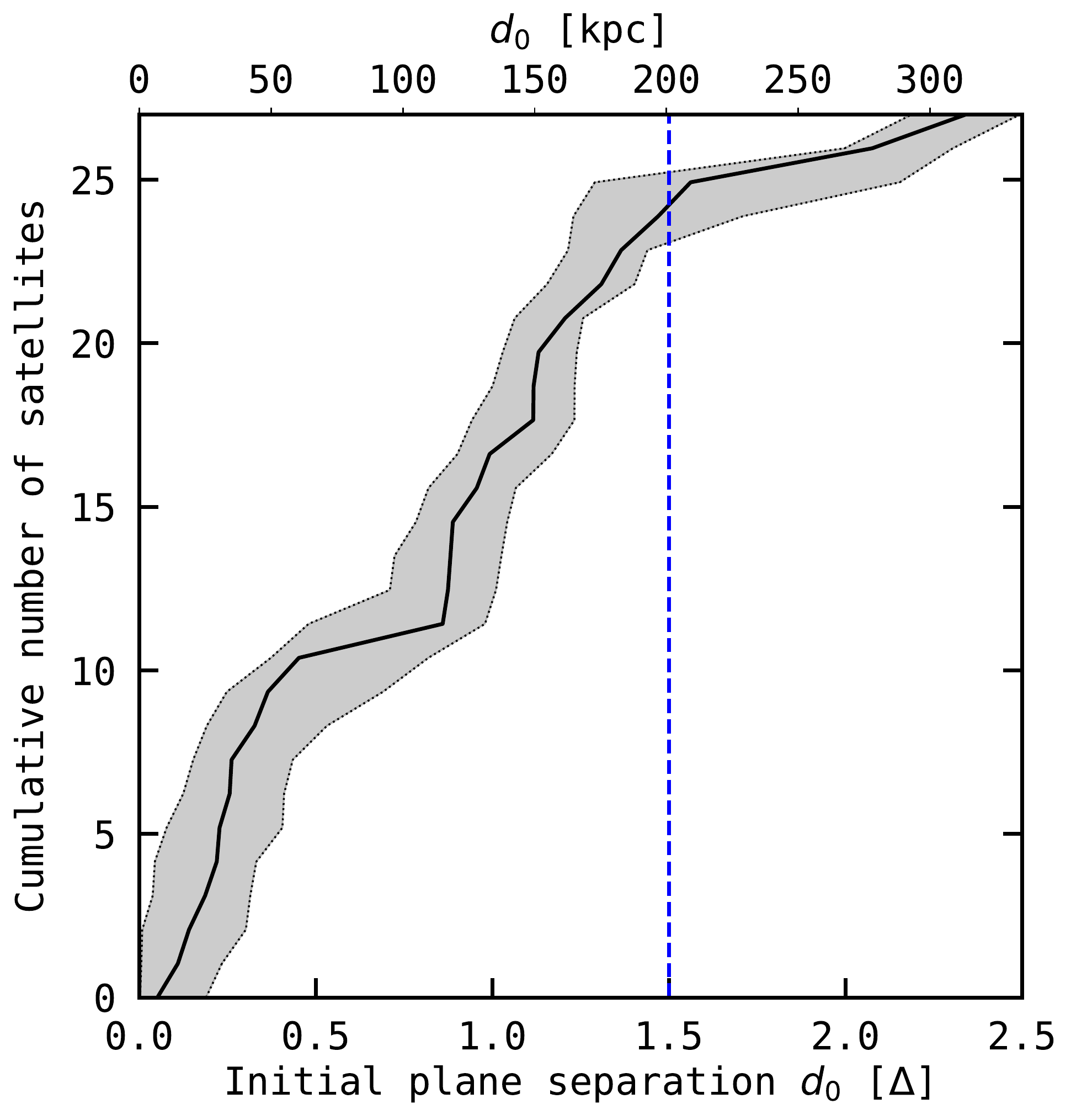}
    \caption{The three-dimensional orthogonal distance $d_0$ between 27 Centaurus A satellites and the CASP, plotted as a cumulative distribution. The grey shaded area corresponds to possible separations within the known heliocentric distance uncertainties. The blue dotted line drawn at $d_0=1.5\Delta$ indicates the threshold required for CASP membership in equation (\ref{eq:s2_maxsep}).}
    \label{fig:s3_iniconds_cdf}
\end{figure}

Uncertainties in the observed line-of-sight velocities in Table~\ref{tab:s2_data} do not significantly influence our results, and are generally overshadowed by the much larger magnitudes of applied TVs. However, the distance uncertainties present are often sufficiently large -- some even comparable to Centaurus A's virial radius -- to directly impact the sense of each satellite's orbit.

To compensate, we mirror the approach taken by \citet{Hodkinson2019m31}. For each satellite, we generate 5 heliocentric distance realisations with
\begin{equation}
    D_i = [\: D-\sigma_{D}, \; D-0.5\sigma_{D}, \; D, \; D+0.5\sigma_{D}, \; D+\sigma_{D} \: ].
    \label{eq:s2_errors}
\end{equation}
Upon testing, any increase in number of distance realisations per satellite past our adopted number did not significantly influence our results. Furthermore, for most satellites, the lower and upper bounds of their distance uncertainty generates the most extreme orbital characteristics, hence making a large number of realisations between these bounds unnecessary. Overall, we generate a total of 6561 initial transverse velocities (corresponding to a $1\,\mu\mathrm{as}\,\mathrm{yr}^{-1}$ resolution per axis) for each of 5 distance realisations for a total of 32,805 orbit integrations.

\begin{figure*}
	\includegraphics[width=0.7\textwidth]{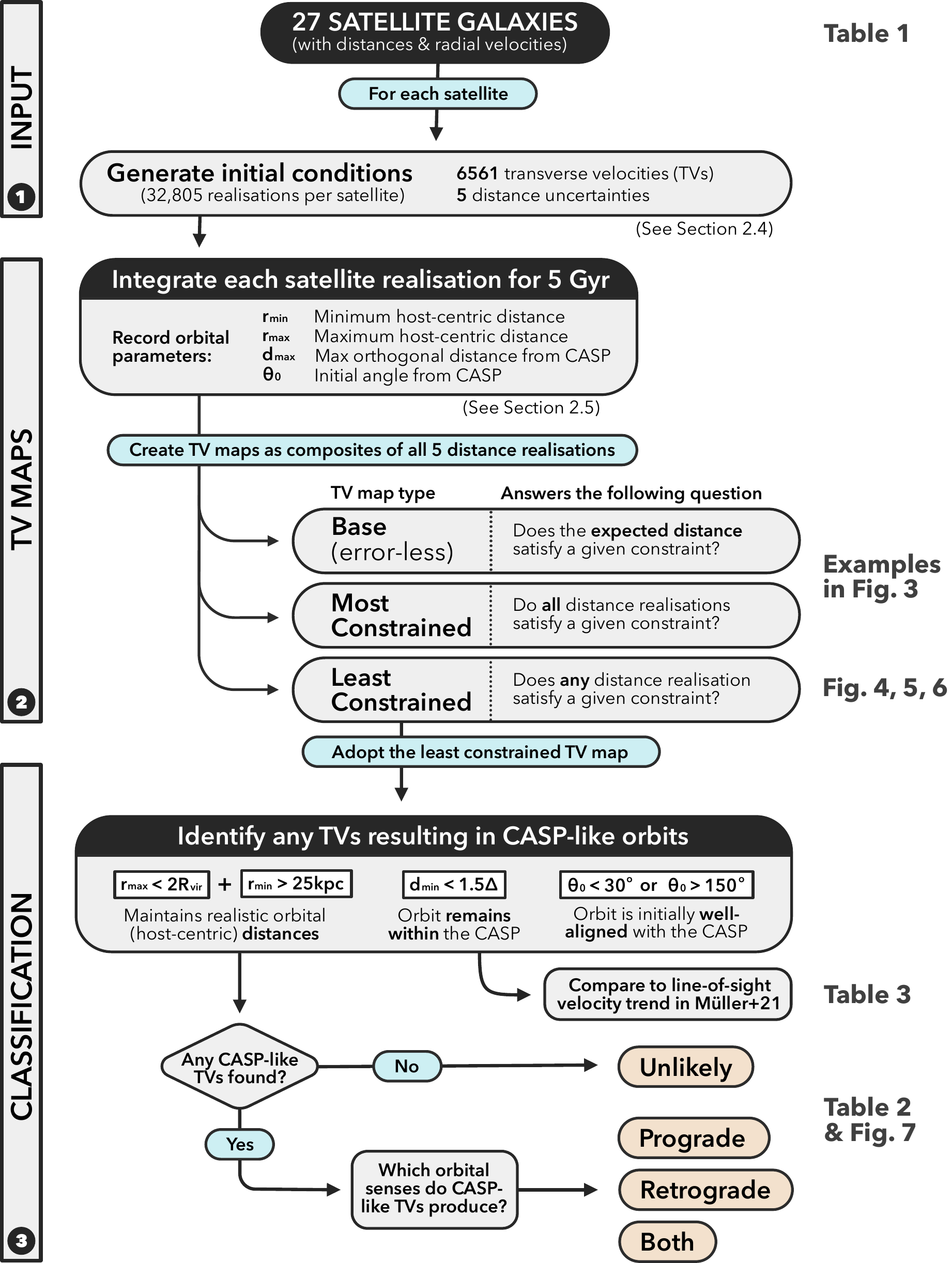}
	\caption{Flowchart of the methodology adopted in this paper. From the top, we: (1) generate an array of transverse velocities with varying distance realisations per satellite, (2) integrate all realisations individually to create TV maps for each satellite, and (3) identify any TVs consistent with CASP membership and their resulting orbital senses to classify Centaurus A's satellite galaxies (yellow bubbles).}
	\label{fig:s2_flowchart}
\end{figure*}

\begin{figure*}
	\includegraphics[width=0.3\textwidth]{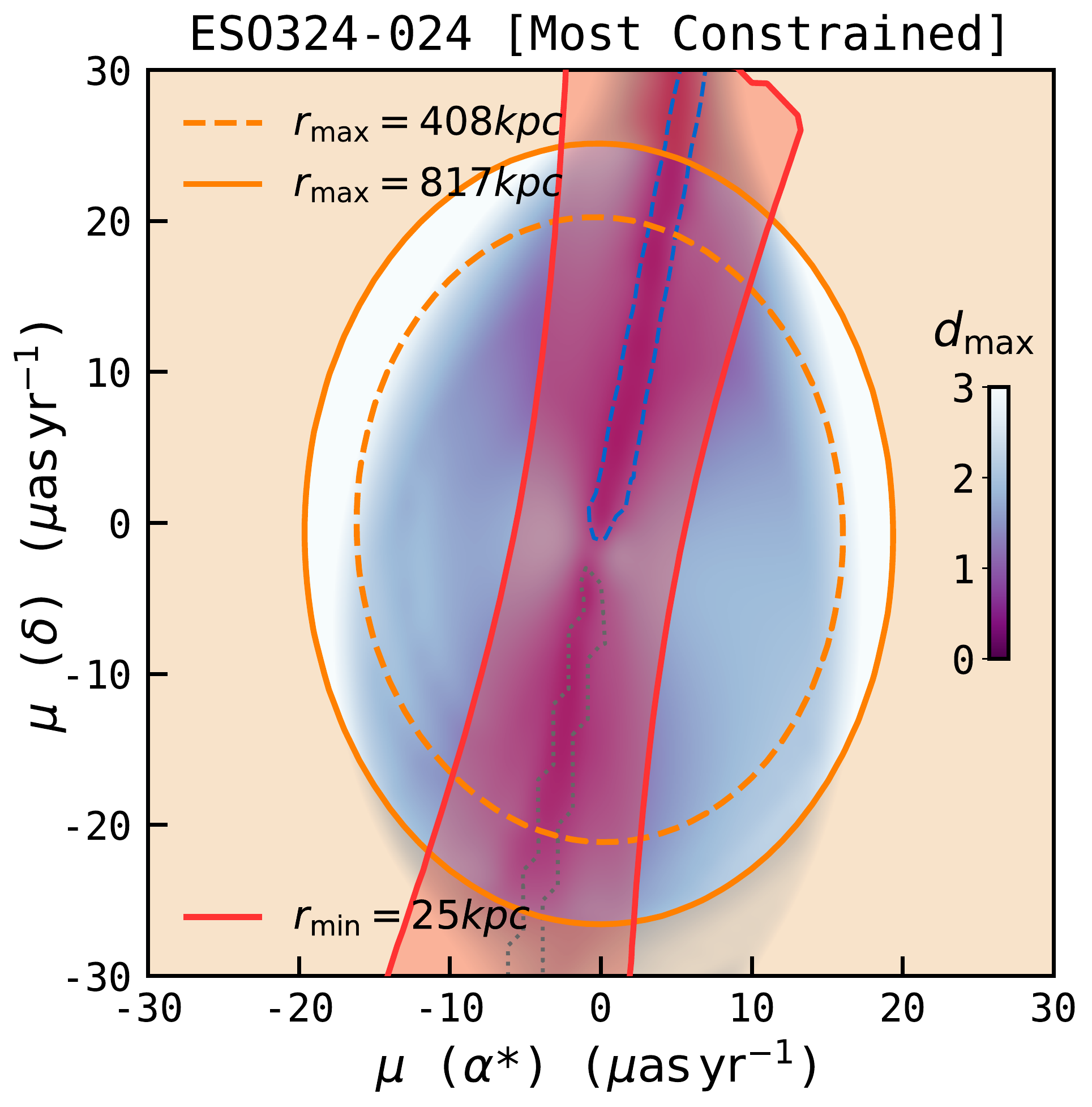}
	\includegraphics[width=0.3\textwidth]{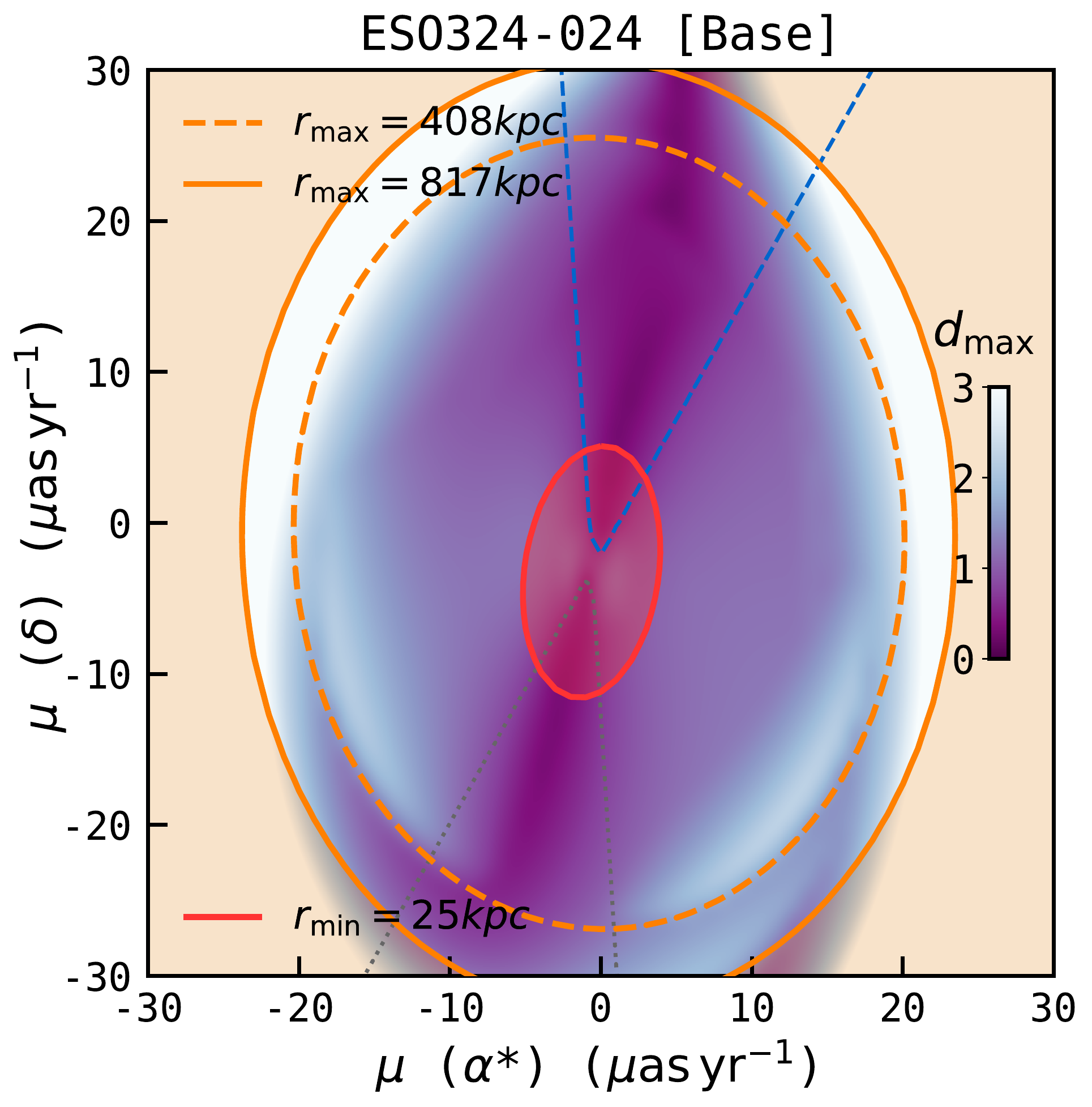}
	\includegraphics[width=0.3\textwidth]{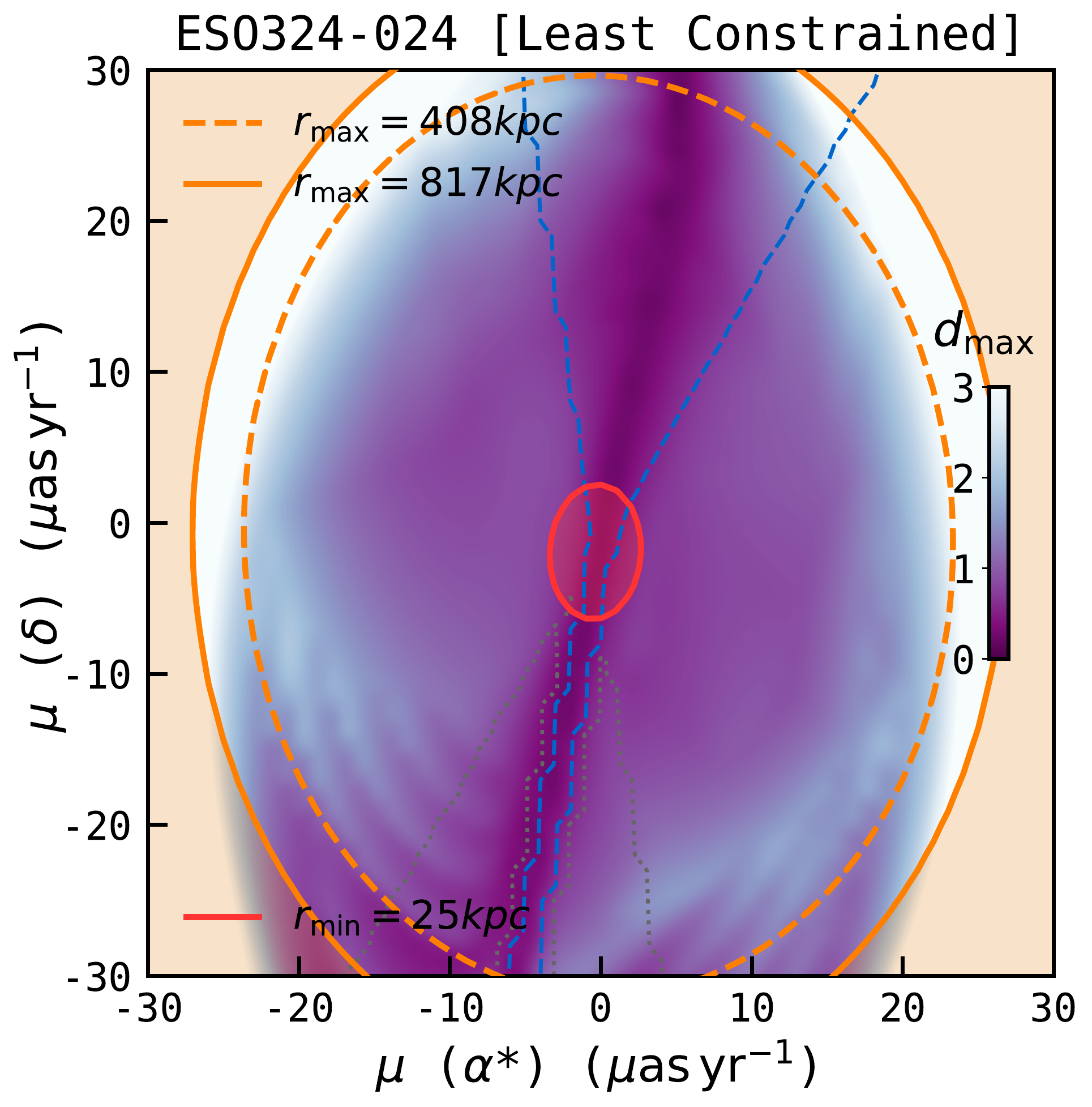}
	\includegraphics[width=0.3\textwidth]{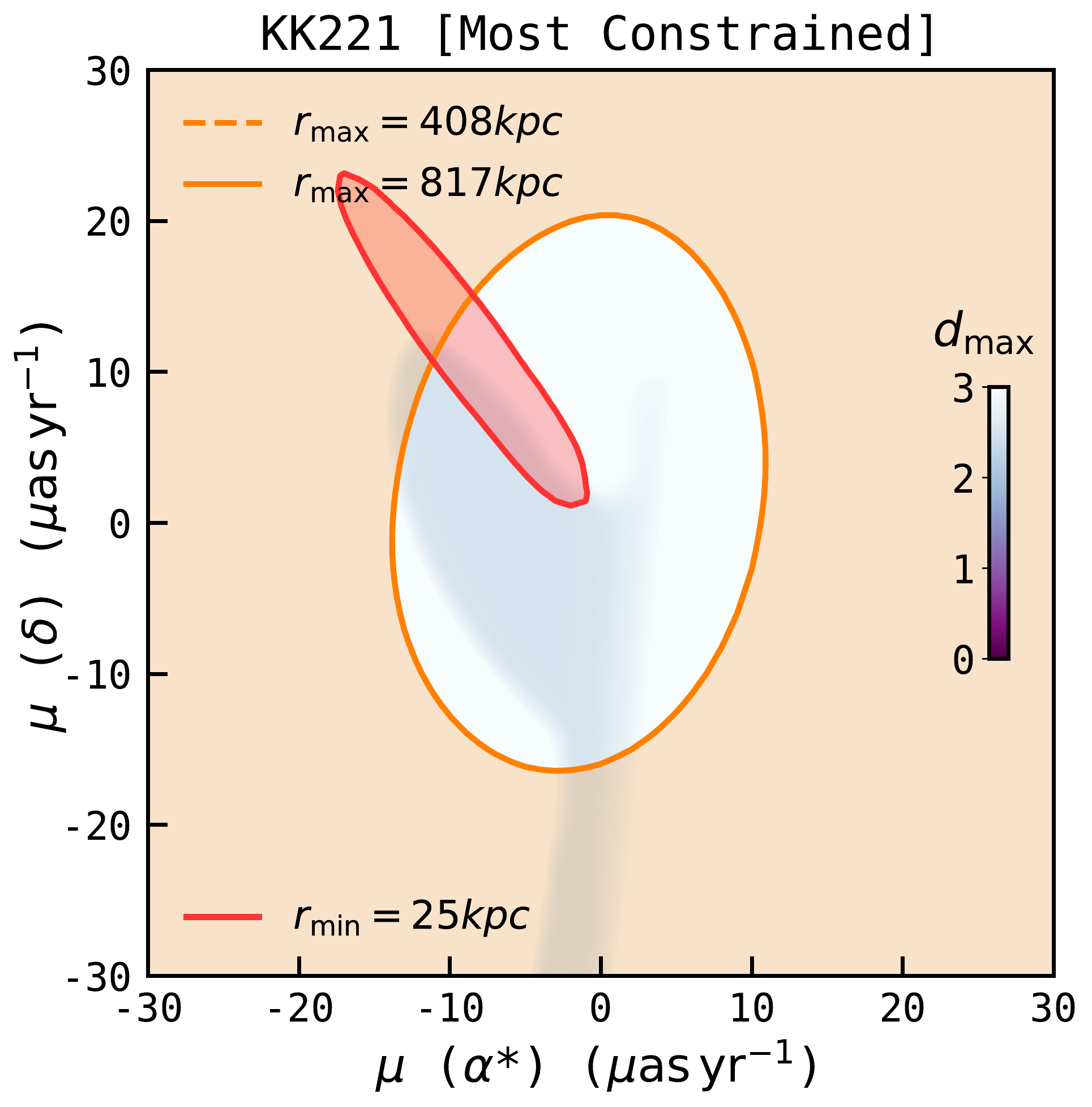}
	\includegraphics[width=0.3\textwidth]{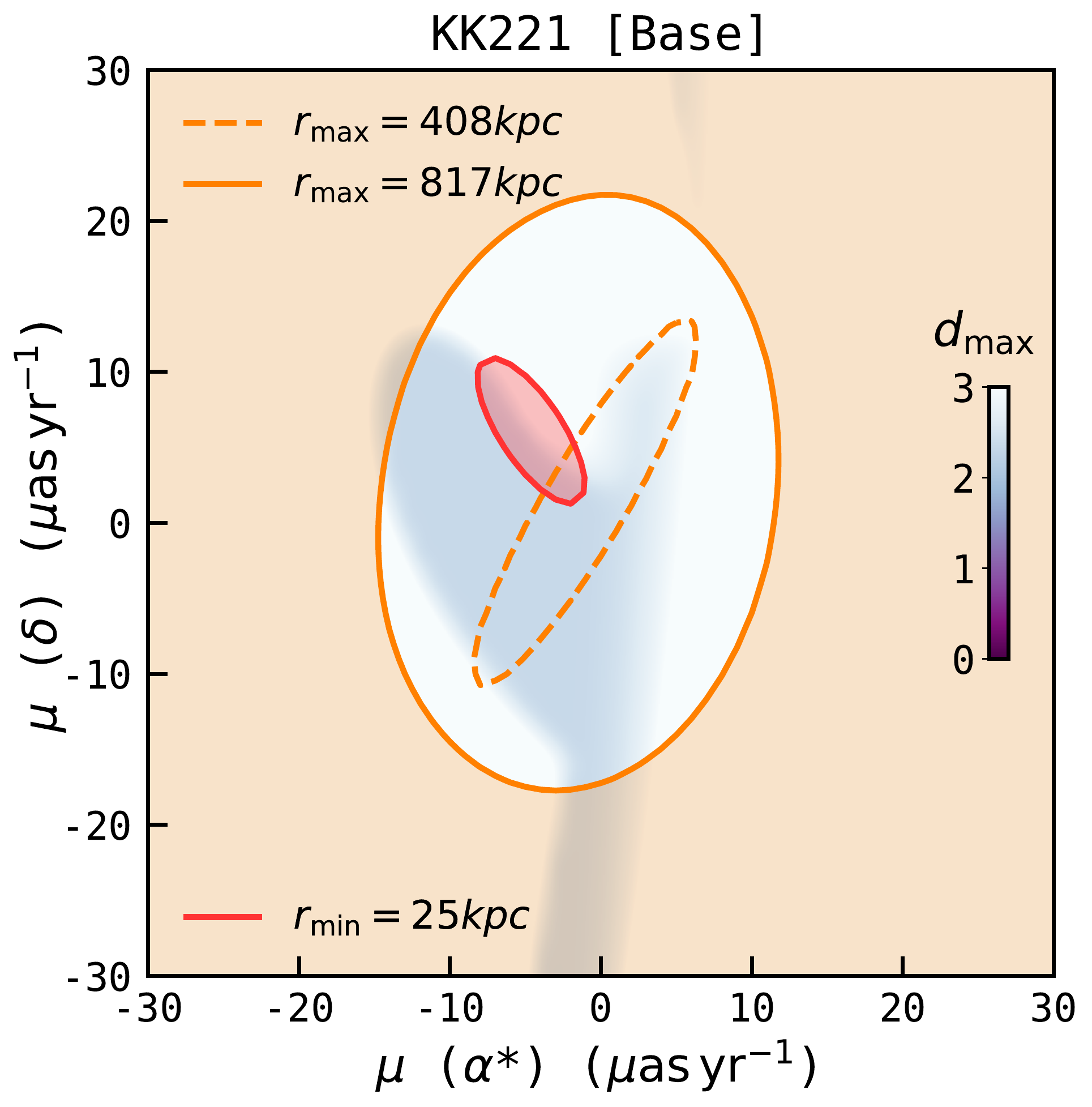}
	\includegraphics[width=0.3\textwidth]{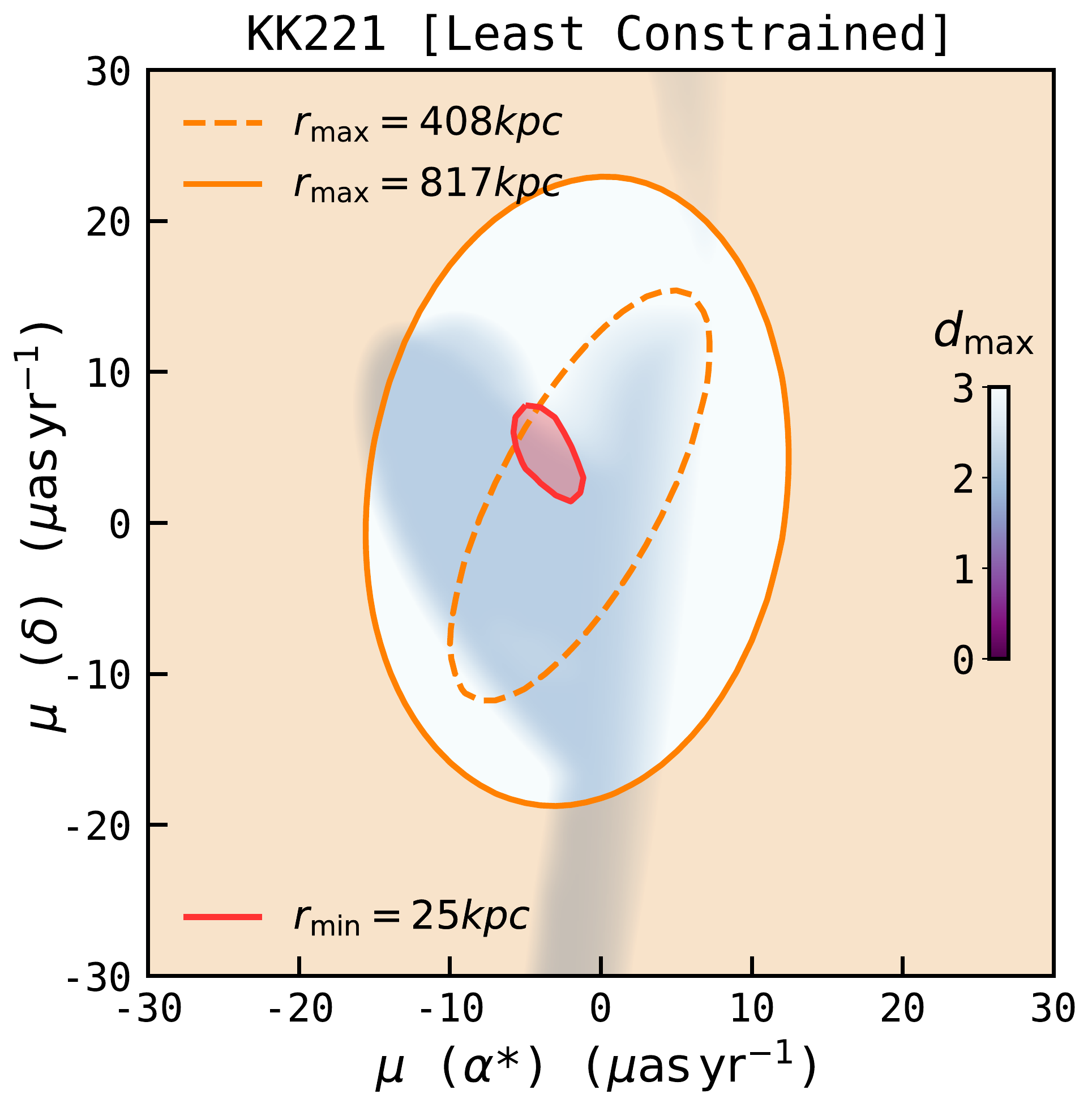}
	\caption{Most-constrained, base, and least-constrained transverse velocity maps of ESO324-024 and KK221 as examples of highly likely and unlikely CASP members respectively, integrated in a spherical NFW halo with a central galaxy potential for 5 Gyr. The horizontal and vertical axes represent TVs expressed in angular velocity units $\mu\mathrm{as}\,\mathrm{yr}^{-1}$ along right ascension ($\mu_{\alpha *}$) and declination ($\mu_{\delta}$) respectively relative to Centaurus A. Orange and red shaded regions represents TVs which result in satellite orbits that travel beyond the maximum and minimum distance constraints of $2 R_{\mathrm{vir}} = 817\,\mathrm{kpc}$ and $25\,\mathrm{kpc}$ respectively. The maximum orthogonal plane separation $d_{\mathrm{max}}$ per TV realisation is indicated by the heatmap, while the region within the blue dashed (grey dotted) line results in prograde (retrograde) orbits with angular momenta aligned to $30^{\circ}$ of the plane's normal. The base map does not take distance uncertainties into account. For each initial TV, an orbital constraint for CASP membership (see Section~\ref{sec:s2_constraints}) is considered to be met in the least-constrained maps if at least 1 out of 5 distance realisations per satellite satisfies it, whereas all 5 realisations must satisfy it in the most-constrained maps.}
	\label{fig:s3_maps_examples}
\end{figure*}

The total error $\sigma_D$ is defined for each satellite as its quoted distance uncertainty combined with that of Centaurus A in quadrature,
\begin{equation}
    \sigma_D = \sqrt{\sigma_{\mathrm{sat}}^2 + \sigma_{\mathrm{CenA}}^2},
\end{equation}
in order to always constrain Centaurus A's position to the coordinate origin. Thus, the resulting set of distance realisations may be considered a reasonable range in which the true distance should lie within, though we emphasize that it is by no means exhaustive.

\subsection{Orbital constraints}
\label{sec:s2_constraints}

\citet{Muller2018whirling} and \citet{Muller2021coherent} discovered that the line-of-sight velocities of Centaurus A's satellites demonstrated a velocity trend indicative of co-rotation along the CASP. Furthermore, \citet{Muller2021mass} found that the satellite system has a significant circular velocity component, and may form a rotationally supported structure. If the CASP is indeed stable, its constituents should have bound and long-lived orbits aligned with the defined plane. In order to identify TVs that result in simulated satellite orbits obeying these constraints, we define four requirements for CASP membership.

To be bound, orbits should remain firmly within the system's gravitational potential -- this can be simply expressed as a maximum distance an orbit may stray from the host galaxy. \citet{Hodkinson2019m31} adopts one virial radius as a maximum bound, but this is motivated in part due to the Milky Way's location only 800 kpc away. A satellite beyond the M31 halo's virialized region risks capture by the Milky Way's similarly massive dark halo. On the other hand, Centaurus A is well-isolated -- its nearest host neighbour, M83, lies at a distance of 1.2 Mpc and is significantly less massive. Therefore, we adopt a relaxed maximum distance of
\begin{equation}
    r_{\mathrm{max}} < 2 R_{\mathrm{vir}} = 817 \, \mathrm{kpc}.
    \label{eq:s2_maxdist}
\end{equation}
Close approaches to a satellite's host galaxy can result in the tidal stripping and disruption of a satellite's content. While such a scenario is inevitable due to the loss of energy from dynamical friction, radial orbits which pass by the system's central region are especially unlikely to be long-lived. Here, we specify a minimum distance of
\begin{equation}
    r_{\mathrm{min}} > 25 \, \mathrm{kpc},
    \label{eq:s2_mindist}
\end{equation}
at which point we assume tidal effects are sufficiently destructive \citep[e.g.][]{Diemand2007tidal, Errani2022tidal}. We motivate the $25\,\mathrm{kpc}$ threshold by scaling the $15\,\mathrm{kpc}$ adopted for the M31 system by \citet{Hodkinson2019m31} by Centaurus A's virial radius -- this also corresponds to an approximate Roche limit for a satellite of mass of order $10^{7}M_{\odot}$ of size $1\,\mathrm{kpc}$ orbiting an enclosed mass of order $10^{11}M_{\odot}$. $r_{\mathrm{min}}$ must also exceed the radius of the Centaurus A galaxy, as closer approaches would result in merger events. Furthermore, the density of a NFW halo diverges at its centre, potentially affecting very close approaches in an unphysical manner. Satellite realisations must maintain a host-centric distance between $r_{\mathrm{min}}$ and $r_{\mathrm{max}}$ for the entirety of their orbital integration.

\begin{table*}
	\centering
	\caption{Orbital metrics and properties for 27 Centaurus A satellites after integration about a spherical NFW halo with a baryonic core for 5 Gyr. The \emph{Morphology} column lists each satellite's de Vancouleurs morphological type as published in \citet{Karachentsev2004catalog, Karachentsev2013catalog}'s Local Volume catalog. $r_0$ represents each satellite's radial distance from Centaurus A at present time, while $d_0$ denotes their initial orthogonal separation from the best-fitting plane. Our results are calculated from the base TV maps without heliocentric distance errors, while the positive and negative uncertainties are found from the least-constrained and most-constrained TV maps respectively. $f(d_{\mathrm{max}})$ and $f(\theta_0)$ represent the percentage of TVs resulting in realistic orbital distances that additionally obey the plane separation (equation \ref{eq:s2_maxsep}) and orbital pole alignment requirements (equation \ref{eq:s2_limalign}) respectively, while $f(\mathrm{CASP})$ corresponds to the fraction that satisfy both requirements simultaneously. Out of the maximum plane separations $d_{\mathrm{max}}$ per TV, $\mathrm{min}(d_{\mathrm{max}})$ indicates the minimized value that also obeys the orbital pole alignment requirement -- for satellites for which no TVs resulting in realistic orbital distances satisfies the alignment requirement, we show the minimum plane separation disregarding alignment in brackets. We also include our classification based on CASP membership potential in the \emph{Type} column.}
	\renewcommand{\arraystretch}{1.2}
	\begin{tabular}{lllllllll}
	    \hline
		Name & Morphology & $r_0$ & $d_0$ & $f(d_{\mathrm{max}})$ & $f(\theta_0)$ & $f(\mathrm{CASP})$ & $\mathrm{min}(d_{\mathrm{max}})$ & Type \\
		& & (kpc) & (kpc) & ($\%$) & ($\%$) & ($\%$) & ($\Delta$) \\
		\hline
        ESO269-037 & dIrr & $616\;^{+88}_{-83}$ & $269\;^{+10}_{-10}$ & $0\;^{+0}_{-0}$ & $27.0\;^{+0}_{-7.4}$ & $0\;^{+0}_{-0}$ & $2.0$ & \textbf{Unlikely} \\
        NGC4945 & Scd & $474\;^{+12}_{-5}$ & $156\;^{+7}_{-7}$ & $5.3\;^{+3.2}_{-2.4}$ & $1.9\;^{+3.1}_{-1.9}$ & $0.7\;^{+2.1}_{-0.7}$ & $1.15$ & Both \\
        ESO269-058 & dIrr & $317\;^{+18}_{-10}$ & $129\;^{+6}_{-6}$ & $30.1\;^{+4.1}_{-5.7}$ & $20.0\;^{+3.6}_{-4.6}$ & $17.4\;^{+2.6}_{-4.5}$ & $0.94$ & Retrograde \\
        KK189 & dSph & $560\;^{+171}_{-165}$ & $113\;^{+20}_{-20}$ & $43.9\;^{+0}_{-12.5}$ & $24.4\;^{+7.1}_{-0}$ & $21.1\;^{+0.2}_{-0}$ & $0.87$ & Both \\
        ESO269-066 & dSph & $201\;^{+28}_{-14}$ & $124\;^{+7}_{-7}$ & $37.6\;^{+6.5}_{-11.6}$ & $0\;^{+11.6}_{-0}$ & $0\;^{+8.8}_{-0}$ & $0.9$ & Prograde \\
        NGC5011C & Tr & $154\;^{+28}_{-9}$ & $145\;^{+7}_{-7}$ & $38.3\;^{+2.0}_{-4.7}$ & $0\;^{+0}_{-0}$ & $0\;^{+0}_{-0}$ & $(1.06)$ & \textbf{Unlikely} \\
        KKs54 & Tr & $726\;^{+30}_{-13}$ & $170\;^{+12}_{-12}$ & $19.9\;^{+4.9}_{-2.8}$ & $18.9\;^{+3.2}_{-1.2}$ & $13.4\;^{+2.6}_{-1.5}$ & $1.21$ & Retrograde \\
        KK196 & dIrr & $315\;^{+112}_{-102}$ & $7\;^{+18}_{-6}$ & $30.3\;^{+14.2}_{-8.0}$ & $32.0\;^{+0.9}_{-3.3}$ & $28.3\;^{+3.0}_{-6.2}$ & $0.09$ & Both \\
        NGC5102 & Sa & $420\;^{+210}_{-7}$ & $111\;^{+49}_{-49}$ & $16.7\;^{+10.2}_{-13.1}$ & $15.3\;^{+13.5}_{-13.4}$ & $13.7\;^{+9.3}_{-12.3}$ & $0.48$ & Both \\
        KK197 & dSph & $168\;^{+62}_{-59}$ & $34\;^{+9}_{-9}$ & $61.4\;^{+6.7}_{-19.8}$ & $33.3\;^{+0.1}_{-4.2}$ & $30.5\;^{+0.2}_{-3.7}$ & $0.19$ & Both \\
        KKs55 & Sph & $175\;^{+84}_{-81}$ & $28\;^{+12}_{-12}$ & $64.5\;^{+6.9}_{-21.8}$ & $33.4\;^{+}0_{-3.5}$ & $31.8\;^{+0}_{-3.0}$ & $0.12$ & Both \\
        dw1322-39 & dIrr & $752\;^{68+}_{-68}$ & $177\;^{+10}_{-10}$ & $22.3\;^{+2.1}_{-0}$ & $29.0\;^{+0.5}_{-0}$ & $19.6\;^{+1.2}_{-0}$ & $1.29$ & Both \\
        dw1323-40b & dSph & $272\;^{+584}_{-128}$ & $18\;^{+88}_{-13}$ & $37.4\;^{+14.2}_{-0}$ & $29.6\;^{+0}_{-5.1}$ & $27.7\;^{+0}_{-5.4}$ & $0.14$ & Both \\
        dw1323-40a & dSph & $154\;^{+102}_{-9}$ & $33\;^{+23}_{-23}$ & $45.1\;^{+17.5}_{-40.8}$ & $14.6\;^{+15.3}_{-14.6}$ & $13.8\;^{+13.9}_{-13.8}$ & $0.08$ & Both \\
        KK203 & Tr & $183\;^{+205}_{-32}$ & $47\;^{+42}_{-42}$ & $41.0\;^{+21.9}_{-37.1}$ & $17.5\;^{+15.5}_{-17.1}$ & $15.9\;^{+15.2}_{-15.5}$ & $0.12$ & Both \\
        ESO324-024 & Sdm & $144\;^{+85}_{-42}$ & $14\;^{+16}_{-14}$ & $59.6\;^{+5.0}_{-48.0}$ & $28.8\;^{+0.8}_{-28.8}$ & $21.4\;^{+0}_{-21.4}$ & $0.16$ & Both \\
        NGC5206 & S0- & $568\;^{+42}_{-40}$ & $29\;^{+10}_{-10}$ & $20.8\;^{+2.4}_{-2.5}$ & $27.1\;^{+1.1}_{-2.0}$ & $20.0\;^{+1.7}_{-1.9}$ & $0.16$ & Both \\
        NGC5237 & BCD & $376\;^{+50}_{-49}$ & $59\;^{+10}_{-10}$ & $26.1\;^{+2.8}_{-1.9}$ & $31.5\;^{+2.5}_{-5.5}$ & $24.3\;^{+1.3}_{-2.6}$ & $0.37$ & Both \\
        NGC5253 & Sdm & $750\;^{+6}_{-2}$ & $42\;^{+10}_{-10}$ & $41.3\;^{+2.3}_{-6.7}$ & $39.9\;^{+2.4}_{-2.5}$ & $36.8\;^{+2.1}_{-4.5}$ & $0.25$ & Prograde \\
        dw1341-43 & dSph & $243\;^{+43}_{-32}$ & $147\;^{+13}_{-13}$ & $29.9\;^{+7.2}_{-13.8}$ & $0\;^{+0}_{-0}$ & $0\;^{+0}_{-0}$ & $(1.03)$ & \textbf{Unlikely} \\
        KKs57 & Sph & $253\;^{+413}_{-59}$ & $203\;^{+96}_{-96}$ & $0\;^{+28.6}_{-0}$ & $0\;^{+21.0}_{-0}$ & $0\;^{+16.3}_{-0}$ & $0.82$ & Both \\
        KK211 & Sph & $238\;^{+47}_{-0}$ & $191\;^{+31}_{-31}$ & $4.0\;^{+18.5}_{-4.0}$ & $0\;^{+0}_{-0}$ & $0\;^{+0}_{-0}$ & $(1.23)$ & \textbf{Unlikely} \\
        dw1342-43 & Tr & $801\;^{+135}_{-134}$ & $24\;^{+29}_{-23}$ & $21.9\;^{+13.4}_{-0}$ & $14.7\;^{+20.8}_{-0}$ & $10.6\;^{+16.9}_{-0}$ & $0.26$ & Both \\
        ESO325-011 & dIrr & $365\;^{+55}_{-49}$ & $145\;^{+15}_{-15}$ & $26.5\;^{+6.1}_{-5.4}$ & $21.4\;^{+5.3}_{-21.4}$ & $14.3\;^{+4.4}_{-14.3}$ & $1.00$ & Both \\
        KKs58 & dSph & $574\;^{+63}_{-47}$ & $115\;^{+23}_{-23}$ & $18.9\;^{+10.6}_{-5.6}$ & $20.7\;^{+3.5}_{-3.6}$ & $13.9\;^{+7.1}_{-2.5}$ & $0.71$ & Both \\
        KK221 & dIrr & $400\;^{+41}_{-26}$ & $304\;^{+20}_{-20}$ & $0\;^{+0}_{-0}$ & $0\;^{+0}_{-0}$ & $0\;^{+0}_{-0}$ & $(2.19)$ & \textbf{Unlikely} \\
        ESO383-087 & Sdm & $698\;^{+39}_{-36}$ & $114\;^{+13}_{-13}$ & $11.4\;^{+20.1}_{-0}$ & $54.3\;^{+0}_{-1.2}$ & $11.4\;^{+20.1}_{-0}$ & $0.93$ & Prograde \\
		\hline
	\end{tabular}
	\renewcommand{\arraystretch}{1}
	\label{tab:s3_metrics}
\end{table*}

The extent to which a satellite galaxy orbits along the CASP can be expressed by its maximum orthogonal distance $d_{\mathrm{max}}$ from the best-fitting plane. To be considered a participant of the CASP as defined in Section~\ref{sec:s2_casp}, we require satellite orbits to remain within a plane separation of
\begin{equation}
    d_{\mathrm{max}} < 1.5 \Delta = 201 \,\mathrm{kpc},
    \label{eq:s2_maxsep}
\end{equation}
given in terms of the CASP's rms plane height, $\Delta=134\,\mathrm{kpc}$. This threshold is motivated by the plane separations $d_0$ of Centaurus A satellites at present time, shown in Fig.~\ref{fig:s3_iniconds_cdf}. The slope of the cumulative distribution becomes shallow around $d_0 \sim 1.5\Delta$, suggesting the presence of outlying satellites.

However, compact orbits with trajectories heavily misaligned with the best-fitting plane may still appear to be CASP members with the above criterion alone, despite not participating in the (assumed to be) rotationally supported plane. The alignment of a satellite's orbit with the Centaurus A plane can be expressed by the angle $\theta$ between its angular momentum vector, or orbital pole, and the normal vector $\hat{n}$ of the CASP ($\theta=0$ represents a fully prograde orbit with respect to the satellite majority). Perfectly planar orbits are only expected around a spherical potential -- a satellite's orbital pole orientation will precess as it evolves around our triaxial halo model. For simplicity, we take the initial angle $\theta_0 = \theta (t=0)$ as our metric of orbital alignment.

Thus, in conjunction with equation~(\ref{eq:s2_maxsep}), we require CASP participants to also demonstrate an initial orbital pole alignment of
\begin{equation}
    \theta_0 < 30^{\circ} \; \mathrm{(prograde)} \; \mathrm{or} \; \theta_0 > 150^{\circ}  \; \mathrm{(retrograde)},
    \label{eq:s2_limalign}
\end{equation}
where $0^{\circ} \leq \theta_0 \leq 180^{\circ}$. This criterion mirrors \citet{Hodkinson2019m31}'s choice of alignment angle for the M31 system, and is slightly stricter than \citet{Pawlowski2013predict, Pawlowski2014predict}'s choice of $\theta=37^{\circ}$ for the Milky Way. Since the Centaurus A plane is much thicker than both the Milky Way and M31 planes, we also explore an alternative, relaxed criterion of $\theta_0 < 45^{\circ}$ or $\theta_0 > 135^{\circ}$.

Individual combinations of a given transverse velocity and satellite distance realisation is considered to result in a CASP-like orbit if the four orbital constraints ($r_{\mathrm{min}}$, $r_{\mathrm{max}}$, $d_{\mathrm{max}}$, and $\theta_0$) are satisfied for the entirety of their orbit integration. A flowchart summarising our methodology in generating satellite realisations, checking their orbital parameters, and classifying their consistency with CASP membership can be seen in Fig.~\ref{fig:s2_flowchart}.

\begin{figure*}
	\includegraphics[width=0.28\textwidth]{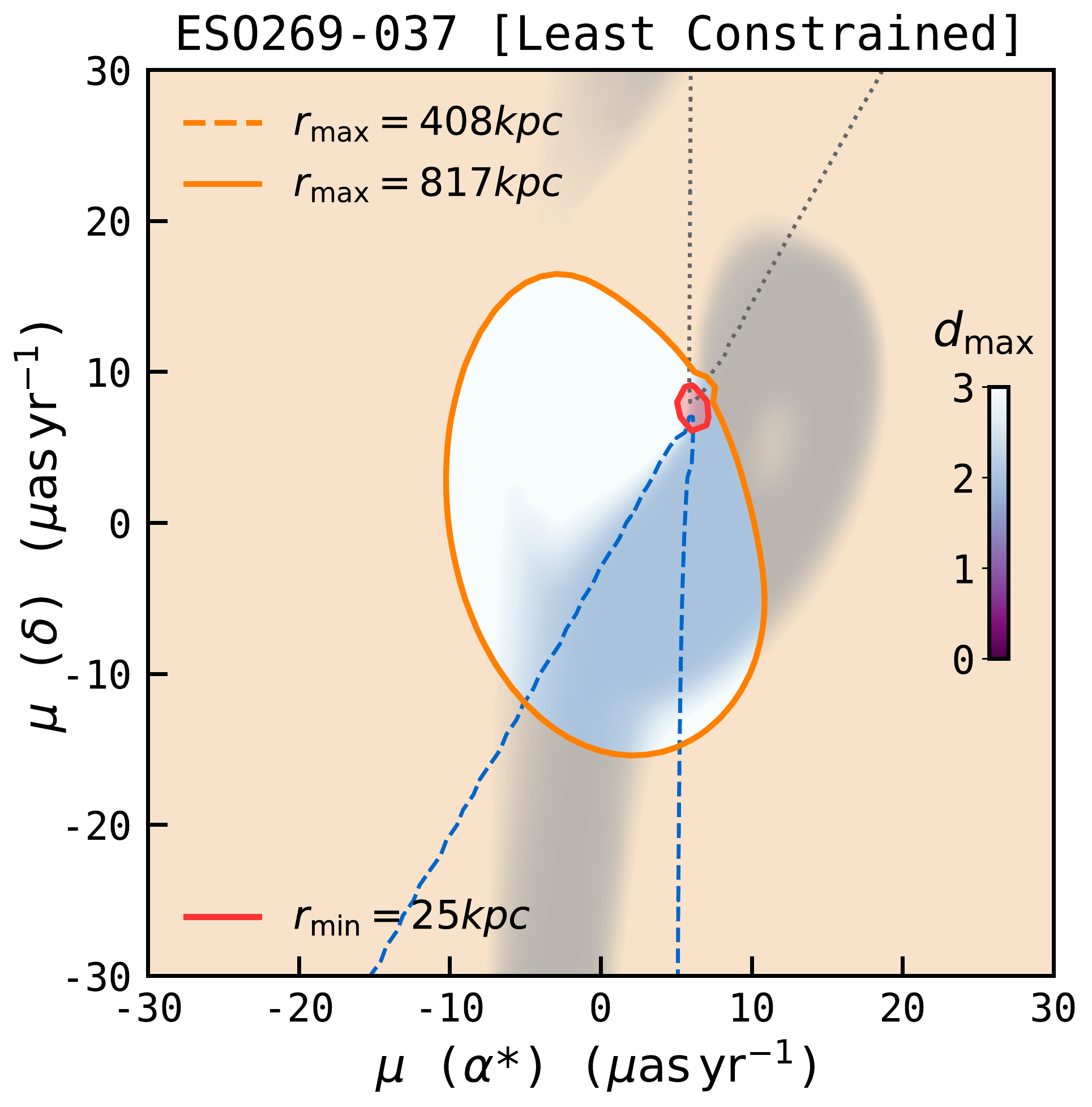}
	\includegraphics[width=0.28\textwidth]{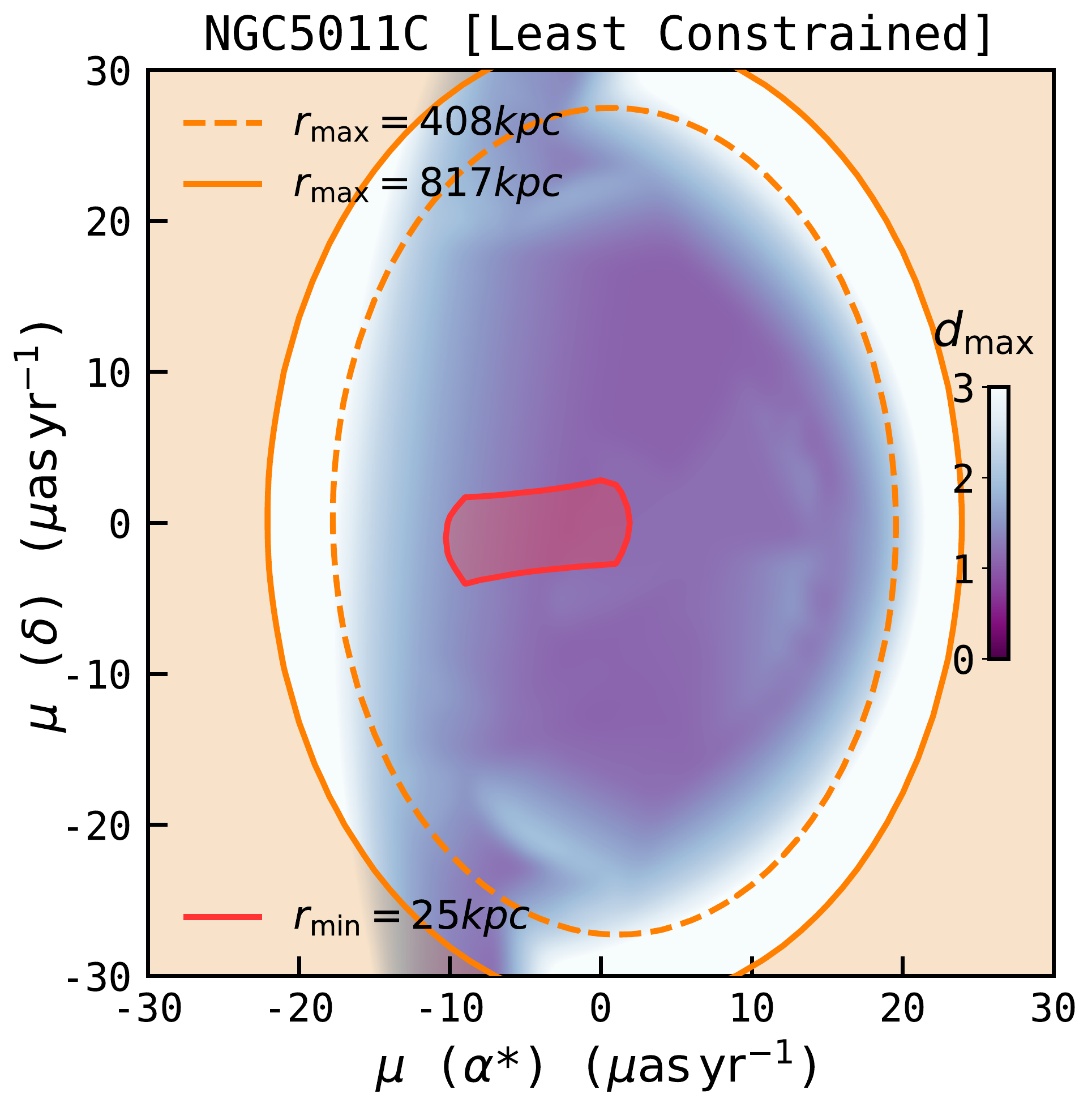}
	\includegraphics[width=0.28\textwidth]{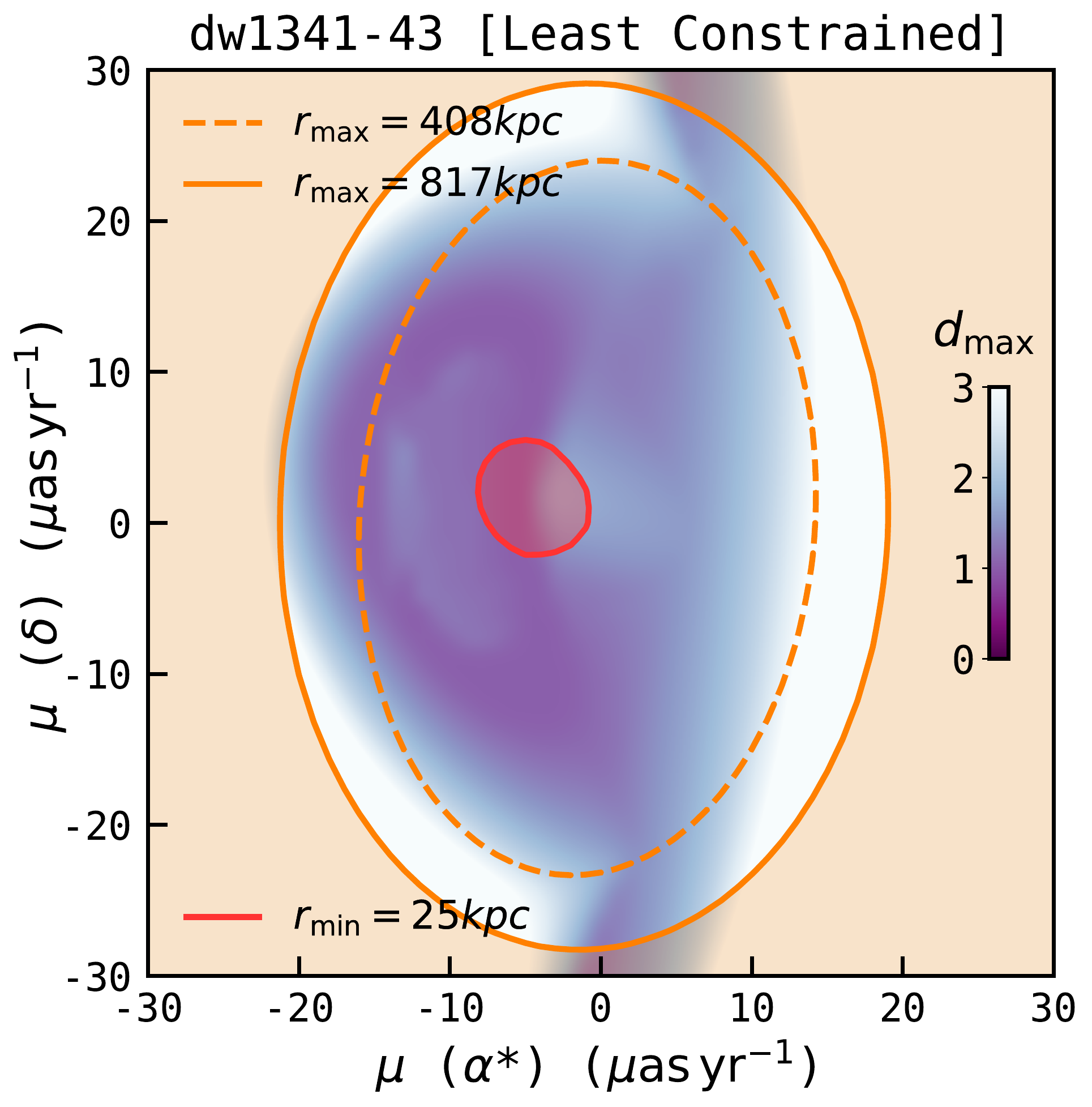}\\
	\includegraphics[width=0.28\textwidth]{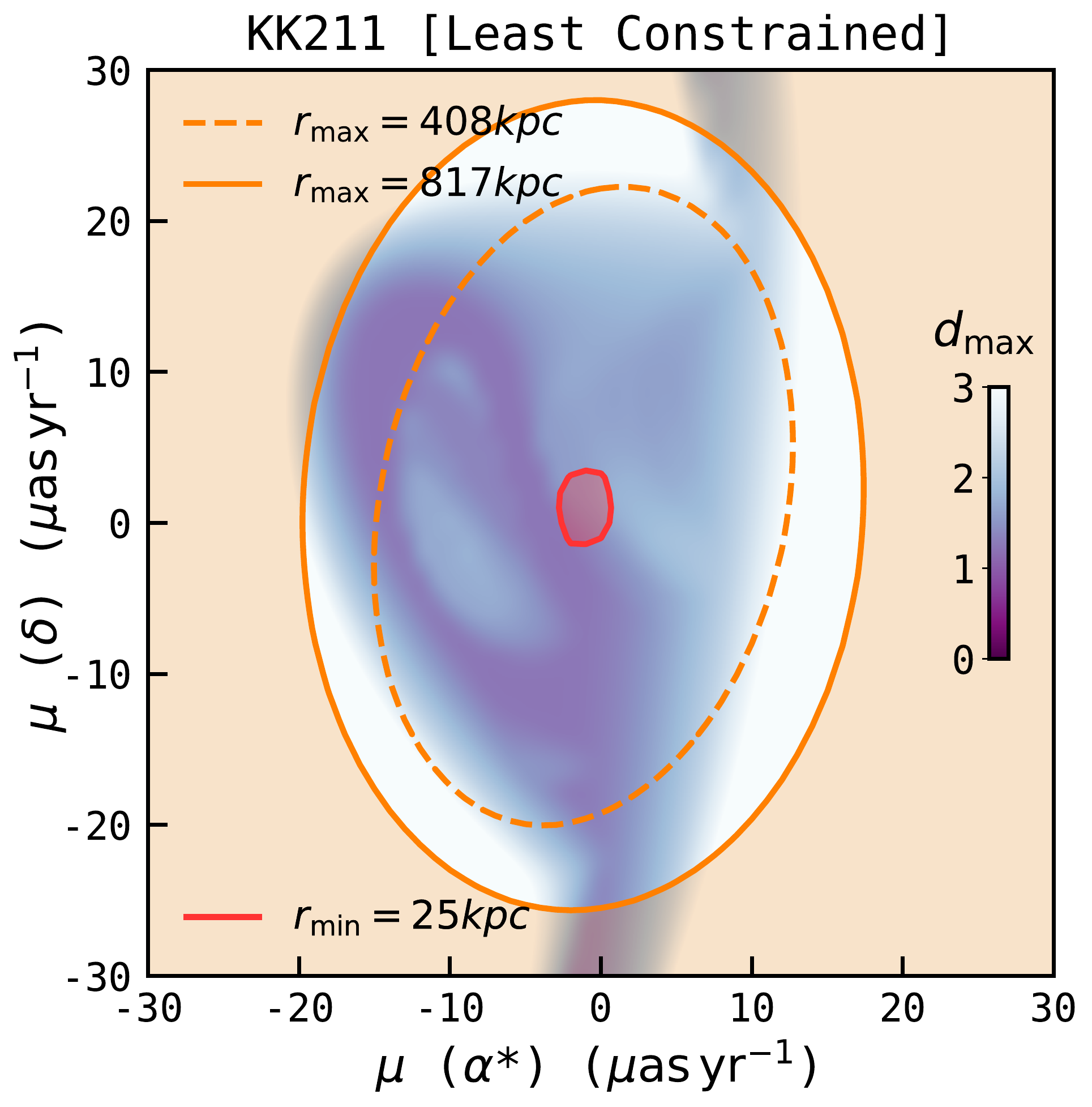}
	\includegraphics[width=0.28\textwidth]{map/map_heat_KK221_best.pdf}
    \caption{Least-constrained transverse velocity (TV) constraints for 5 \emph{Unlikely} satellites inconsistent with CASP membership, using the same representation as Fig.~\ref{fig:s3_maps_examples}. TV maps of satellites classified as \emph{Unlikely} are characterised by a lack of TVs resulting in orbits that obey the maximum and minimum distance requirements (i.e. avoiding the orange and red-shaded region) which additionally remain within an orthogonal separation of $1.5\Delta$ from the CASP (shaded in purple) and demonstrate initial angular momenta aligned to within $30^{\circ}$ of the CASP (drawn within the dashed blue or dotted grey lines for prograde and retrograde senses respectively).}
    \label{fig:s3_maps_unlikely}
\end{figure*}

\section{Results and Discussion}
\label{sec:s3}

\subsection{Transverse velocity constraints}
\label{sec:s3_maps}

In Fig.~\ref{fig:s3_maps_examples}, we present transverse velocity maps for two Centaurus A satellites ESO324-024 and KK221 -- integrated forward within a spherical NFW halo with a central baryonic component for 5 Gyr -- as examples of likely and unlikely CASP members respectively. Each map is drawn in angular velocity space in proper motion units with a resolution of $1\,\mu\mathrm{as}\,\mathrm{yr}^{-1}$ along each axis. The horizontal axis represents TVs in right ascension compensated for declination $\mu_{\alpha*}$, while the vertical axis gives TVs in declination $\mu_{\delta}$. The origin corresponds to Centaurus A's transverse velocity, which should be propagated to each satellite in order to recover accurate TV constraints. Despite integrating all TVs in the range $[-40,\,40]\,\mu\mathrm{as}\,\mathrm{yr}^{-1}$ along both axes, we only display those within $[-30,\,30]\,\mu\mathrm{as}\,\mathrm{yr}^{-1}$ since a majority of the relevant features are contained therein. All quantitative results are calculated from the full range of integrated TVs.

The four orbital constraints defined in Section~\ref{sec:s2_constraints} are visualised as follows. Firstly, the solid orange line indicates the threshold at which satellites stray further than $2R_{\mathrm{vir}} = 817\,\mathrm{kpc}$ from Centaurus A at one point during their integration, and the shaded orange region covers TVs which result in orbits exceeding this limit. The dotted orange line similarly represents the threshold at which the satellite reaches a maximum distance of $R_{\mathrm{vir}} = 409\,\mathrm{kpc}$. For satellite realisations where an orbit remaining within the latter limit is not possible for any given TV, the dashed line may not be displayed.

On the other hand, the solid red line indicates the threshold at which satellites achieve a minimum distance of 25 kpc from Centaurus A, beyond which we assume they are at a heightened risk of tidal disruption. The corresponding red-shaded region represents TVs resulting in satellite orbits passing through this central region. If no transverse velocities result in an orbit entering this minimum limit, these elements may not be present, and the legend will instead indicate the minimum distance ever achieved.

The heatmap represents the maximum orthogonal satellite-plane distance $d_{\mathrm{max}}$ achieved over the full integration period. Purple regions satisfy the orthogonal distance constraint in equation~(\ref{eq:s2_maxsep}) in order to be considered a potential member of the CASP, while the blue regions represent maximum separations between $1.5\Delta$ and $3\Delta$ -- the latter threshold roughly equivalent to Centaurus A's virial radius.

Finally, the blue dashed line indicates the region corresponding to initial orbital pole alignments of within $30^{\circ}$ to the CASP's normal vector, resulting in prograde orbits with respect to the satellite majority as suggested by line-of-sight velocities. The grey dotted line indicates the same threshold, but resulting in retrograde orbits. The blue and grey lines may be absent from TV maps in cases where no TVs satisfy their respective orbital pole alignment criteria.

We also compensate for distance uncertainties by generating \emph{base}, \emph{least-constrained}, and \emph{most-constrained} TV maps. 
Base TV maps are generated using expected satellite positions, and do not take distance uncertainties into account.
In the latter two maps, we examine the characteristics of the 5 orbits arising from the same number of distance realisations for each initial TV -- the resulting least-constrained and most-constrained maps are composites of these 5 orbital realisations.
For a given satellite, its least-constrained map shows whether a given TV is consistent with a CASP-like orbit. On the other hand, its most-constrained shows whether the satellite with a given TV is guaranteed -- at least, at present time -- to lie on a CASP-like orbit.

\begin{figure*}
	\includegraphics[width=0.28\textwidth]{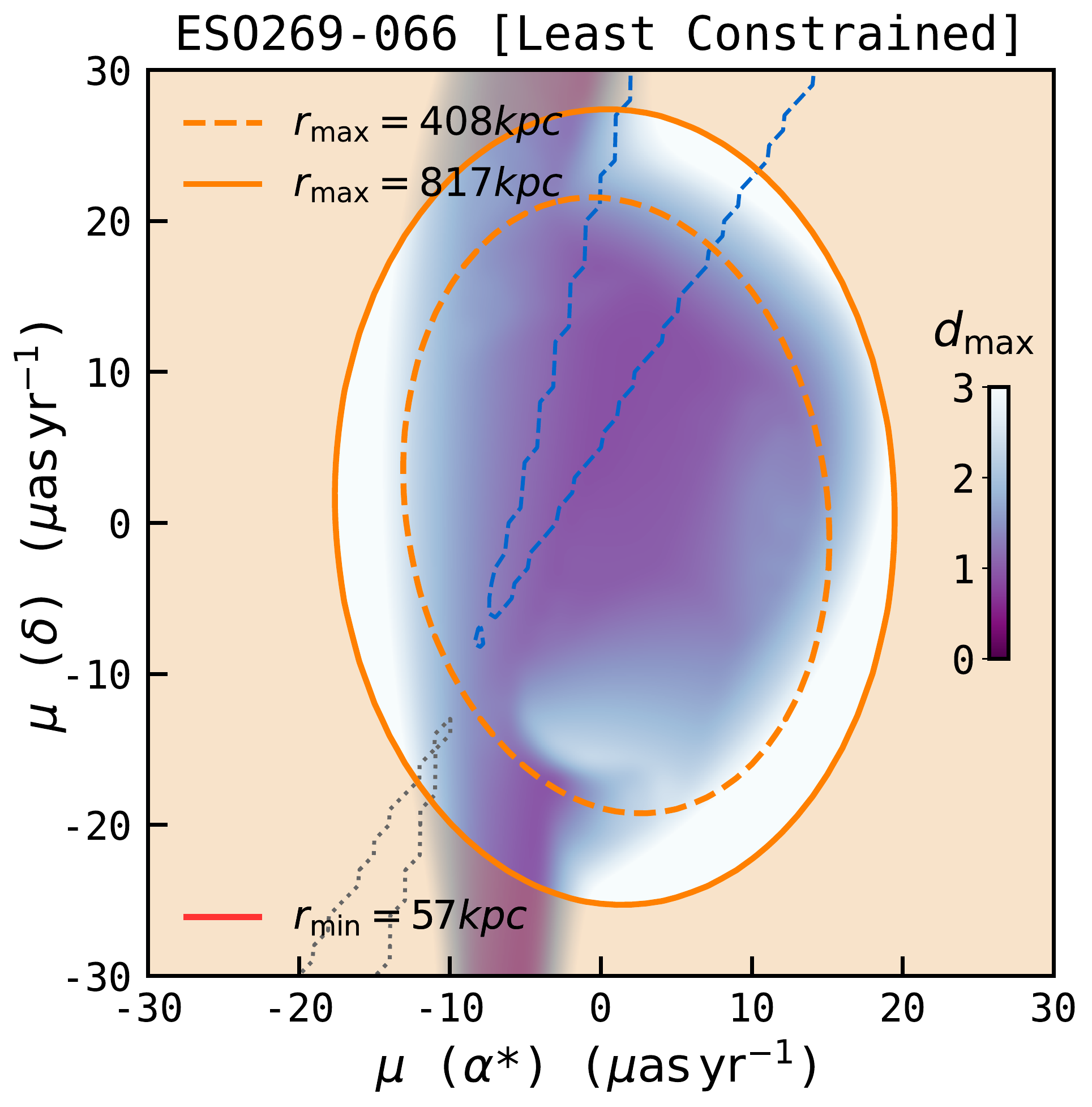}
	\includegraphics[width=0.28\textwidth]{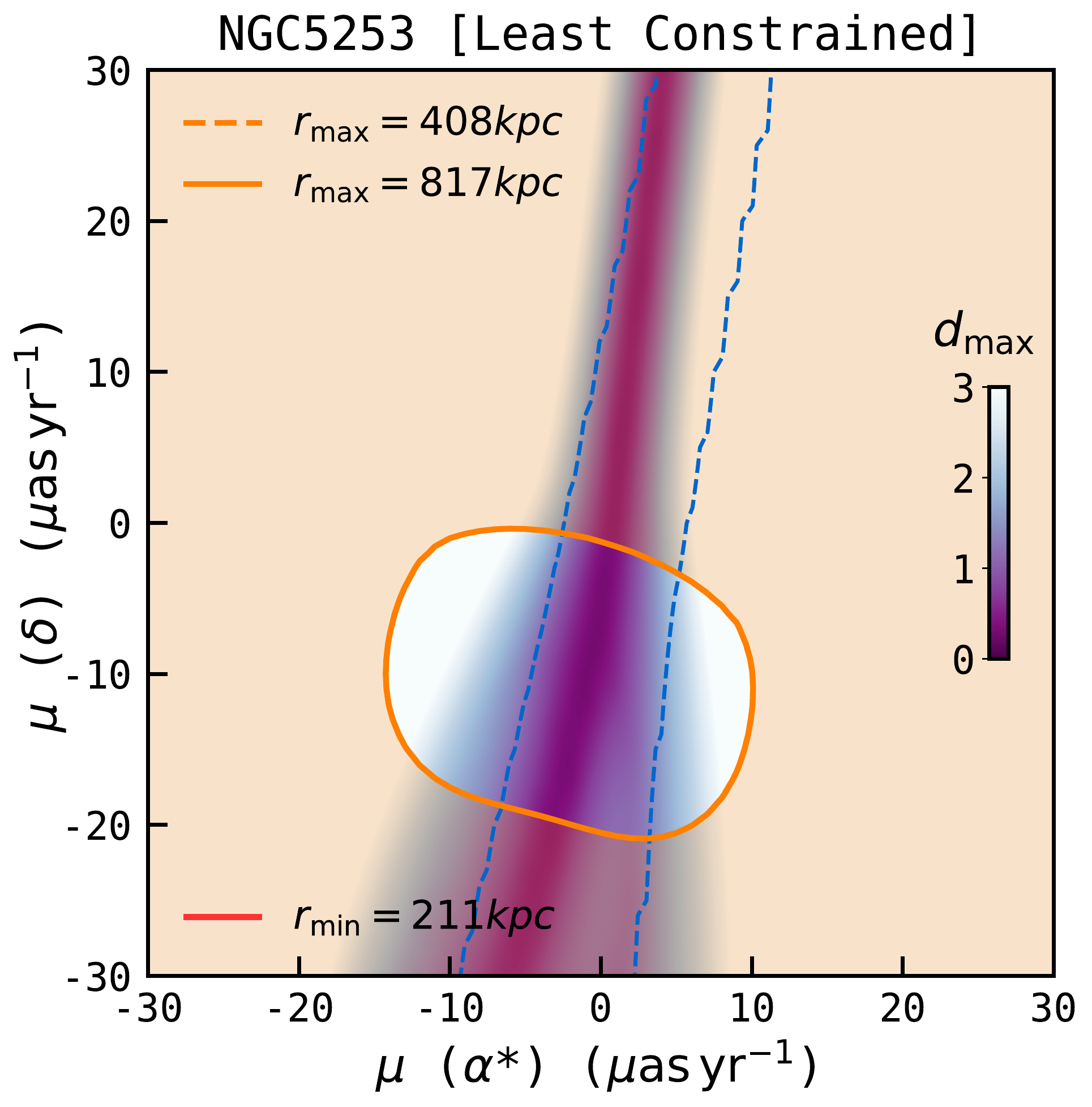}
	\includegraphics[width=0.28\textwidth]{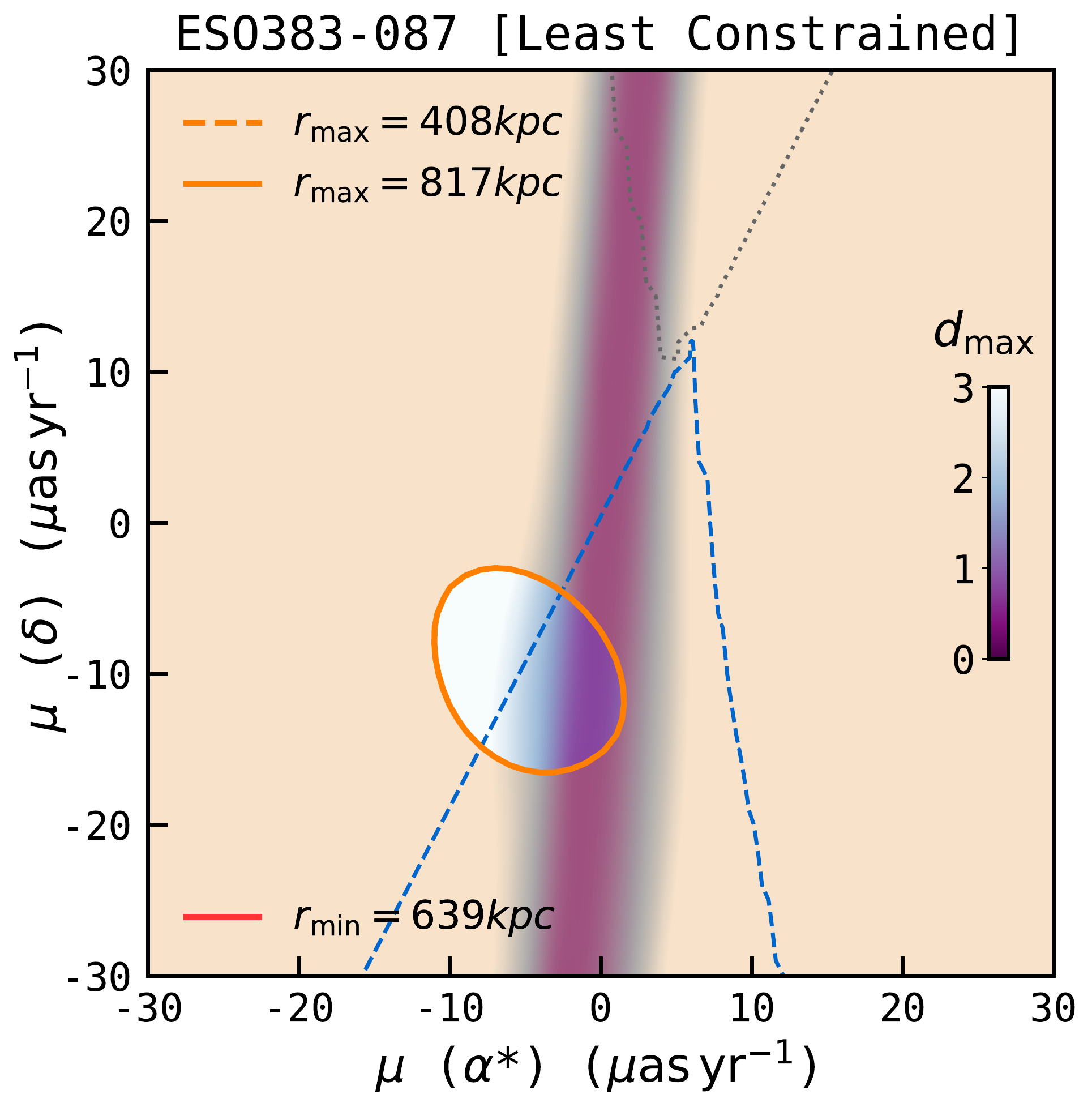}\\
	\includegraphics[width=0.28\textwidth]{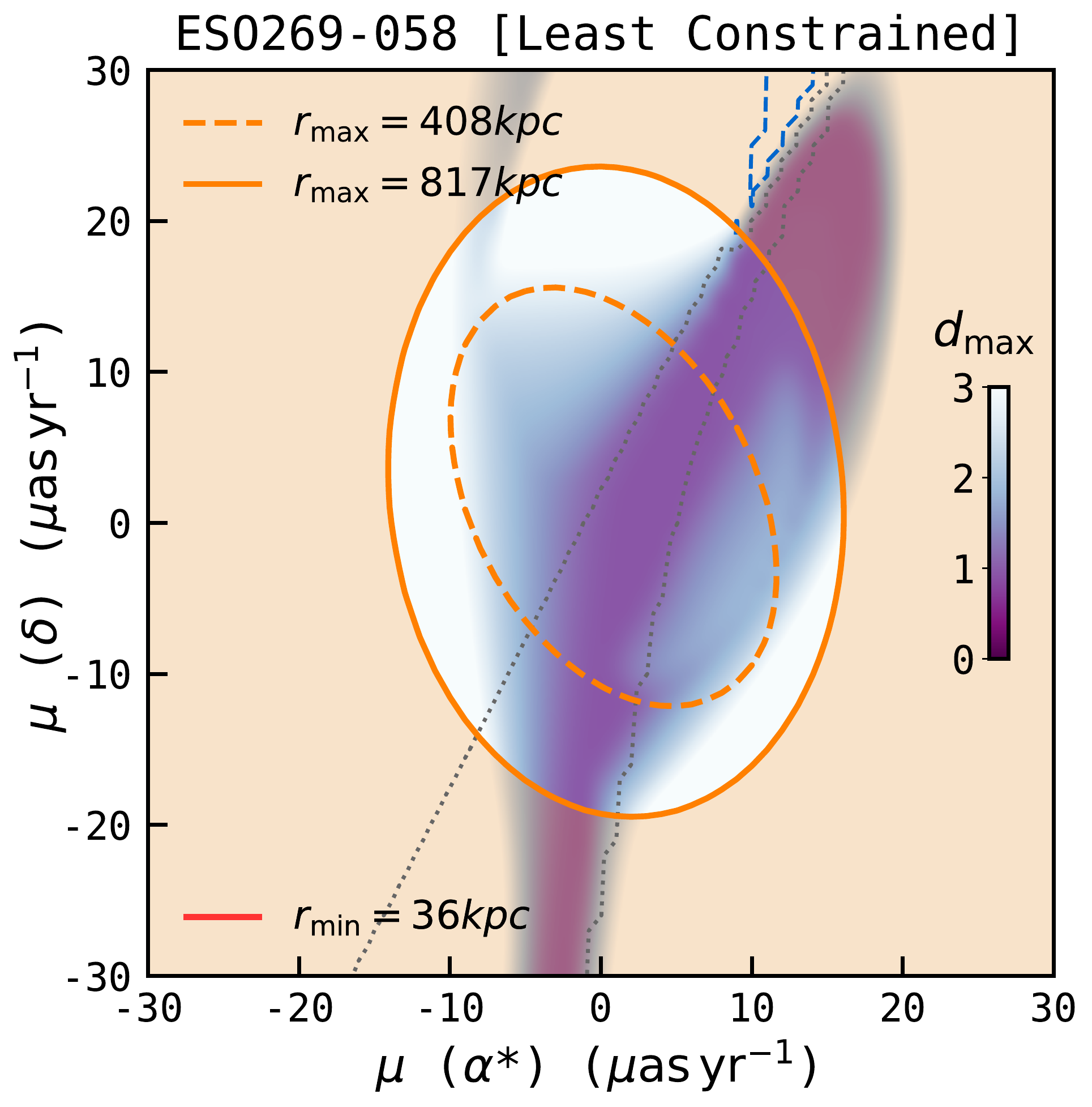}
	\includegraphics[width=0.28\textwidth]{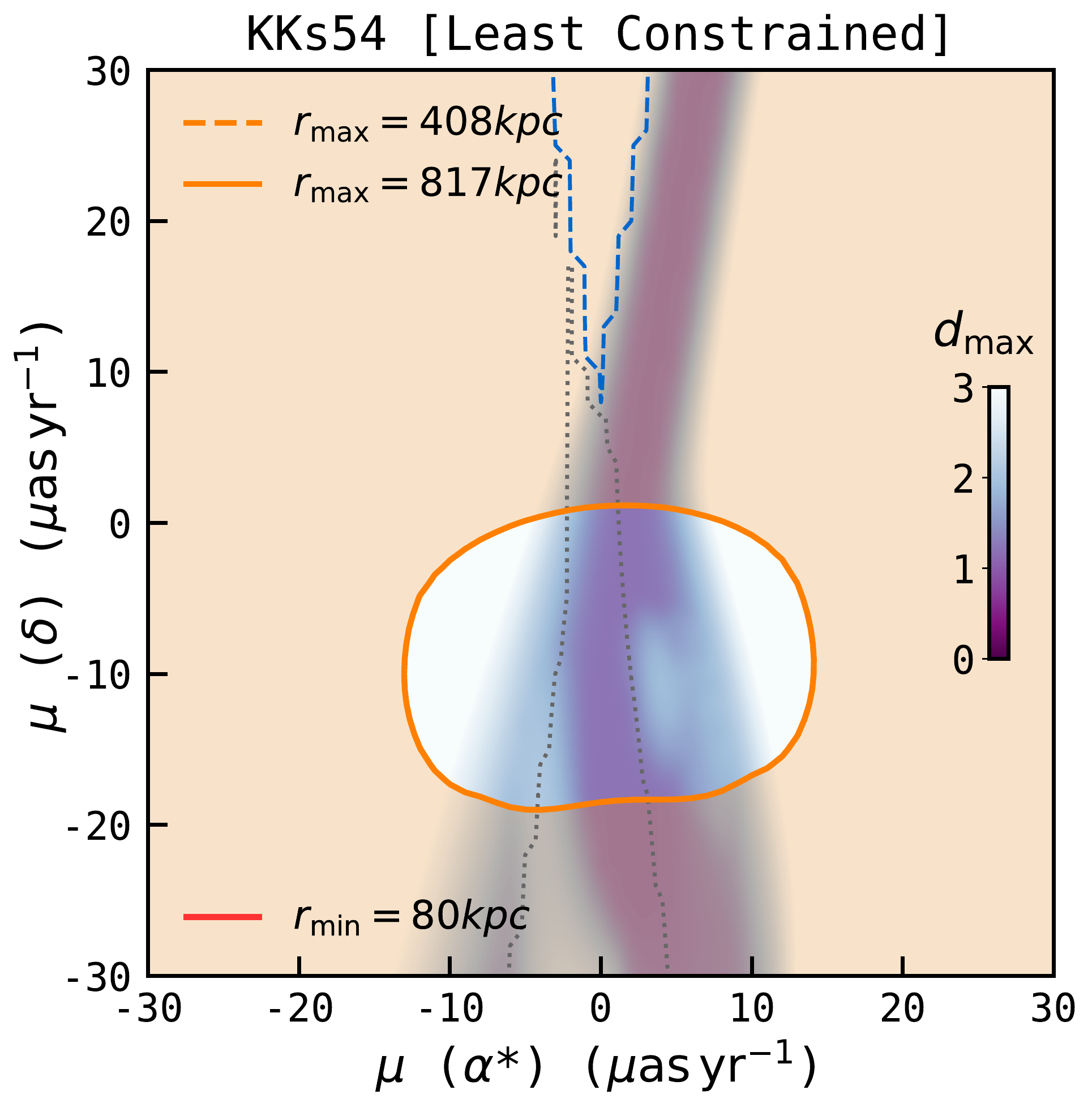}
    \caption{Same as Fig.~\ref{fig:s3_maps_unlikely}, but for 5 \emph{Prograde} (upper panels) and \emph{Retrograde} (lower panels) satellites only consistent with a single orbital sense if orbiting within the Centaurus A plane. In these satellites' least-constrained maps, TVs resulting in realistic orbital distances that remain within $1.5\Delta$ of the CASP (i.e. TVs shaded in purple that avoid the orange and red regions) can only be found within either the blue dashed or grey dotted contours (indicating prograde and retrograde orbits respectively). The four satellites other than ESO269-066 remain consistent with only prograde or retrograde motion even when disregarding the orthogonal separation criterion (see equation~\ref{eq:s2_maxsep}).}
    \label{fig:s3_maps_proret}
\end{figure*}

For a given transverse velocity to meet an orbital constraint in the least-constrained map, at least 1 out of 5 realisations needs to satisfy the constraint. For instance, if one orbit realisation manages to remain within an orthogonal distance of $d_{\mathrm{max}}=1.5\Delta$ of the CASP while the other 4 exceed it, the corresponding TV is still drawn within the purple region -- put simply, we show the minimum $d_{\mathrm{max}}$ amongst the 5 distance realisations for each TV point. This approach is useful in demonstrating which TVs are consistent with orbiting along the CASP given the distance uncertainties. Note that prograde TVs are 'preferred' over retrograde TVs -- if both orbital senses are possible for a given TV over its set of distance realisations, that TV is marked as prograde in the least-constrained TV map. Hence, for each satellite, its corresponding least-constrained map answers the question: how low of a $d_{\mathrm{max}}$ value can ever be achieved within the distance uncertainties while satisfying the remaining three orbital constraints?

Conversely, all 5 realisations must satisfy an orbital constraint in the most-constrained map. For each satellite, its corresponding most-constrained map answers the question: how low of a $d_{\mathrm{max}}$ value can be ensured within the distance uncertainties while satisfying the remaining three orbital constraints? Hence, a transverse velocity indicated in the most-constrained map as resulting in a CASP-aligned orbit is highly likely to do so regardless of the satellite's true distance.

\subsection{Classifying potential plane members}
\label{sec:s3_classify}

In the absence of direct proper motion measurements, constrained transverse velocities can provide insights into the kinematics of satellite galaxies that are otherwise difficult to obtain from line-of-sight velocities alone. These insights are dependent on the assumptions we make regarding their orbital characteristics. 
From our assumption of bound and long-lived orbits, we motivate a subsample of transverse velocities resulting in realistic orbital distances -- specifically, those that produce orbits obeying the maximum and minimum distance constraints in equations (\ref{eq:s2_maxdist}) and (\ref{eq:s2_mindist}) respectively.
The fraction of TVs resulting in realistic orbital distances that also obey the plane separation and orbital pole alignment requirements in equations (\ref{eq:s2_maxsep}) and (\ref{eq:s2_limalign}) -- while its value does not directly constitute a corresponding probability, -- can be considered to be a rough tracer of the satellite's likelihood of participating in (the rotationally supported) Centaurus A plane.

To illustrate our argument, we look to TV maps for satellites ESO324-024 and KK221 in Fig.~\ref{fig:s3_maps_examples}. In its base map, ESO324-024 demonstrates a relatively large number of TVs which result in orbits remaining within $d_{\mathrm{max}} < 1.5\Delta$ of the CASP. This is further accentuated in the least-constrained TV map -- over $50$ per cent of TVs resulting in realistic orbital distances have the potential to generate $1.5\Delta$ orbits. (This fraction is, of course, much lower in the most-constrained map, as only a reduced number of TVs can guarantee a $d_{\mathrm{max}} < 1.5\Delta$ orbit regardless of the distance realisation adopted.) While this is in part due to the satellite's initial proximity to Centaurus A, a notable fraction of $1.5\Delta$ TVs also demonstrate orbital pole alignments of $30^{\circ}$ or better, fulfilling our requirements to be considered a CASP member. However, also note that distance uncertainties heavily contribute to ESO324-024's orbital volatility -- in the most-constrained map, all TVs resulting in realistic orbital distances that also ensure a $1.5\Delta$ orbit additionally risk tidal disruption from a close encounter with Centaurus A.

\begin{figure*}
	\includegraphics[width=0.80\textwidth]{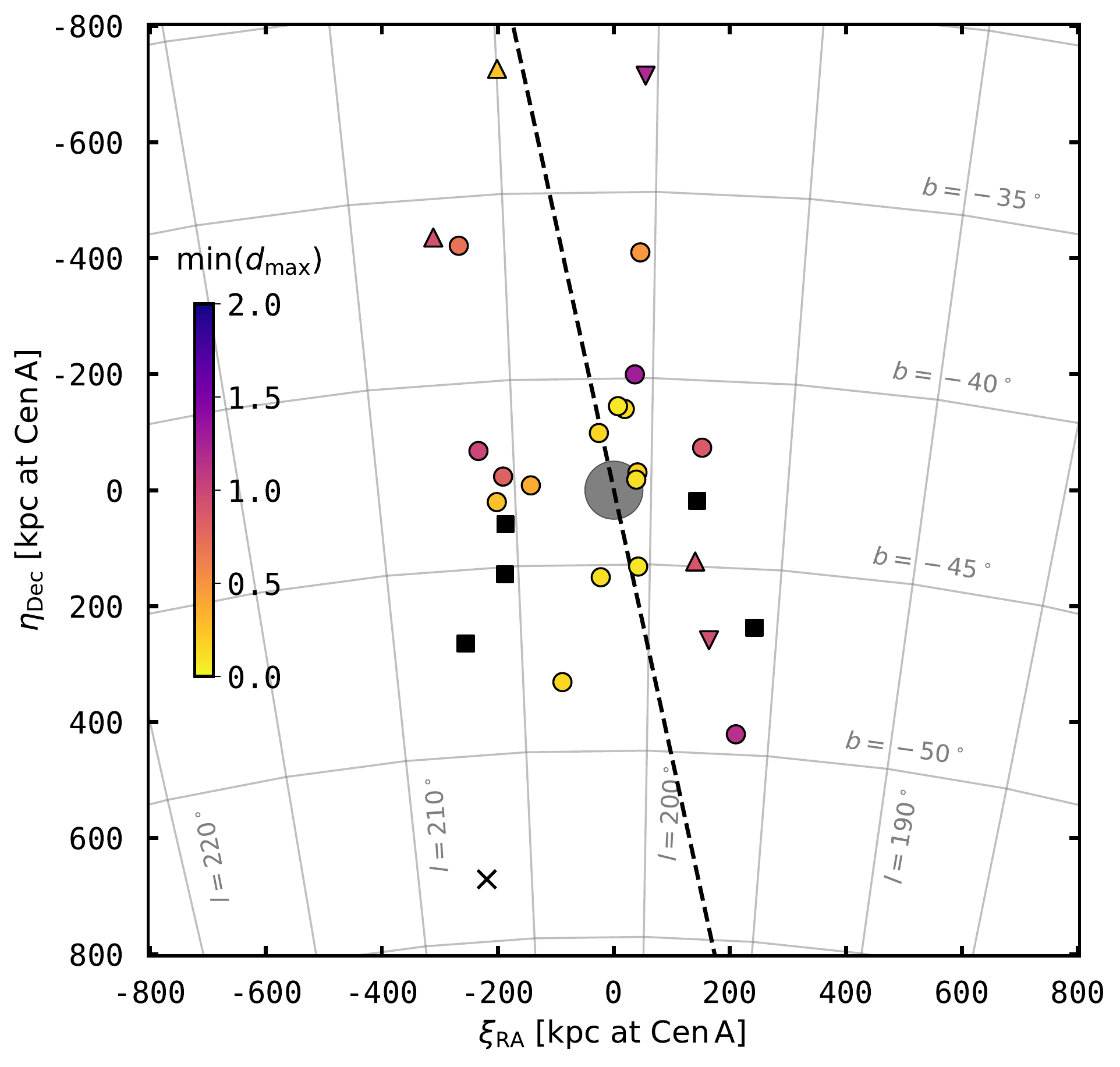}
    \caption{The on-sky distribution of the Centaurus A system, consisting of 27 satellite galaxies with available distances and radial velocity measurements. The horizontal and vertical axes $\xi$ and $\eta$ are aligned with equatorial coordinates at Centaurus A's location (drawn as a grey circle), and are measured in kpc at Centaurus A's distance of $3.68\,\mathrm{Mpc}$. Satellites are drawn with different shapes according to their classification in Section~\ref{sec:s3_classify}. Black squares are \emph{Unlikely} to participate in the CASP, \emph{Prograde} (co-rotating) satellites are indicated with upwards triangles, \emph{Retrograde} satellites with downwards triangles, and satellites consistent with both orbital senses (\emph{Both}) are drawn as circles. In addition, satellites consistent with CASP membership are coloured by their minimum plane separation achieved by transverse velocities which result in bound and long-lived orbits, measured in terms of the CASP's rms plane height $\Delta=134\,\mathrm{kpc}$. The black dashed line represents the line in which the CASP intersects with our $\xi-\eta$ plane. KKs59, rejected from our working sample due to only having a velocity measurement, is drawn for reference as a black cross.}
    \label{fig:s3_skymap}
\end{figure*}

On the other hand, KK221 is a satellite that only manages to orbit within $3\Delta$ of the CASP. In addition, no transverse velocities result in orbital pole alignments of within $45^{\circ}$ in the least-constrained TV map, and KK221 can be unequivocally rejected as a CASP member. This is a direct consequence of its initial position being nearly orthogonal to the CASP, and can be inferred without referring to our TV constraints. However, a satellite's consistency with a CASP-like orbit becomes nuanced when the satellite demonstrates an initial orthogonal distance from the plane, $d_0$, which is slightly below our defined threshold of $1.5\Delta$. As later seen in Fig.~\ref{fig:s3_skymap} (wherein many satellites found to be inconsistent with CASP membership share similar initial plane separations as those consistent with CASP membership), whether such a satellite has a TV resulting in a CASP-like orbit is influenced by the magnitude of its heliocentric distance uncertainty, its line-of-sight velocity and associated energy, and assumptions regarding the potential of its host halo.

In Table~\ref{tab:s3_metrics}, we present a number of useful metrics and orbital properties for all 27 Centaurus A satellites with TRGB distances and heliocentric velocity measurements in our sample. Notable here is $f(\mathrm{CASP})$, which represents the fraction of TVs resulting in realistic orbital distances that additionally produce orbits satisfying both the plane separation and orbital pole alignment requirements -- that is, they satisfy all four criteria to be considered likely CASP participants. Upper and lower uncertainties are derived from least-constrained and most-constrained TV maps respectively.

\begin{figure*}
	\includegraphics[width=0.24\textwidth]{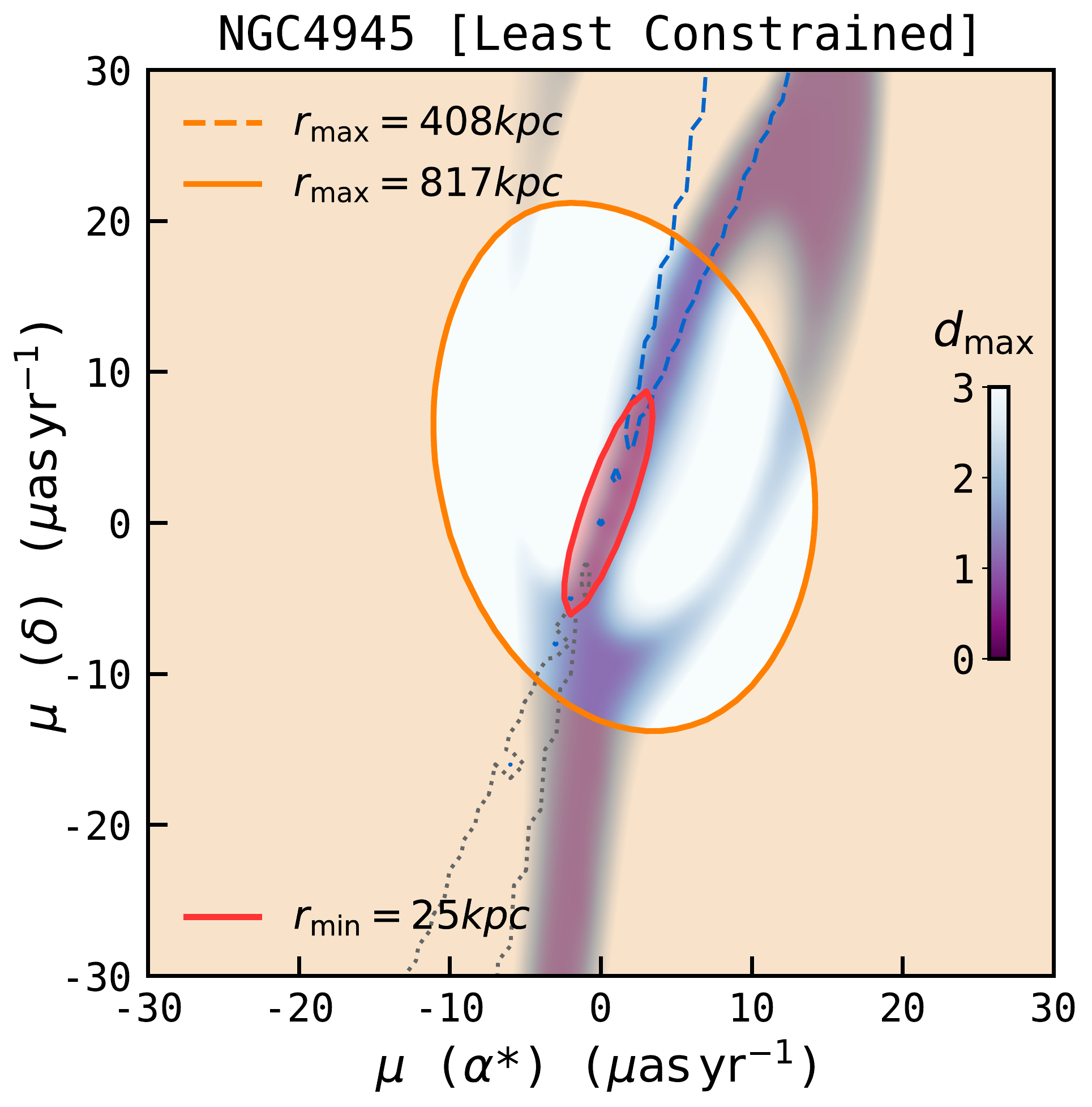}
	\includegraphics[width=0.24\textwidth]{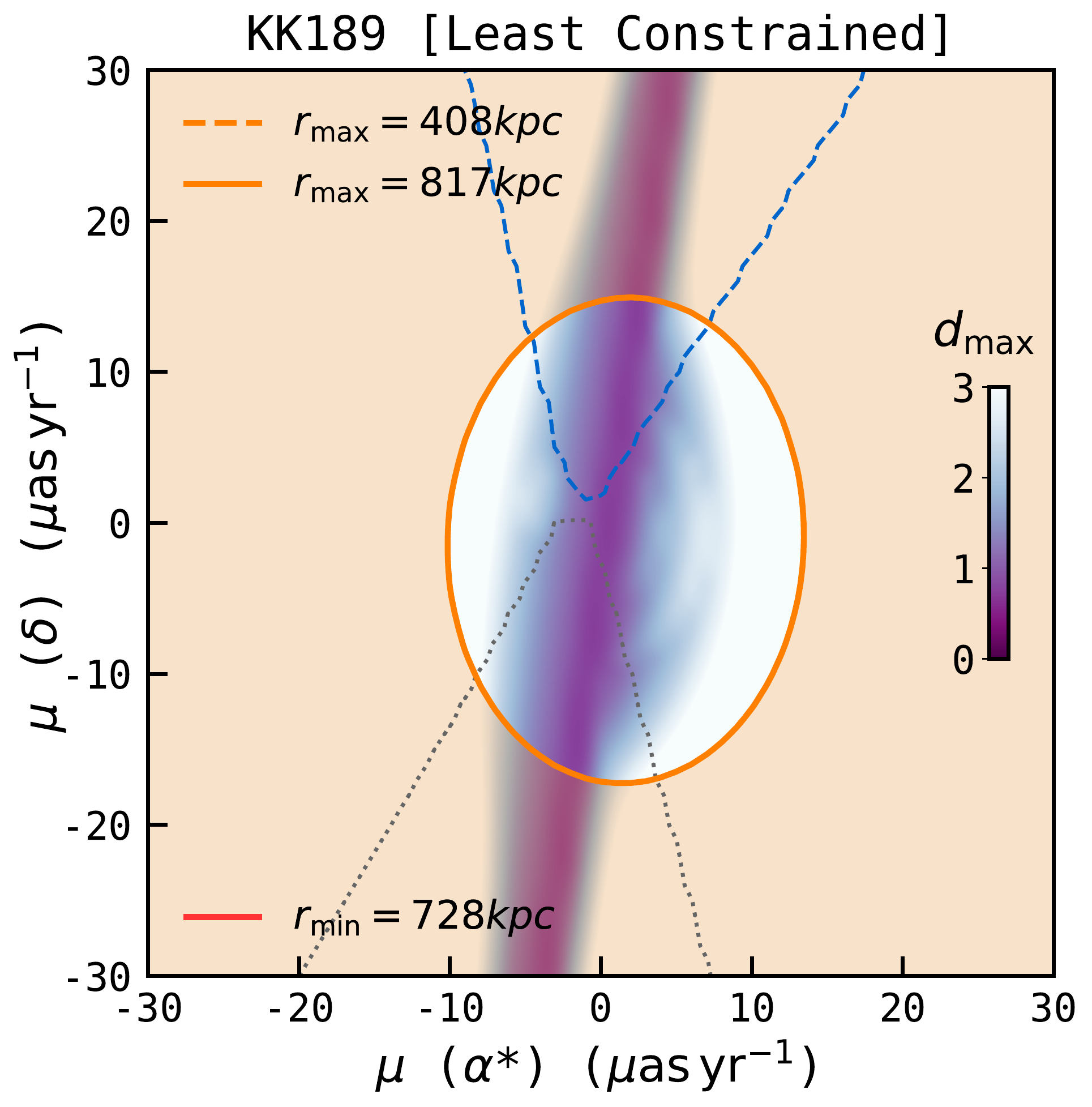}
	\includegraphics[width=0.24\textwidth]{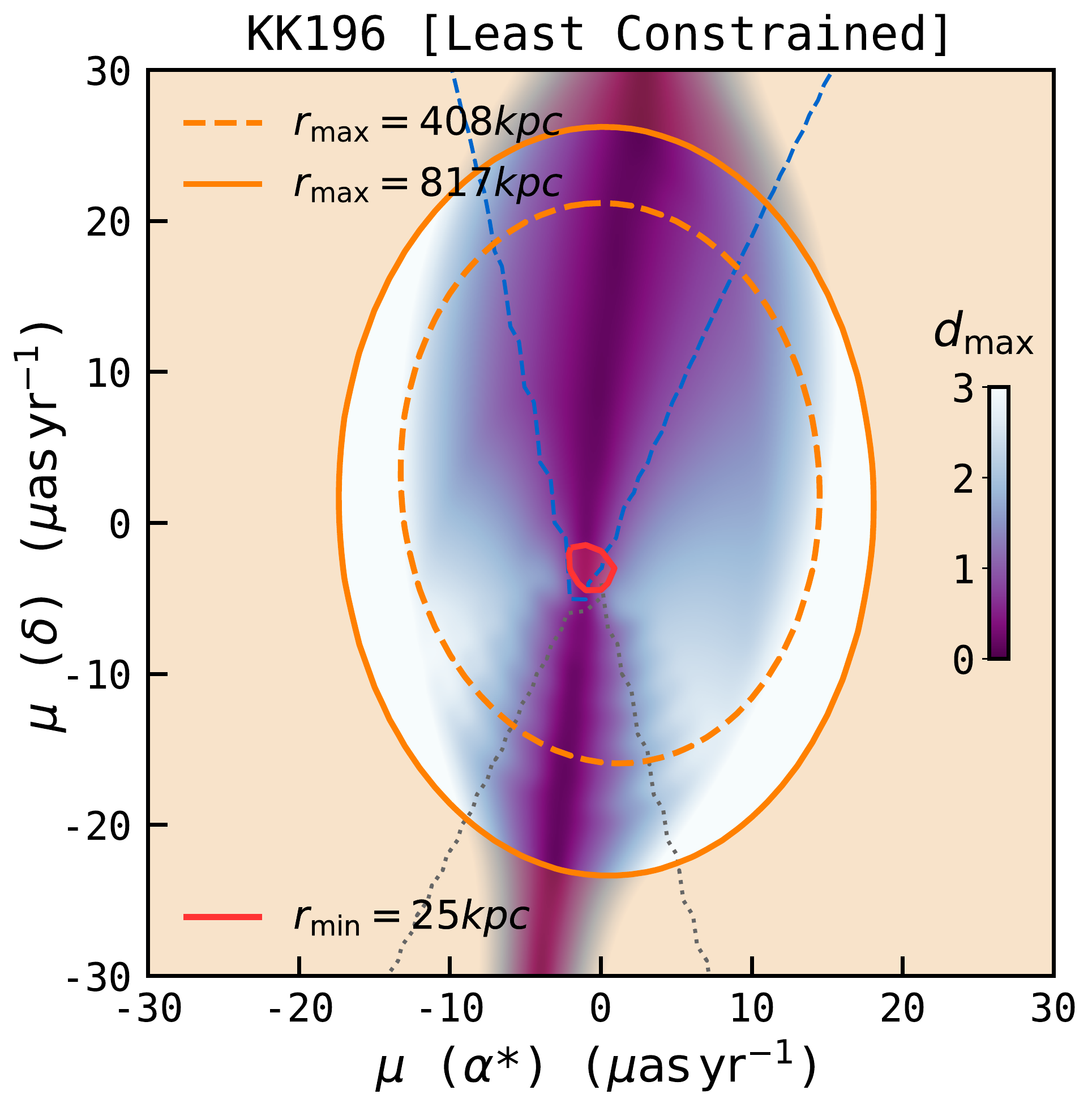}\\
	\includegraphics[width=0.24\textwidth]{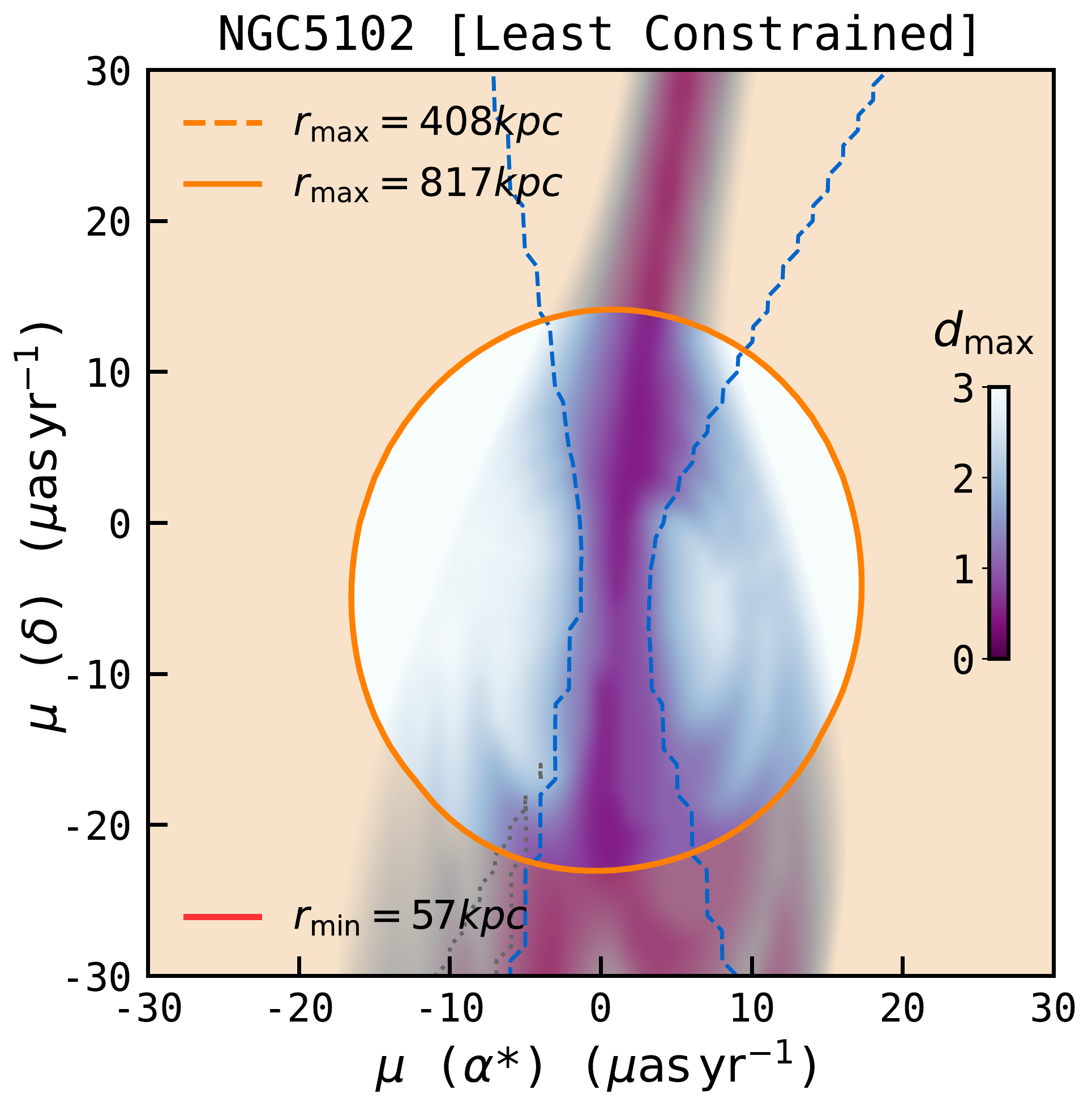}
	\includegraphics[width=0.24\textwidth]{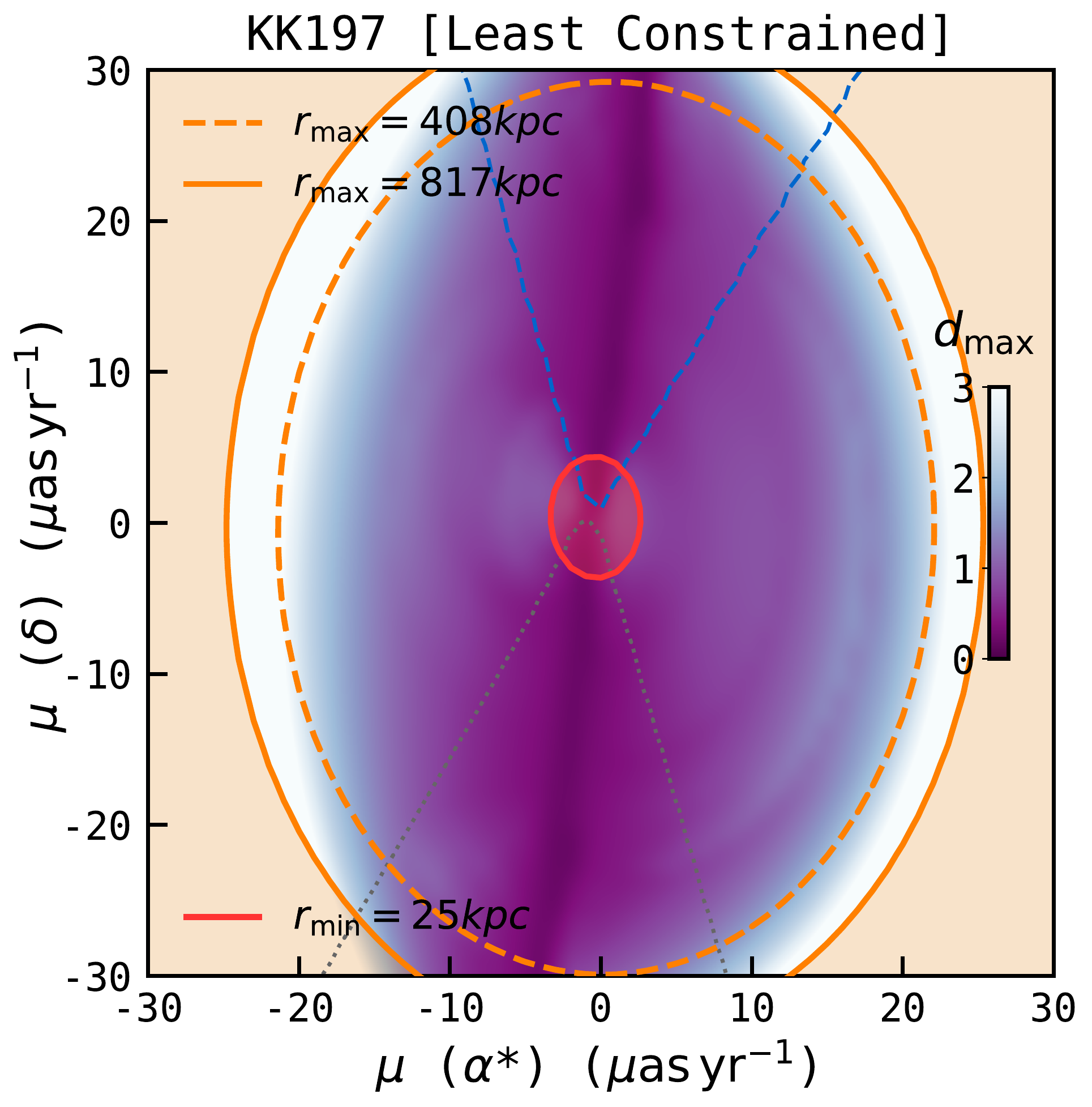}
	\includegraphics[width=0.24\textwidth]{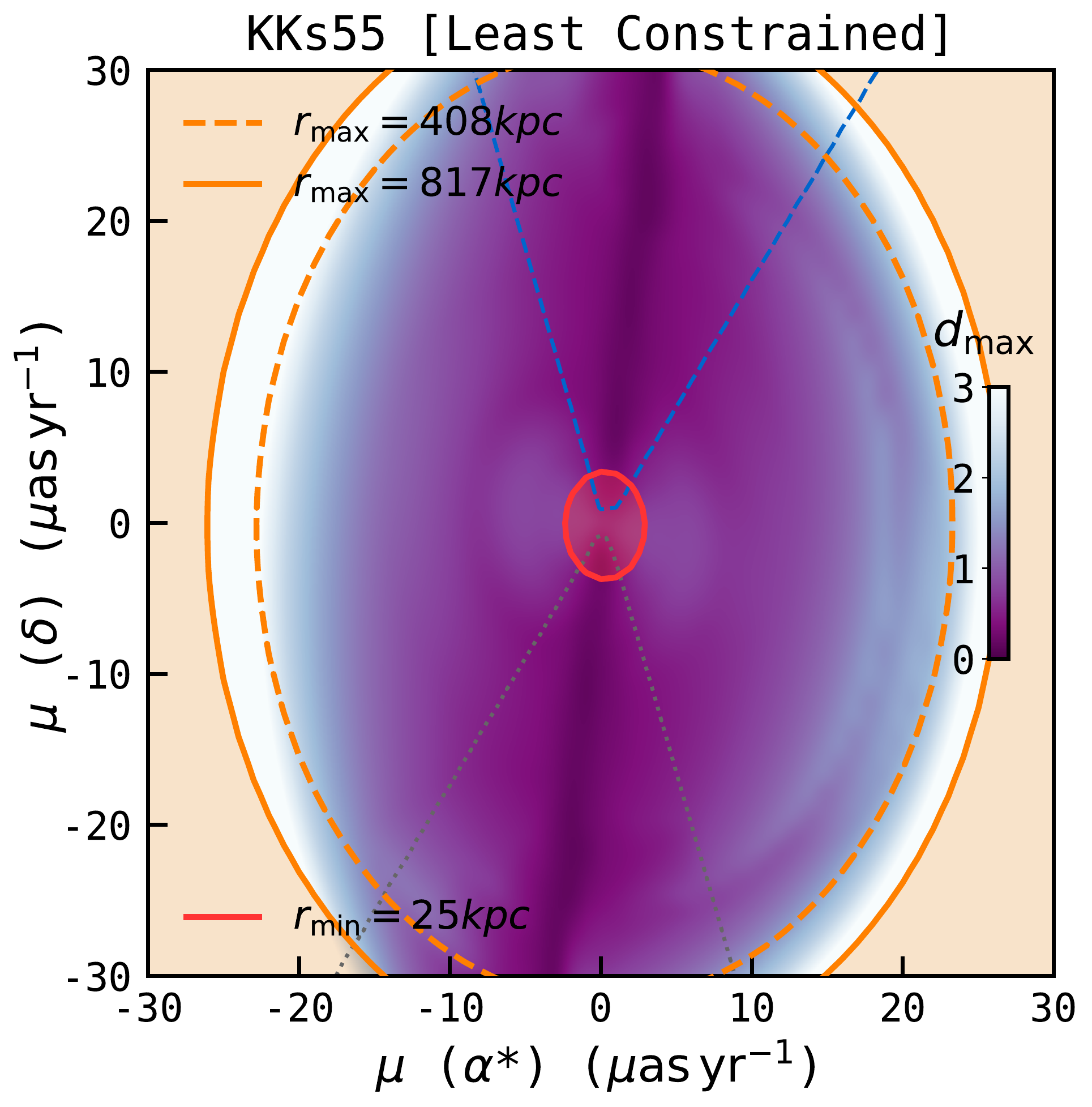}
	\includegraphics[width=0.24\textwidth]{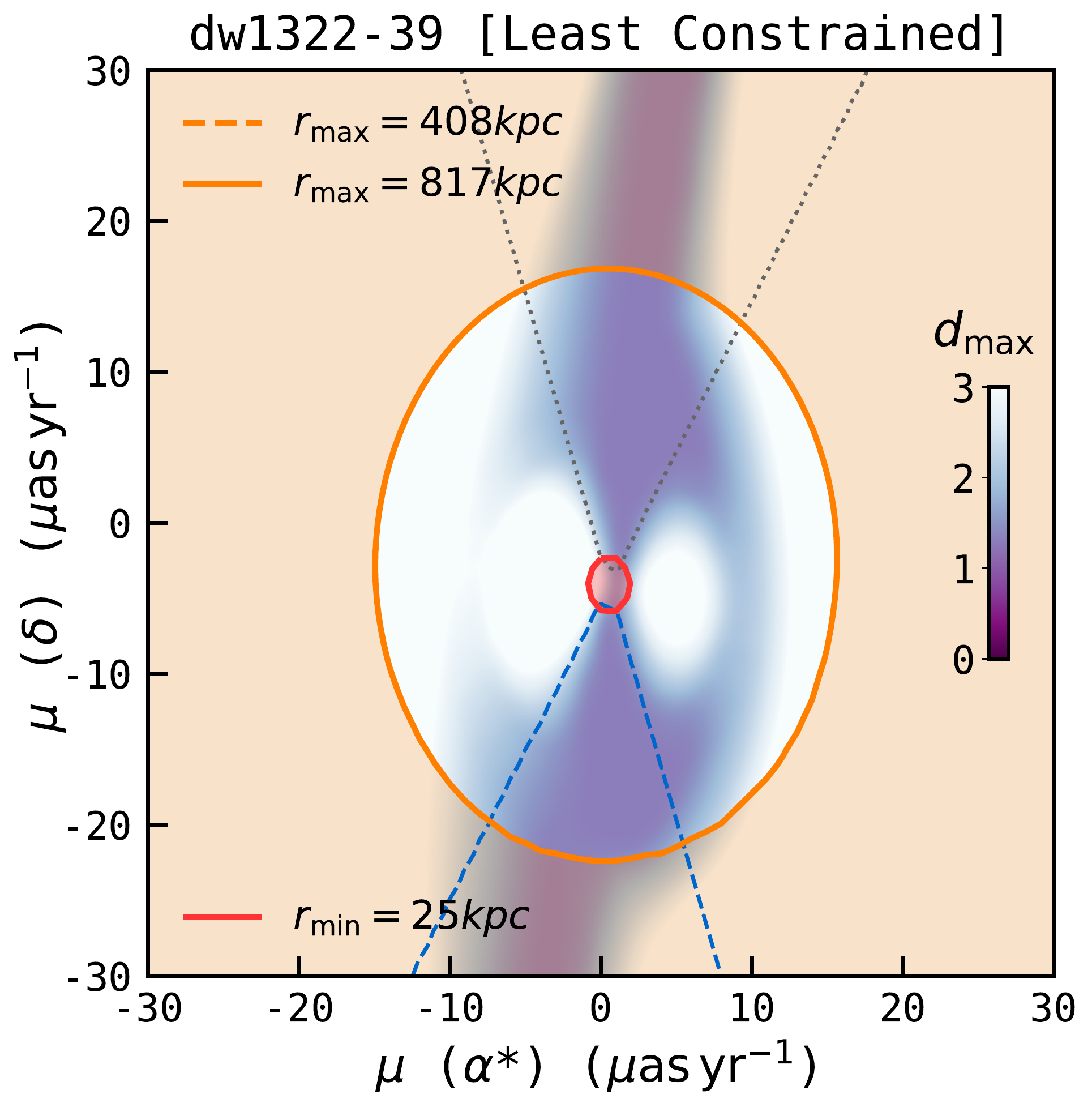}\\
	\includegraphics[width=0.24\textwidth]{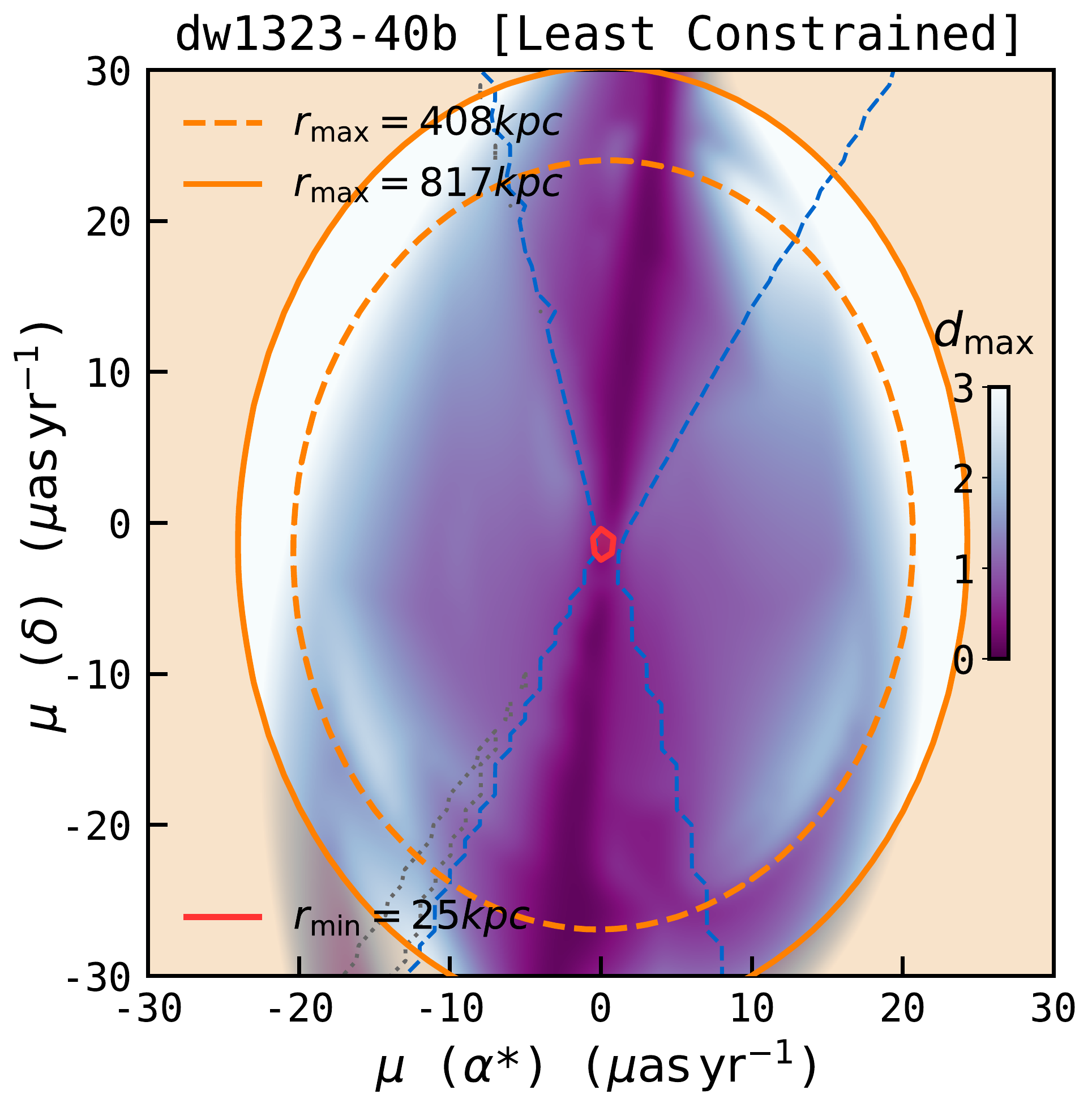}
	\includegraphics[width=0.24\textwidth]{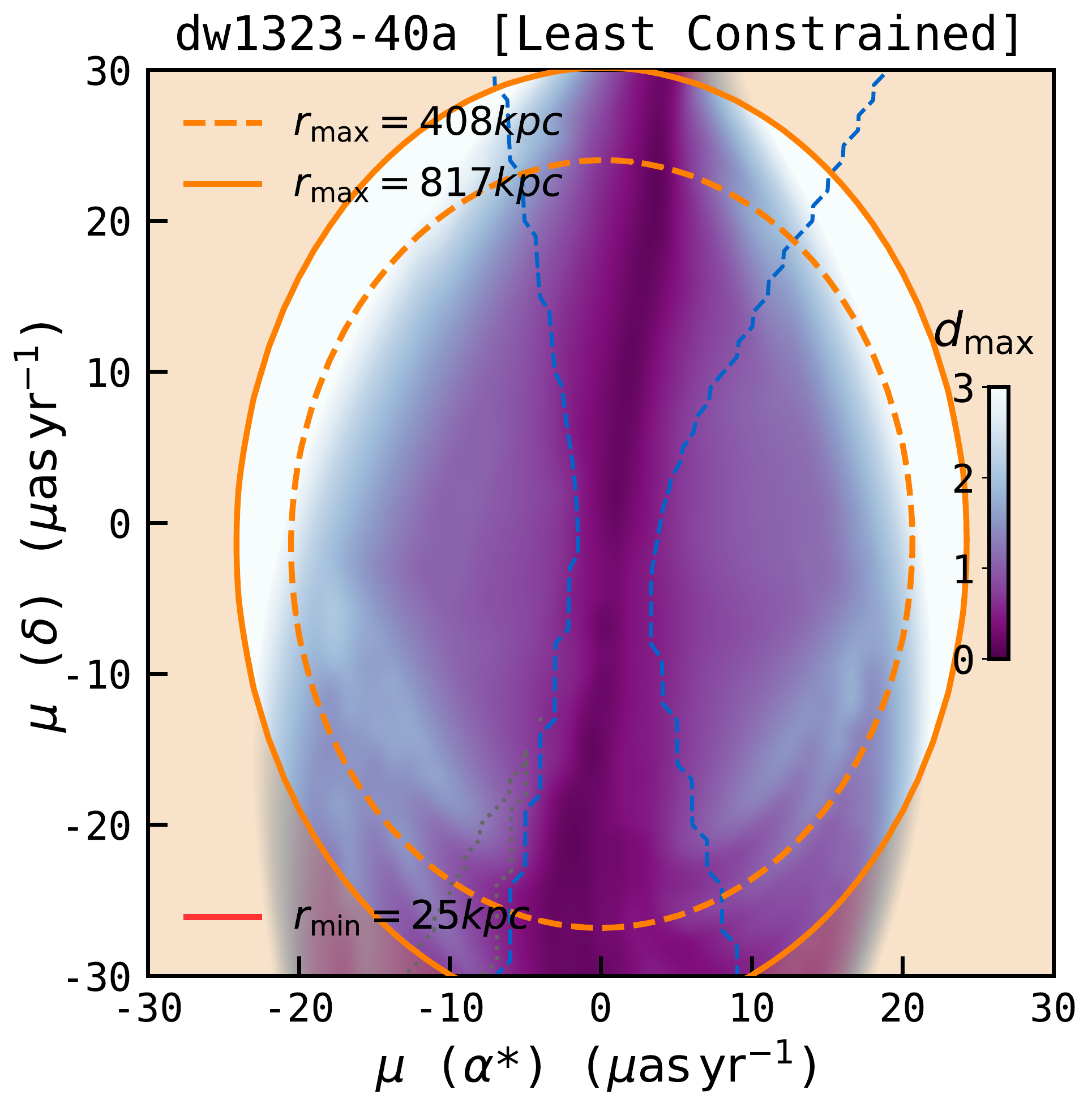}
	\includegraphics[width=0.24\textwidth]{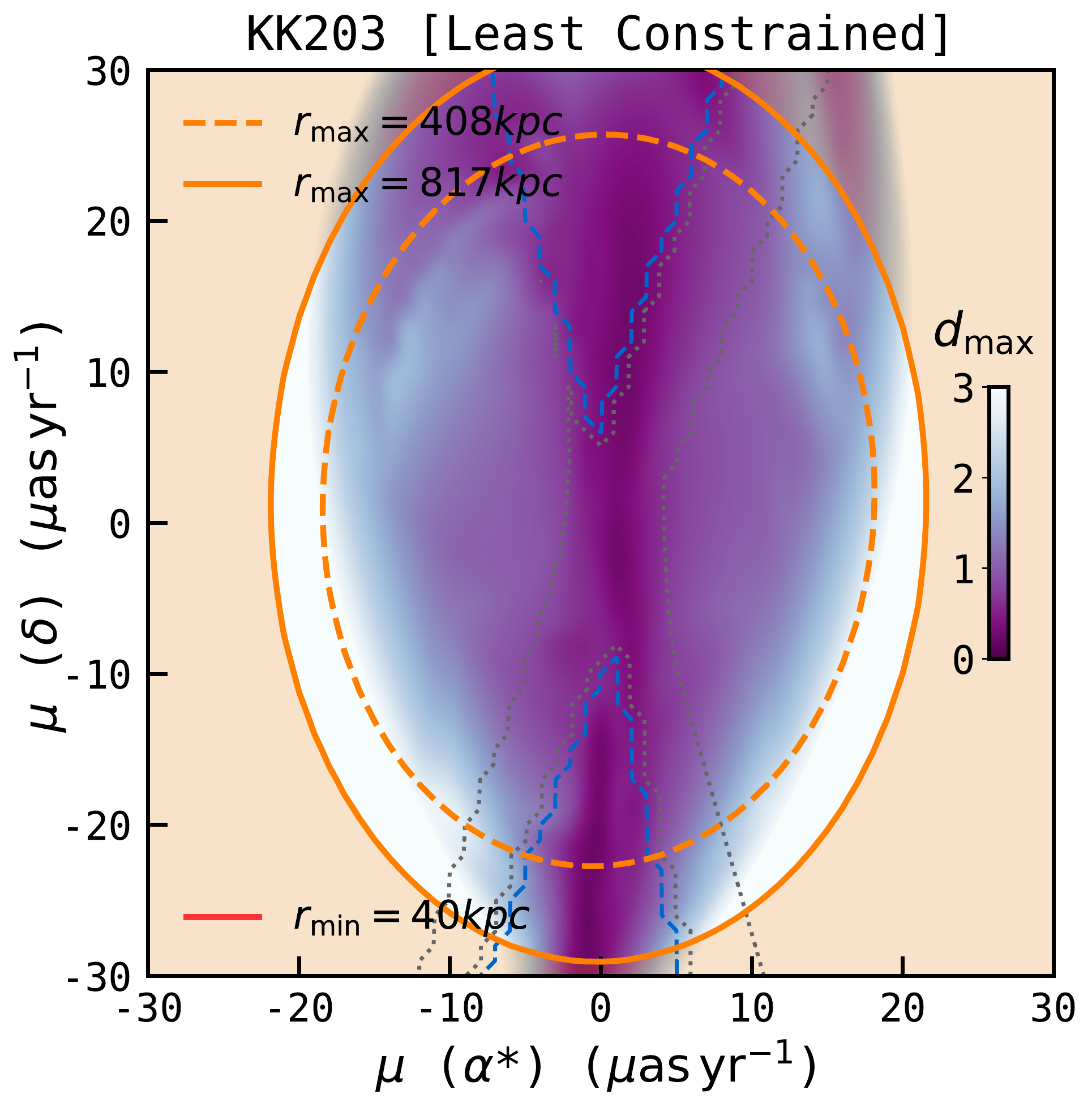}
	\includegraphics[width=0.24\textwidth]{map/map_heat_ESO324-024_best.pdf}\\
	\includegraphics[width=0.24\textwidth]{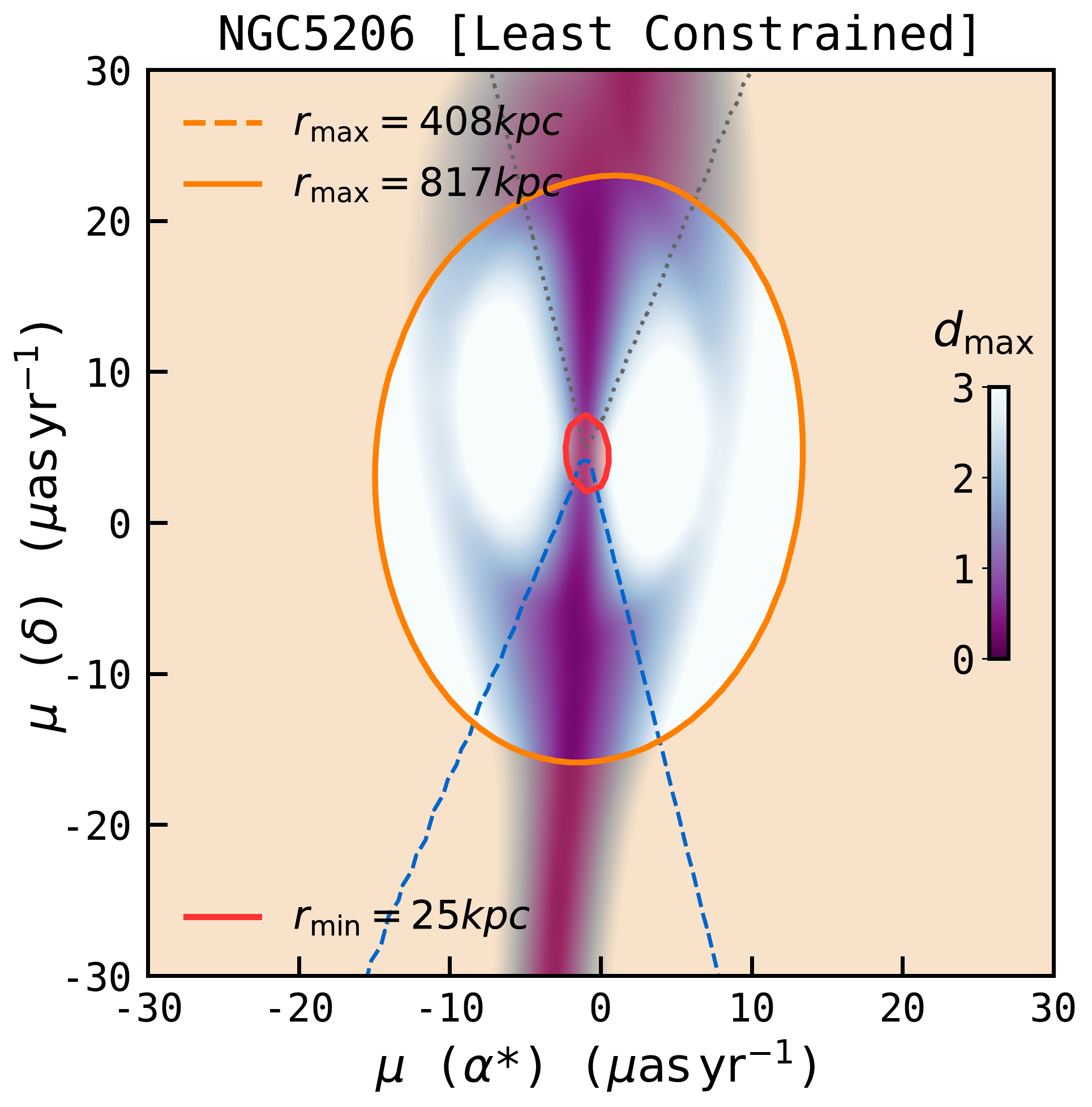}
	\includegraphics[width=0.24\textwidth]{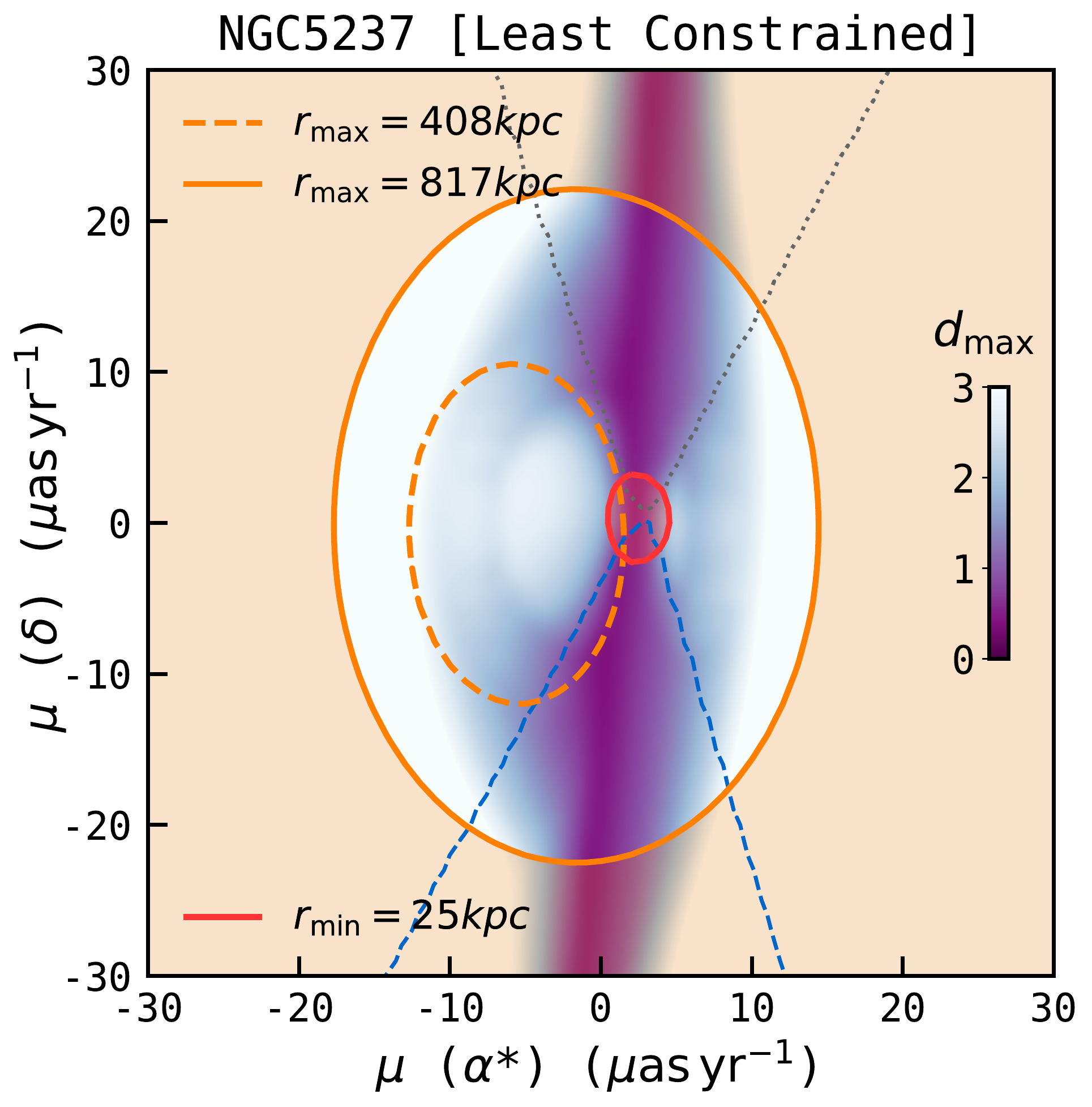}
	\includegraphics[width=0.24\textwidth]{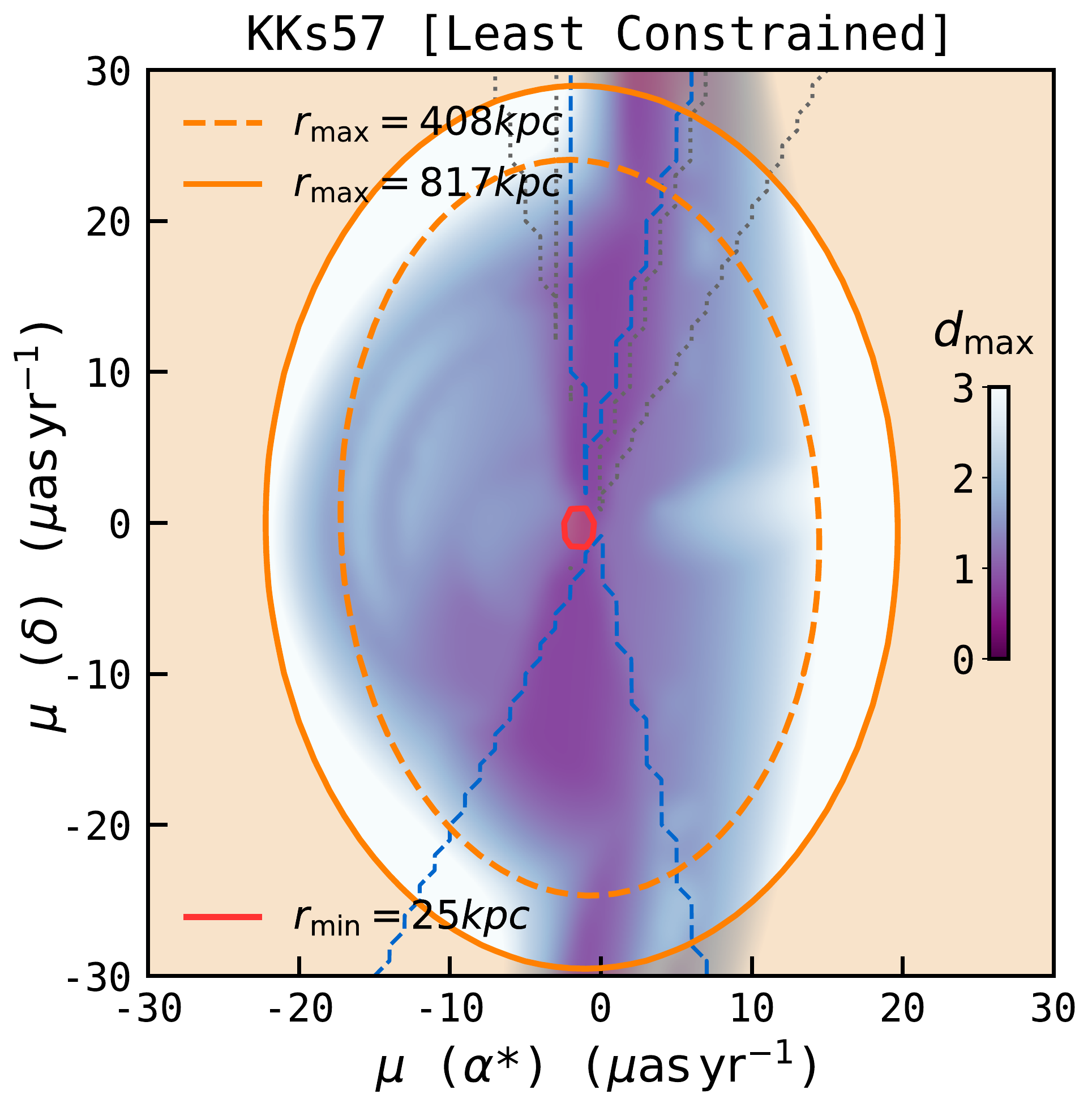}
	\includegraphics[width=0.24\textwidth]{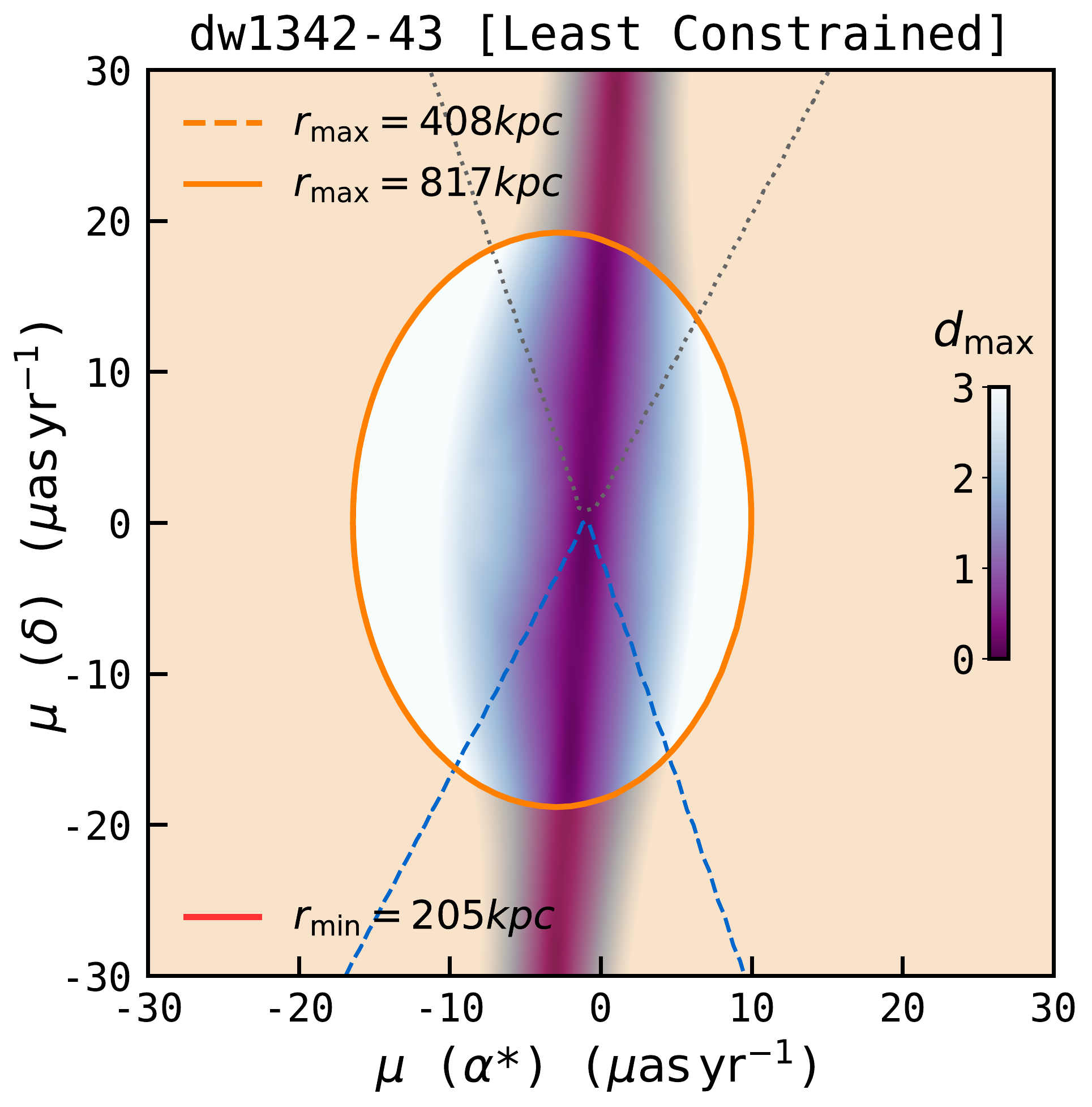}\\
	\includegraphics[width=0.24\textwidth]{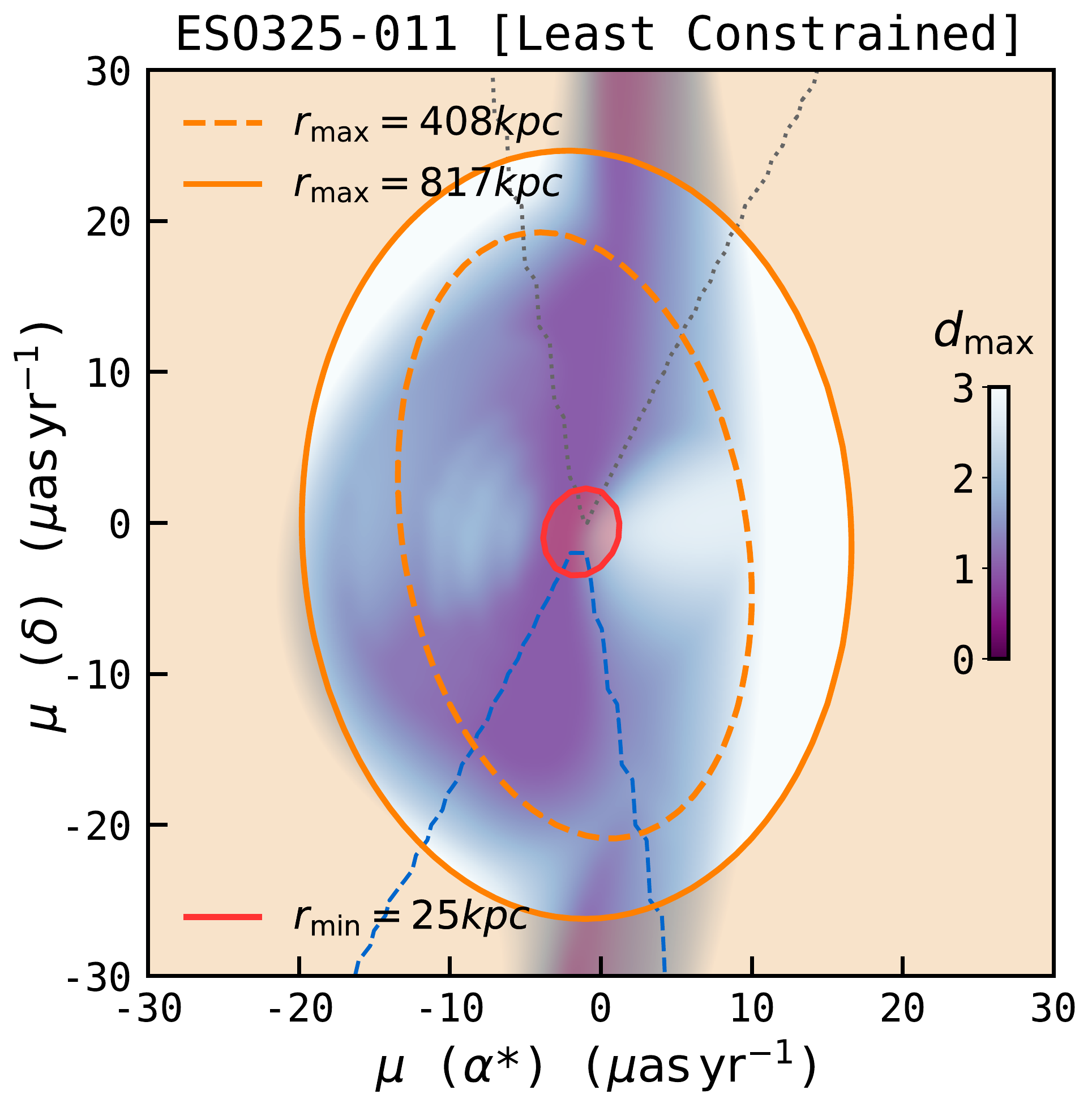}
	\includegraphics[width=0.24\textwidth]{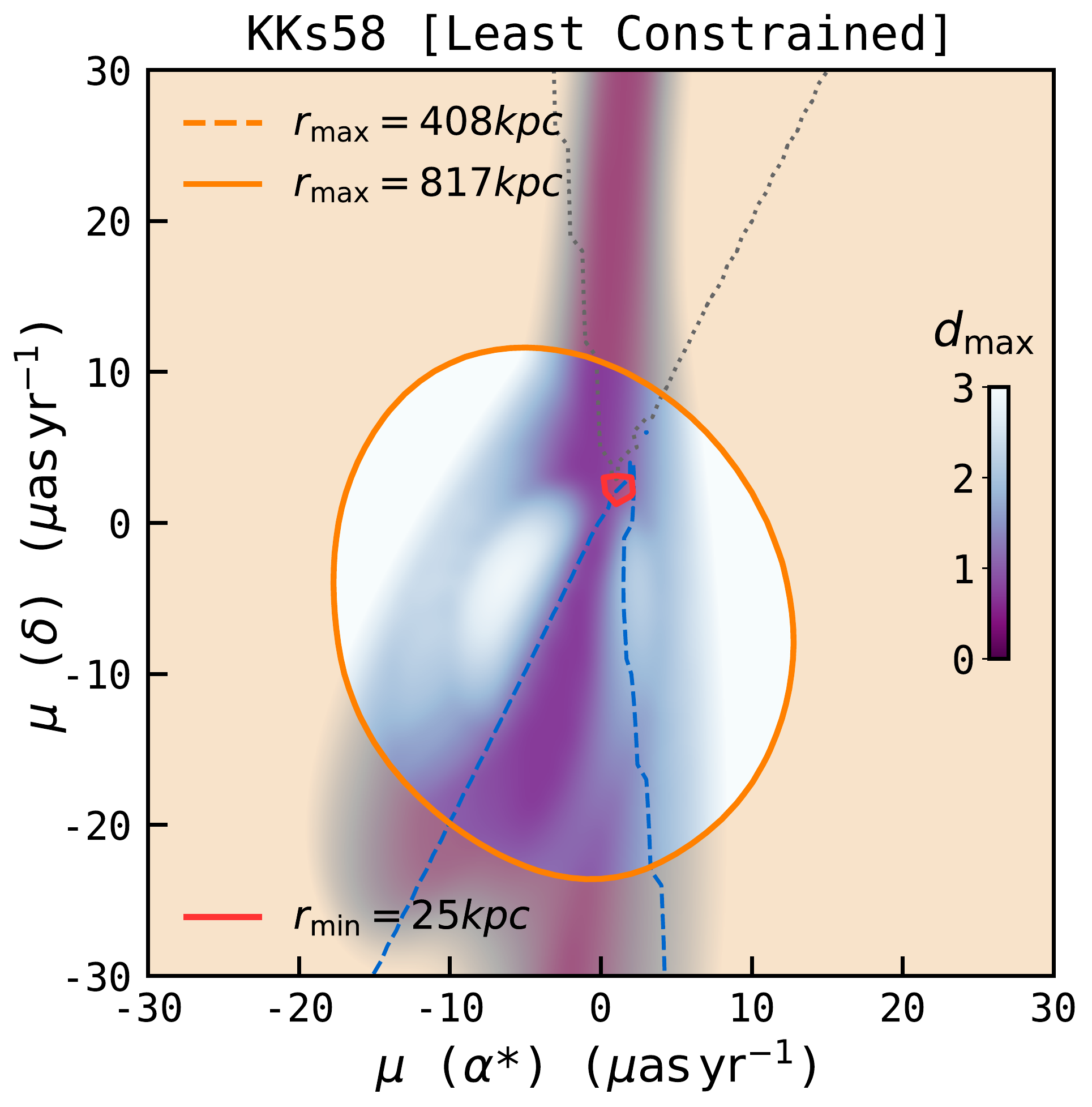}
    \caption{Same as Fig.~\ref{fig:s3_maps_unlikely}, but for 17 satellites consistent with CASP membership with \emph{Both} prograde and retrograde orbits. Unlike the \emph{Prograde} and \emph{Retrograde} satellites in Fig.~\ref{fig:s3_maps_proret}, purple-shaded TVs outside the orange and red regions can be found within both the blue dashed and grey dotted contours.}
    \label{fig:s3_maps_both}
\end{figure*}

Based on this metric, we classify the satellites into 2 types. The 5 satellite galaxies in the \emph{Unlikely} category lacks any TVs in the least-constrained map that satisfy the four requirements for CASP membership -- assuming our model for the Centaurus A potential is accurate, these satellites cannot orbit along the CASP regardless of the distance realisation adopted. Least-constrained TV maps for these off-plane satellites are shown in Fig.~\ref{fig:s3_maps_unlikely}. The remaining 22 CASP member candidates are then subdivided into 3 categories. For the 3 \emph{Prograde} satellites, all possible orbits that satisfy the separation and alignment requirements co-orbit with respect to the satellite majority; the 2 \emph{Retrograde} satellites counter-orbit instead. The 17 satellites classified as \emph{Both} may demonstrate either orbital sense depending on the distance realisation and TV selected. TV maps for satellites classified as \emph{Prograde}/\emph{Retrograde} and \emph{Both} are shown in Fig.~\ref{fig:s3_maps_proret} and Fig.~\ref{fig:s3_maps_both} respectively.

These satellite classifications are visualised in Fig.~\ref{fig:s3_skymap}, which colours satellites by the minimum $d_{\mathrm{max}}$ they demonstrate throughout all TVs resulting in realistic orbital distances. It is important to note that the fraction of TVs that produce CASP-like orbits do not necessarily constitute a corresponding probability of plane membership -- hence, whether any TVs resulting in realistic orbital distances permits the satellite to orbit along the CASP (making the satellite \emph{consistent} with CASP membership) is much more relevant than the distribution of plane alignments within the TV sample.

While some satellites can be ruled out as CASP members due to their initial orthogonal distance from the plane, Fig.~\ref{fig:s3_skymap} demonstrates a surprising degree of nuance in identifying satellites consistent and inconsistent with plane membership (as per our criteria in Section~\ref{sec:s2_constraints}). For instance, several of the \emph{Unlikely} satellites appear to be closer to or at the same orthogonal distance to the CASP as those which achieve a minimum $d_{\mathrm{max}}$ smaller than the CASP's plane height.

\begin{table}
	\centering
	\caption{A comparison between \citet{Muller2021coherent}'s predictions of the Centaurus A satellites' orbital senses based on their line-of-sight velocity trend (denoted M+21) and our results. $f(\mathrm{prograde})$ represents the percentage of TVs resulting in realistic orbital distances for which $\theta_0 < 90^{\circ}$ holds, corresponding to prograde orbits with respect to the satellite majority. $f(\mathrm{prograde})_{\mathrm{CASP}}$ additionally requires the orbit to satisfy the plane separation and orbital pole alignment requirements for CASP membership, given in equations (\ref{eq:s2_maxsep}) and (\ref{eq:s2_limalign}) respectively -- blank entries indicate satellites with no TVs consistent with CASP membership, classified as \emph{Unlikely} in Table~\ref{tab:s3_metrics}.}
	\renewcommand{\arraystretch}{1.2}
	\begin{tabular}{llll}
	    \hline
		Name & $f(\mathrm{prograde})$ & $f(\mathrm{prograde})_{\mathrm{CASP}}$ & M+21 \\
		& ($\%$) & ($\%$) \\
		\hline
        ESO269-037 & $94.1\;^{+0}_{-1.5}$ & & Co-rotating \\
        NGC4945 & $59.7\;^{+7.7}_{-9.3}$ & $100.0\;^{+0}_{-100.0}$ & Co-rotating \\
        ESO269-058 & $0\;^{+5.4}_{-0}$ & $0\;^{+0}_{-0}$ & Counter-rotating \\
        KK189 & $38.9\;^{+5.1}_{-0}$ & $28.9\;^{+11.2}_{-0}$ & Counter-rotating \\
        ESO269-066 & $94.7\;^{+5.3}_{-11.1}$ & $0\;^{+100.0}_{-0}$ & Co-rotating \\
        NGC5011C & $56.7\;^{+22.3}_{-5.7}$ & & Counter-rotating \\
        KKs54 & $0\;^{+0}_{-0}$ & $0\;^{+0}_{-0}$ & Counter-rotating \\
        KK196 & $67.9\;^{+8.7}_{-0}$ & $76.2\;^{+23.8}_{-2.4}$ & Co-rotating \\
        NGC5102 & $99.5\;^{+0.5}_{-44.7}$ & $100.0\;^{+0}_{-0}$ & Co-rotating \\
        KK197 & $48.4\;^{+0.6}_{-1.5}$ & $46.3\;^{+1.0}_{-3.1}$ & Co-rotating \\
        KKs55 & $49.8\;^{+0.4}_{-0}$ & $49.3\;^{+0.7}_{-0}$ & Co-rotating \\
        dw1322-39 & $46.1\;^{+0}_{-0.5}$ & $46.3\;^{+0.6}_{-0}$ & Co-rotating \\
        dw1323-40b & $55.5\;^{+26.5}_{-0}$ & $55.6\;^{+35.0}_{-}$ & Co-rotating \\
        dw1323-40a & $82.1\;^{+17.0}_{-44.9}$ & $100.0\;^{+0}_{-100.0}$ & Co-rotating \\
        KK203 & $20.3\;^{+34.1}_{-20.3}$ & $4.8\;^{+18.4}_{-4.8}$ & Counter-rotating \\
        ESO324-024 & $55.4\;^{+18.9}_{-18.4}$ & $57.3\;^{+0}_{-57.3}$ & Co-rotating \\
        NGC5206 & $52.7\;^{+0.7}_{-1.3}$ & $57.3\;^{+2.1}_{-3.0}$ & Co-rotating \\
        NGC5237 & $53.8\;^{+0.2}_{-0.1}$ & $56.7\;^{+0.7}_{-2.0}$ & Co-rotating \\
        NGC5253 & $100.0\;^{+0}_{-0}$ & $100.0\;^{+0}_{-0}$ & Co-rotating \\
        dw1341-43 & $52.1\;^{+0.7}_{-0}$ & & Counter-rotating \\
        KKs57 & $52.5\;^{+34.2}_{-52.5}$ & $0\;^{+54.1}_{-0}$ & Co-rotating \\
        KK211 & $60.0\;^{+34.3}_{-48.8}$ & & Co-rotating \\
        dw1342-43 & $51.9\;^{+0}_{-1.8}$ & $64.7\;^{+0}_{-13.5}$ & Co-rotating \\
        ESO325-011 & $50.6\;^{+0.4}_{-0.3}$ & $45.2\;^{+3.9}_{-45.2}$ & Co-rotating \\
        KKs58 & $79.3\;^{+2.9}_{-0}$ & $96.1\;^{+3.9}_{-3.0}$ & Co-rotating \\
        KK221 & $29.8\;^{+15.8}_{-24.8}$ & & Counter-rotating \\
        ESO383-087 & $100.0\;^{+0}_{-0}$ & $100.0\;^{+0}_{-0}$ & Co-rotating \\
		\hline
	\end{tabular}
	\renewcommand{\arraystretch}{1}
	\label{tab:s3_rotation}
\end{table}

If instead adopting an alternate, stricter plane separation requirement of $1\Delta$, the number of \emph{Unlikely} satellites increases to 9 with the addition of NGC4945, KKs54, dw1322-39, and ESO325-011. These satellites are characterised by highly extended orbits as well as orbital pole alignments below $30^{\circ}$, and are kinematically consistent with the CASP despite appearing to lie outside it at present time. On the other hand, relaxing the required orbital pole alignment to $45^{\circ}$ results in NGC5011C, dw1341-43, and KK211's removal from the set of \emph{Unlikely} satellites, instead being classified as \emph{Both}. These three satellites demonstrate an initial proximity to Centaurus A and may visually appear to participate in the CASP, but have orbits fundamentally misaligned from the satellite plane. The outlying satellites ESO269-037 and KK221 remain as robust non-members regardless of the specific criteria for CASP membership we adopt.

\subsection{Orbital senses of Centaurus A satellites}
\label{sec:s3_senses}

We now explore the relationship between \citet{Muller2021coherent}'s predictions of satellite orbital senses based on their line-of-sight velocity trend and that from our transverse velocity constraints, shown in Table~\ref{tab:s3_rotation}. 
Under $f(\mathrm{prograde})$, we show the fraction of transverse velocities resulting in realistic orbital distances that generate prograde orbits with respect to the satellite majority, regardless of the goodness of their orbital pole alignments with the CASP's minor axis. On the other hand, $f(\mathrm{prograde})_{\mathrm{CASP}}$ denotes the fraction of TVs consistent with CASP membership that result in prograde orbits, and is hence limited to a $30^{\circ}$ alignment or less. Since some TVs can produce either prograde or retrograde orbits depending on the distance realisation adopted, $f(\mathrm{prograde})$ and $f(\mathrm{prograde})_{\mathrm{CASP}}$ do not necessarily sum to unity with their retrograde counterparts. Nevertheless, these flexible TVs are sufficiently rare such that the fraction of retrograde TVs can be adequately approximated by subtracting the corresponding prograde fraction from unity.

In Table~\ref{tab:s3_metrics}, only 5 satellites are constrained to one orbital sense when requiring orbits that are consistent with CASP membership. This drops to 4 satellites when considering all base-map TVs resulting in realistic orbital distances (see Table~\ref{tab:s3_rotation}), and 3 when additionally taking distance uncertainties into account. Consequently, most satellites in the Centaurus A system are consistent with both prograde and retrograde orbits. Visually, prograde and retrograde satellites as classified in Fig.~\ref{fig:s3_skymap} are distributed evenly to the north and south of Centaurus A, in stark contrast to the dramatic satellite line-of-sight velocity trend reported by \citet{Muller2018whirling} and \citet{Muller2021coherent}.

This consistency with both orbital senses is especially strong in a subsample of 14 'neutral' satellites ($40 < f(\mathrm{prograde}) < 60$ per cent) that lack a strongly preferred orbital sense. According to their line-of-sight velocities, however, 11 of these neutral satellites appear to be co-rotating while 3 counter-rotate \citep{Muller2021coherent}. Assuming prograde and retrograde orbits are equally likely, the probability via the binomial coefficient of such a distribution arising is $P(X \geq 11 \,|\, 14) = 2.9$ per cent. While a slight decrease in significance compared to the full satellite population, $P(X \geq 20 \,|\, 27) = 1.0$ per cent, the neutral satellites' line-of-sight velocities appear to imply a degree of co-rotation far more pronounced than that suggested by our transverse velocity constraints.

\begin{figure*}
	\includegraphics[width=0.32\textwidth]{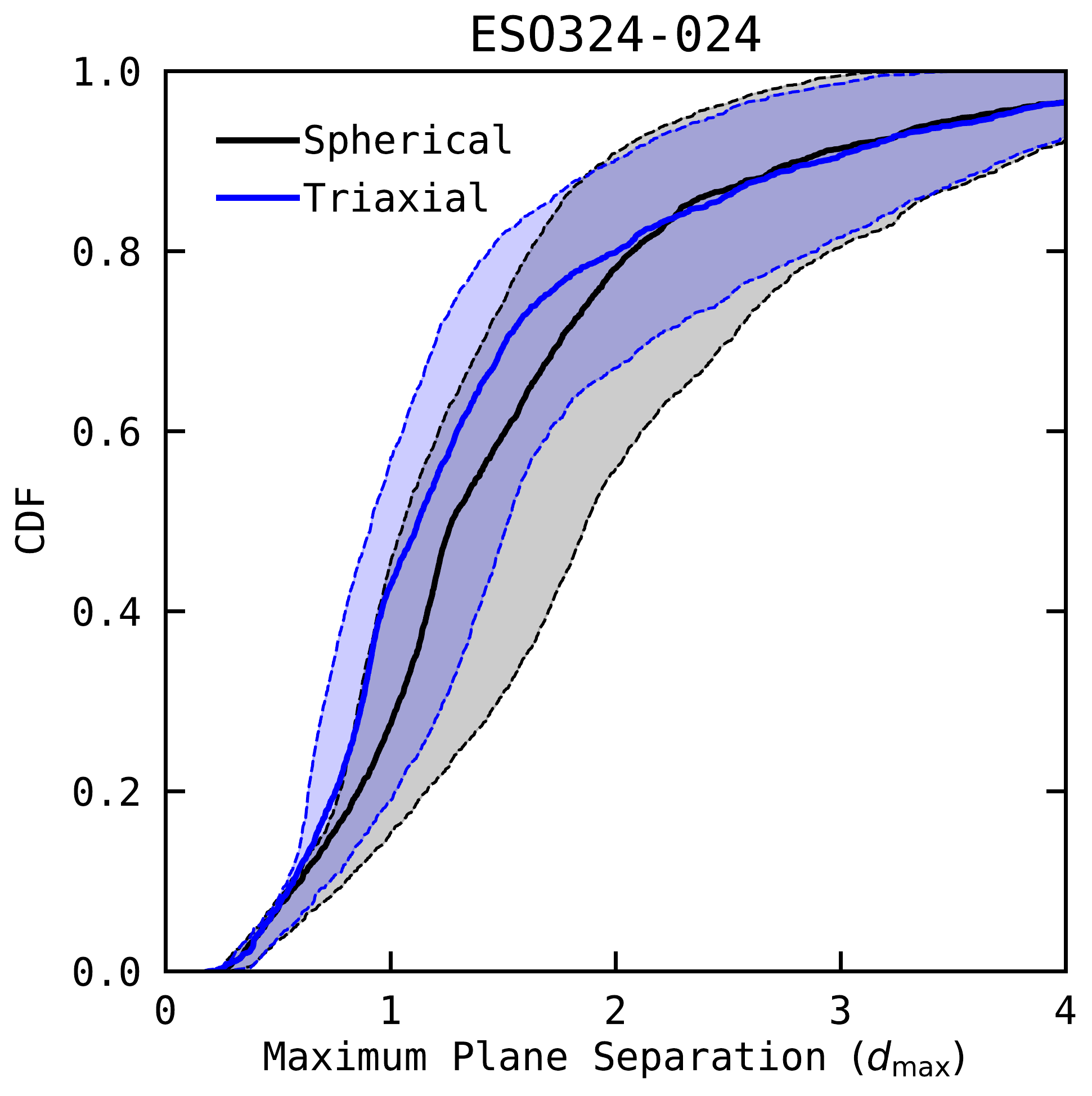}
	\includegraphics[width=0.32\textwidth]{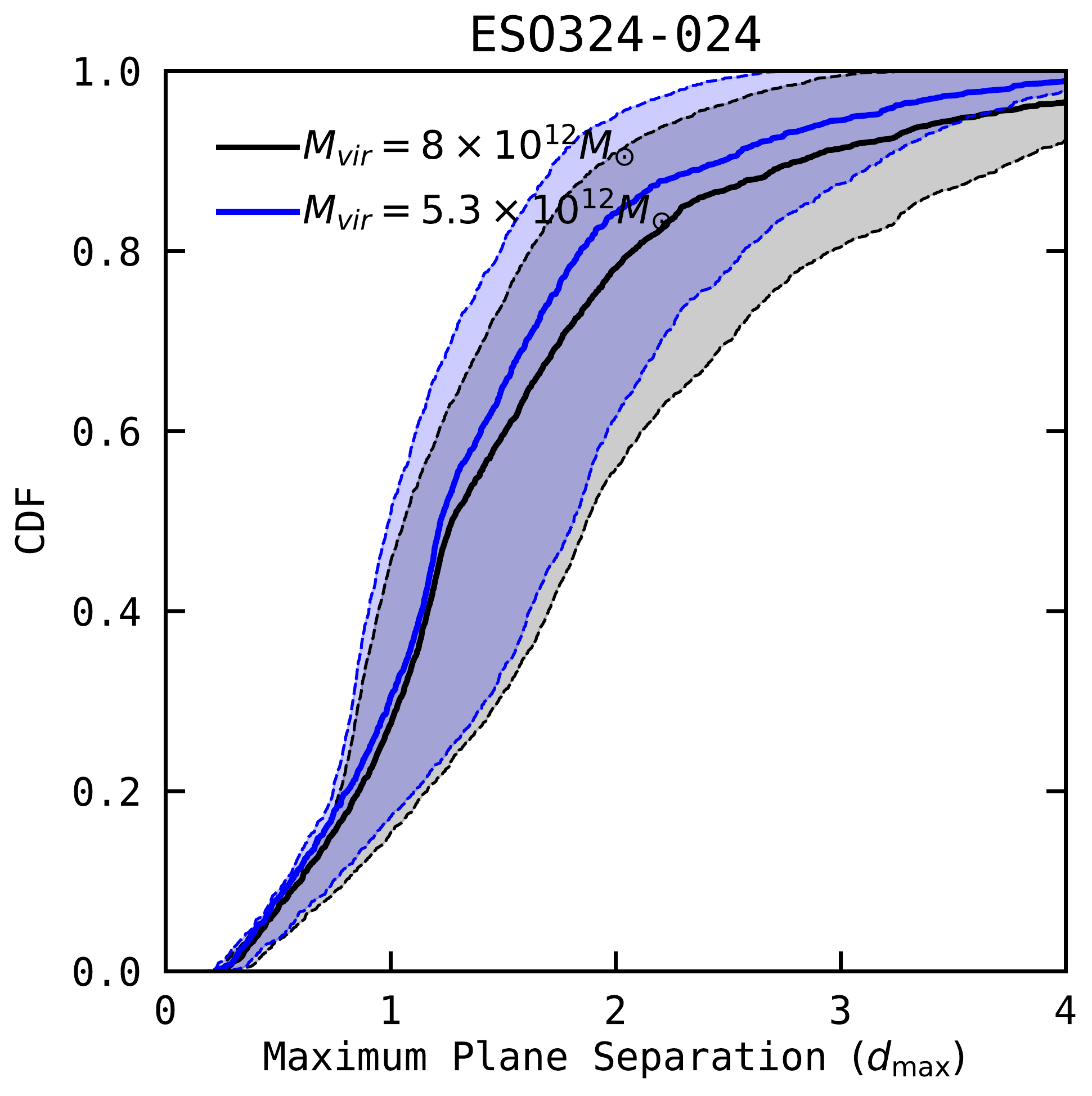}
	\includegraphics[width=0.32\textwidth]{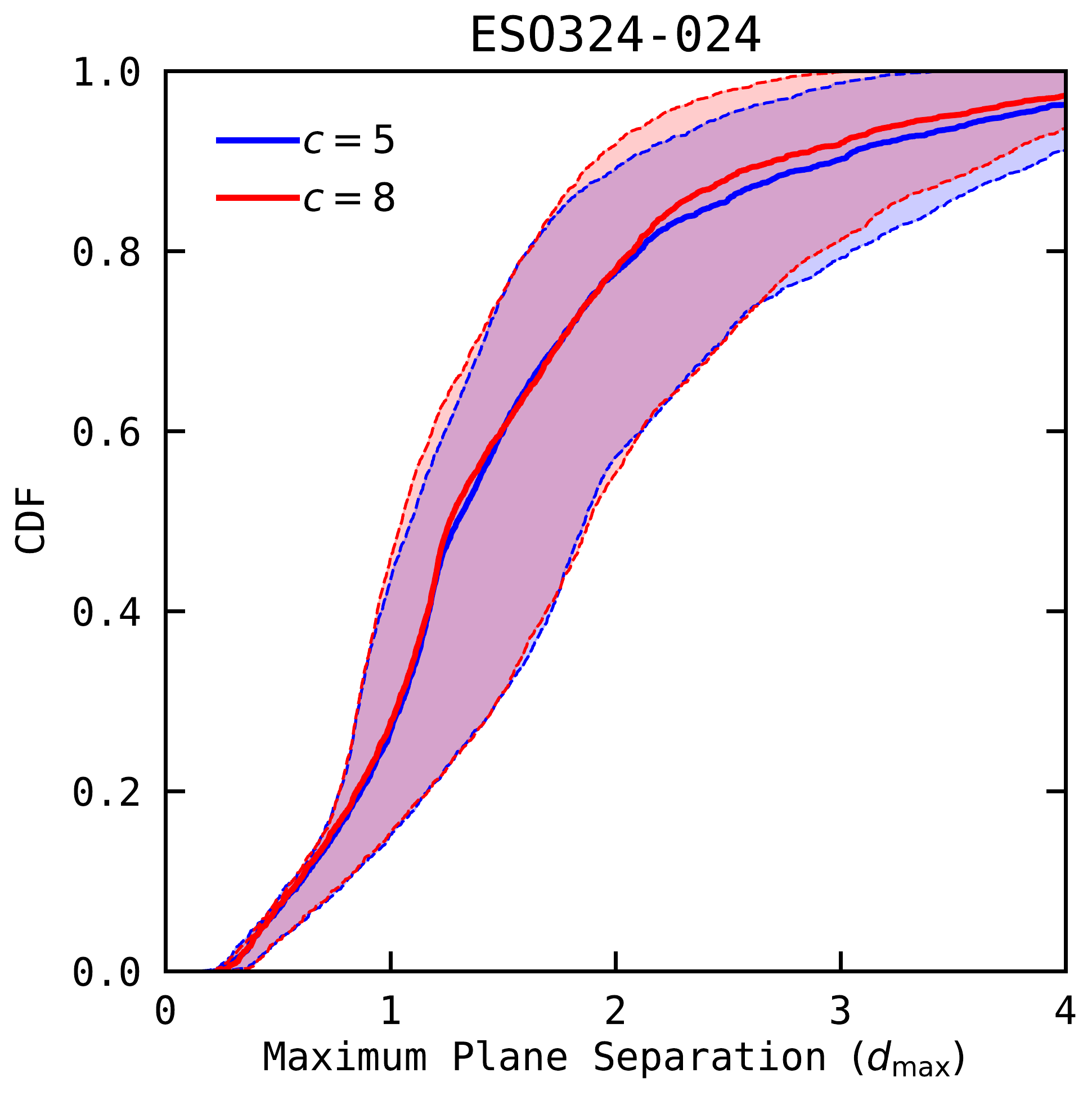}\\
	\includegraphics[width=0.32\textwidth]{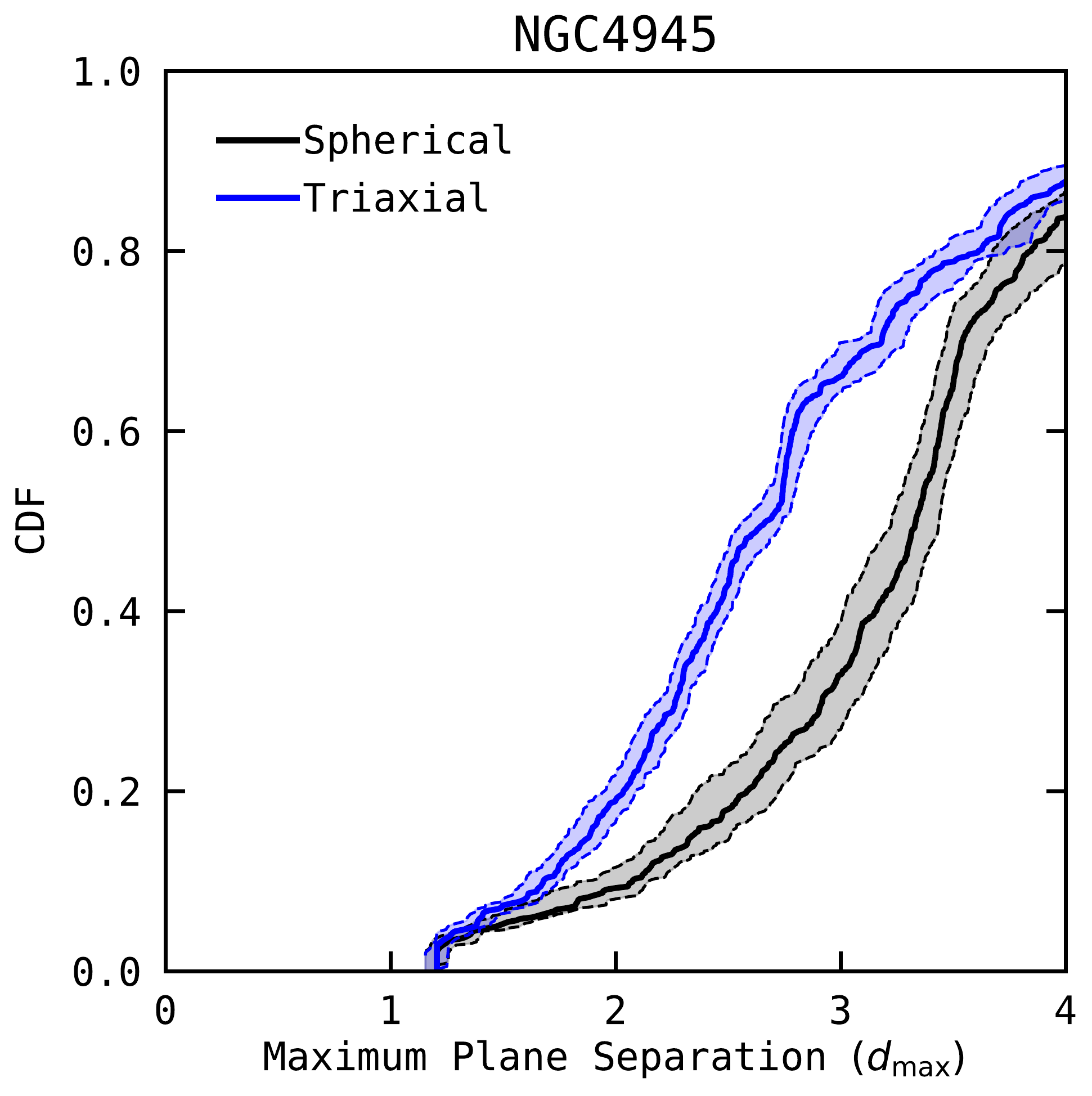}
	\includegraphics[width=0.32\textwidth]{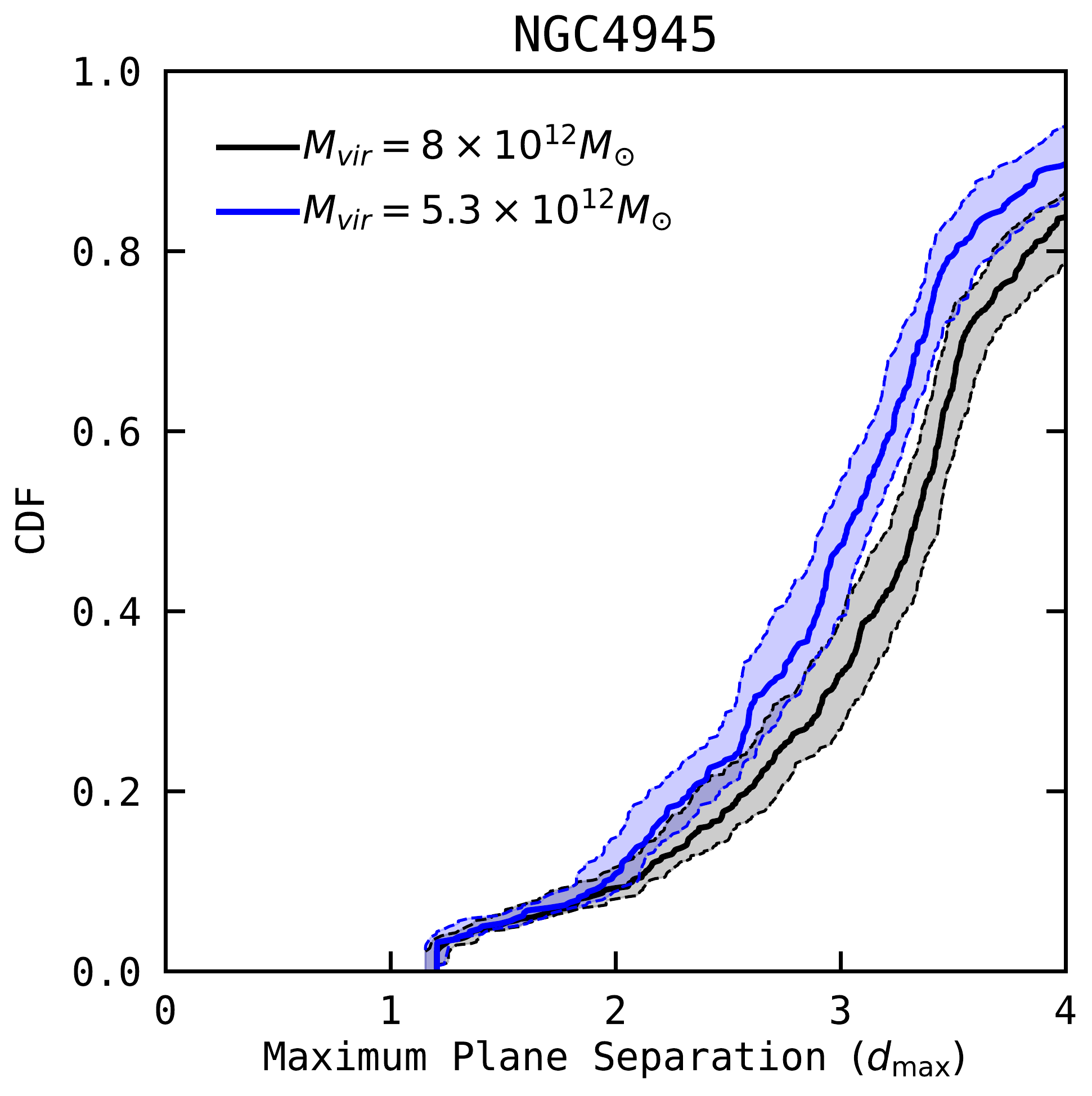}
	\includegraphics[width=0.32\textwidth]{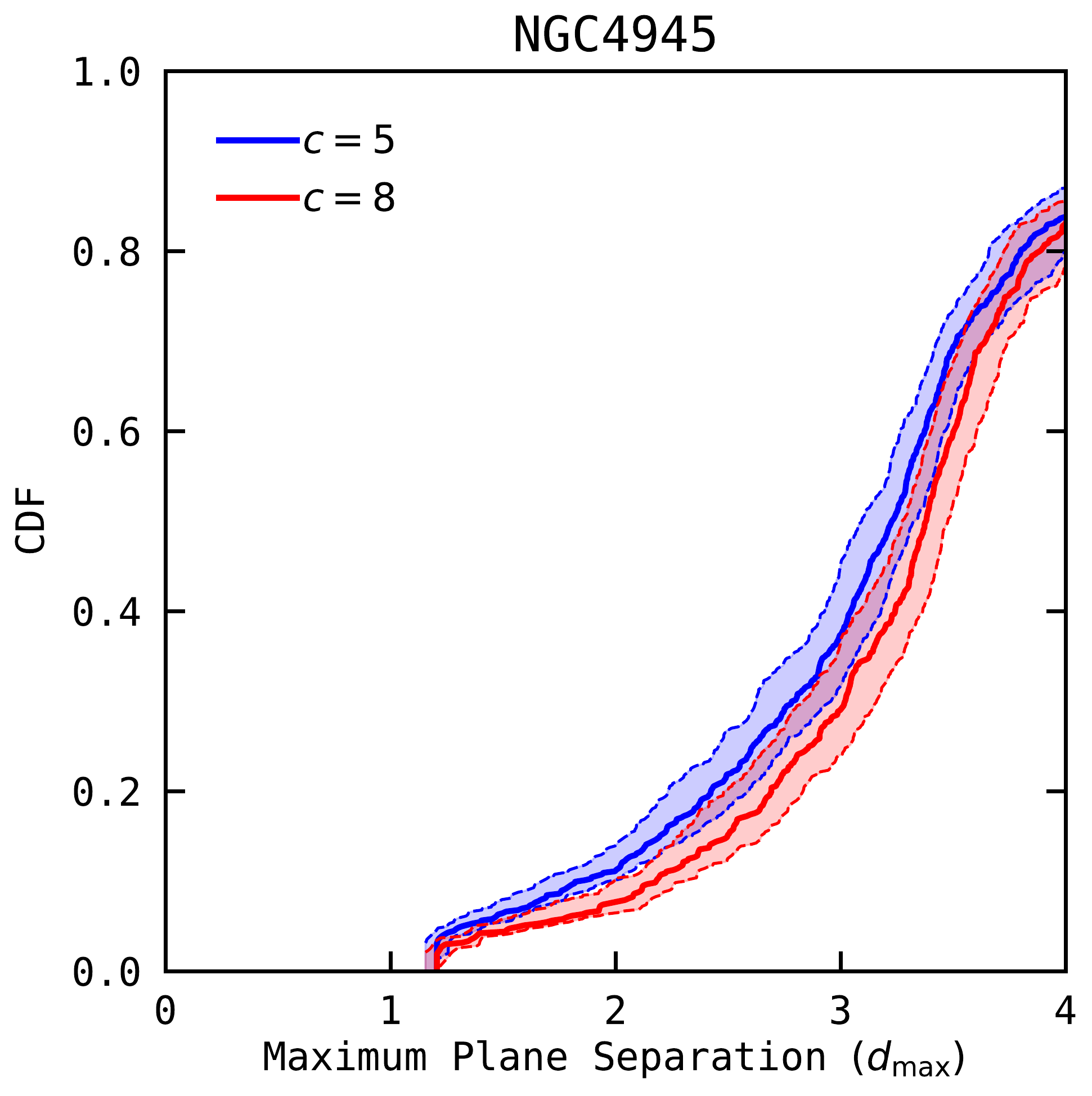}
    \caption{Cumulative distributions of maximum plane separations $d_{\mathrm{max}}$ for all TVs that satisfy the maximum and minimum distance requirements, drawn for ESO324-024 and NGC4945. The solid line corresponds to the base TV constraints (without taking distance errors into account), while the dotted lines represent the maximum and minimum $d_{\mathrm{max}}$ found within 5 distance realisations per TV. From left to right, the plots compare our adopted NFW halo morphology, virial mass, and concentration parameter.}
    \label{fig:s3_cdf}
\end{figure*}

Next, the full sample of 27 satellites is separated into those which tend towards prograde orbits -- characterised by $f(\mathrm{prograde}) > 50$ per cent in our base TV constraints -- and retrograde orbits. $19/27$ satellites are prograde-leaning with a corresponding binomial probability of $P = 2.6$ per cent, compared to $13/22$ satellites with $P = 26.2$ per cent when adopting $f(\mathrm{prograde})_{\mathrm{CASP}}$ as our metric instead. TVs which result in CASP-aligned orbits appear to demonstrate a lower degree of co-rotation. However, note that many satellites demonstrate prograde fractions around $50$ per cent, and are not notably biased towards either orbital sense. When limiting our sample to satellites that show a strong preference for prograde or retrograde orbits as characterised by a $f(\mathrm{prograde})$ below $25$ or above $75$ per cent, $7/10$ satellites tend towards co-rotation -- or $6/9$ satellites with a corresponding probability of $P = 25.4$ per cent when removing \emph{Unlikely} plane members. When using $f(\mathrm{prograde})_{\mathrm{CASP}}$ instead, we obtain a similar result of $7/11$ satellites and $P = 27.4$ per cent. Overall, the Centaurus A system's predicted degree of co-rotation does not notably change when assuming its constituents orbit along the CASP.

We also briefly check whether the satellites' consistency with CASP membership or a given orbital sense is correlated with their morphological type as given in Table~\ref{tab:s3_metrics}. The 5 satellites inconsistent with on-plane orbits consist of a wide variety of morphologies -- as is with the case for potential CASP members -- and we observe no obvious disparity between the populations. Our constraints on satellite orbital senses are minimal, and there appears to be no morphological differences between prograde-only and retrograde-only satellites.

We do feel the need to reiterate, however, that the prograde fraction of a given satellite does not straightforwardly correspond to its probability of orbiting in a prograde sense. Instead, we can only reliably know whether the satellite is consistent with a prograde orbit within the transverse velocities it may have (as per our satellite classifications in Table~\ref{tab:s3_metrics}). 

Interestingly, only 2 out of the 5 satellites classified as \emph{Unlikely} appear to be co-rotating according to their line-of-sight velocities. When limiting our sample to satellites consistent with CASP membership, $18/22$ are argued to be co-rotating in \citet{Muller2021coherent}, with a corresponding binomial probability of $P(X \geq 18 \,|\, 22) = 0.2$ per cent compared to the full sample's $P(X \geq 20 \,|\, 27) = 1.0$ per cent. \emph{Disregarding the 5 off-plane satellites enhances the significance of the CASP's line-of-sight velocity coherence fivefold.}

\section{Robustness with halo parameters}
\label{sec:s4}

We now compare the robustness of our TV constraints and satellite classifications with respect to several key parameters of our adopted halo model -- namely, morphology, virial mass, and concentration. A comparison of cumulative distributions of plane separations $d_{\mathrm{max}}$ for TVs resulting in realistic orbital distances with respect to these parameters are shown in Fig.~\ref{fig:s3_cdf}. We use ESO324-024 and NGC4945 as examples of satellites likely and unlikely to be CASP members respectively.

\subsection{Halo morphology}
\label{sec:s4_robust_shape}

All results in Section~\ref{sec:s3_maps} and \ref{sec:s3_classify} were obtained by assuming a spherically symmetric NFW halo potential. By instead adopting a highly triaxial halo morphology as discussed in Section~\ref{sec:s2_model}, we find that the fraction of TVs resulting in realistic orbital distances that also obey the plane separation and orbital pole alignment requirements is slightly elevated -- though the strength of this increase varies by satellite. The triaxial halo model, by construction, traces the distribution of the 27 satellite galaxies in terms of axis ratios and orientation. This non-spherical potential moves semi-misaligned orbits close to the CASP, as off-plane satellites experience a greater potential gradient towards the satellite plane itself. However, this correcting effect is minimized for heavily misaligned satellites such as KK221, with initial positions nearly orthogonal from the Centaurus A plane -- the triaxial potential appears symmetric when orbiting about one of its defining axes. In the leftmost panels of Fig.~\ref{fig:s3_cdf}, NGC4945 -- a satellite initially located around $1\Delta$ from the CASP -- demonstrates a greater disparity between halo shapes than ESO324-024, which is located closer to the central spherical regime of our triaxial model.

The triaxiality of our alternative potential traces the distribution of luminous satellites perfectly by construction, and thus is more extreme than that expected for most cosmological haloes (see Section~\ref{sec:s2_model}). In spite of this, the adoption of the triaxial halo potential does not modify our satellite classifications, and our results appear to be valid regardless of the halo morphology adopted.

\subsection{Halo mass}
\label{sec:s4_robust_mass}

Thus far, we have taken $M_{\mathrm{vir}} = 8 \times 10^{12} M_{\odot}$ from \citet{Tully2015groups} as our dark halo's mass. Recent work by \citet{Muller2021mass} raised an updated estimate of $M_{\mathrm{vir}} = 5.3 \pm 3.5 \times 10^{12} M_{\odot}$ -- the previous value roughly corresponds to the upper limit in its uncertainty. We now adopt this new mass estimate and a corresponding maximum distance constraint of $d_{\mathrm{max}} < 2\,R_{\mathrm{vir}} = 688\,\mathrm{kpc}$. We find a general decrease in the distribution of maximum plane separations $d_{\mathrm{max}}$ as seen in the middle column in Fig.~\ref{fig:s3_cdf}. This is most likely due to the reduction in the range of TVs which produce under $2R_{\mathrm{vir}}$ orbits -- while the region of TVs that results in $1.5\Delta$ orbits also shrinks, the former effect is still stronger, marginally increasing the $f(d_{\mathrm{max}})$ fraction for the newer mass estimate. However, this also results in KKs54, dw1322-39, NGC5253, and ESO383-087 lacking TVs that allow them to remain within $2R_{\mathrm{vir}}$, and we re-classify them as \emph{Unlikely} plane members. In addition, this reduction removes regions in angular velocity space that enables KKs58 to orbit retrograde within a orbital pole alignment of $30^{\circ}$, and it is thus classified as a \emph{Prograde} satellite.

\subsection{Halo concentration}
\label{sec:s4_robust_conc}

Our adoption of halo concentration $c=6.5$ corresponds to the median concentration for a Centaurus A-mass NFW halo, but a wide range has been found for similar halo masses in CDM simulations \citet{Diemer2019conc}. We test haloes with $c=5$ and $c=8$ in order to determine whether this spread has a significant impact on our results. As seen in the third column of Fig.~\ref{fig:s3_cdf}, the change in $d_{\mathrm{max}}$ distribution is marginal, and whether a higher or lower concentration improves plane separations depends on the satellite in question. The adoption of either concentration does not modify any of the satellite classifications in Table~\ref{tab:s2_data}, and neither does it affect the possible orbital senses for each satellite.

\section{Conclusions}
\label{sec:s5}

The Centaurus A group is one of particular interest when studying correlated systems of satellite galaxies. Its position beyond the Local Group lends it an evolutionary history mostly independent of the Milky Way and M31 planes-of-satellites, while it does not suffer from the lack of reliable distances or radial velocity measurements common in systems at Mpc scales. However, previous analyses \citep{Muller2018whirling, Muller2021coherent} of the Centaurus A satellite plane (CASP) have been limited to that of its projected distribution.

In this work, we have examined the possible kinematics of 27 Centaurus A satellites with TRGB distances and line-of-sight velocity measurements from \citet{Muller2021properties}. Integrating them forward for 5 Gyr around a Centaurus A-like potential, we constrained their transverse velocities in proper motion units under the assumptions of bound, long-lived satellite orbits and a rotationally supported CASP. Using these TV predictions, we further classified satellites by whether they are consistent with being plane members -- remaining within the CASP's plane height and demonstrating a good alignment between their angular momentum vectors and the satellite distribution's minor axis. We identified a set of 5 satellites -- NGC5011C, dw1341-43, KK211, ESO269-037, and KK221 -- as being inconsistent with CASP membership, the latter two of which remain robustly so even when relaxing our criteria. For the remainder of the satellites, we reiterate that the fraction of TVs which result in orbits aligned with the plane does not constitute a probability of CASP membership -- using this methodology, we may only conclude whether a satellite is consistent or inconsistent with participating in the plane.

The line-of-sight velocities of Centaurus A's satellites have been reported to demonstrate a trend indicative of a common co-orbiting motion. The CASP's large fraction of apparently co-rotating satellites is a major contributor to the CASP's statistical rarity in hydrodynamic CDM simulations \citep{Muller2021coherent, Muller2018whirling}. However, by determining the range of possible orbital senses for each satellite, we found that a majority are consistent with both prograde and retrograde orbits. \citet{Muller2018whirling, Muller2021coherent} mock-observed simulated systems to obtain line-of-sight velocities before comparing them to the observed Centaurus A system, such that our comparison has no direct impact on the reported significance of the CASP. Regardless, we demonstrate that the projected veocity trend does not necessarily ensure the system's co-rotation in full phase-space.

On the other hand, the majority of satellites consistent with CASP membership co-rotate according to the velocity trend, while less than half of the off-plane satellites do the same. Removing the 5 off-plane satellites enhances the binomial significance of the remaining galaxies' line-of-sight velocity trend by a factor of 5. This result may possibly hint at an underlying co-rotational motion along the plane-of-satellites, though this is difficult to verify with our orbital alignment constraints alone.

Finally, we evaluated the robustness of our results with respect to our assumed halo potential. Adopting a triaxial halo tracing the luminous satellite distribution improves the satellites' orbital fit to the CASP when compared to our initially spherical potential, but this does not affect our classifications. When testing \citet{Muller2021mass}'s updated virial mass estimate in lieu of the previous result by \citet{Tully2015groups}, weakly bound satellites with an initially high line-of-sight velocity were removed from the sample of CASP member candidates. Our results were stable with varying halo concentration.

In summary, we present a set of robust transverse velocity constraints and corresponding plane membership classifications for Centaurus A's satellites, which are expected to hold as long as our two underlying assumptions of orbit stability and longevity apply. The constraints can also be interpreted as proper motion predictions under the assumption of a rotationally supported CASP. At the distance of Centaurus A, one requires proper motions with associated uncertainties on the order of $5\,\mu\mathrm{as}\,\mathrm{yr}^{-1}$ to place meaningful constraints on its satellite galaxies' orbits. To this end, we searched for data available in the \textit{Mikulski Archive for Space Telescopes} and found that 13 of the 27 satellites listed in Table~\ref{tab:s2_data} have potentially useful HST images that may be used as the first-epoch data. While challenging, future follow-up observations with the HST, JWST, or the Nancy Grace Roman Space Telescope may allow the determination of proper motions to the required accuracies.

\section*{Acknowledgements}

We thank the anonymous referee for their thoughtful and helpful comments. KJK and MSP acknowledge funding via a Leibniz-Junior Research Group (project number J94/2020). MSP also thanks the German Scholars Organization and Klaus Tschira Stiftung for support via a KT Boost Fund. O.M. is grateful to the Swiss National Science Foundation for financial support under the grant number PZ00P2\_202104.

%%%%%%%%%%%%%%%%%%%%%%%%%%%%%%%%%%%%%%%%%%%%%%%%%%
\section*{Data Availability}
 
The data underlying this article will be shared on reasonable request to the corresponding author.



\bibliographystyle{mnras}
\bibliography{bibliography} % if your bibtex file is called example.bib

%\begin{thebibliography}{99}
%\end{thebibliography}

%%%%%%%%%%%%%%%%%%%%%%%%%%%%%%%%%%%%%%%%%%%%%%%%%%

%%%%%%%%%%%%%%%%% APPENDICES %%%%%%%%%%%%%%%%%%%%%

%\appendix



%%%%%%%%%%%%%%%%%%%%%%%%%%%%%%%%%%%%%%%%%%%%%%%%%%

% Don't change these lines
\bsp	% typesetting comment
\label{lastpage}
\end{document}